%% file: georouting.tex
\documentclass{article}

\usepackage{amsthm}
\usepackage{amssymb}
\usepackage{graphicx}
\usepackage[utf8]{inputenc}

\usepackage{subfigure}

\renewcommand{\arraystretch}{1.5}

\newcommand{\englobing}{{\cal E}}

\author{Aubin Jarry}
\title{The Four Principles of Geographic Routing}
\date{April 14, 2014}

\newtheorem{theorem}{Theorem}

\begin{document}
\maketitle

\begin{abstract}
Geographic routing consists in using the position information of nodes to assist in the routing process, and has been a widely studied subject in sensor networks.
One of the outstanding challenges facing geographic routing has been its applicability. Authors either make some broad assumptions on an idealized version of wireless networks which are often unverifiable, or they use costly methods to planarize the communication graph.

The overarching questions that drive us are the following. When, and how should we use geographic routing? Is there a criterion to tell whether a communication network is fit for geographic routing? When exactly does geographic routing make sense?

In this paper we formulate the four principles that define geographic routing and explore their topological consequences. Given a localized communication network, we then define and compute its geographic eccentricity, which measures its fitness for geographic routing. Finally we propose a distributed algorithm that either enables geographic routing on the network or proves that its geographic eccentricity is too high.
\end{abstract}

\section{Introduction}
``When the position of source and destination is known as are the positions of intermediate nodes, this information can be used to assist in the routing process.''  (in: Protocols and Architectures for Wireless Sensor Networks, Holger Karl and Andreas Willig, 2005\cite{bib_book}). The concept of geographic routing, also known as position-based routing, has been an actively studied approach to routing for wireless networks since 1984~\cite{bib_greedy}.
Nonetheless, one of the outstanding challenges that still face geographic routing has been its applicability. Authors either make some broad assumptions on an idealized version of wireless networks (e.g. Unit Disc Graphs~\cite{bib_udg}) which are often at best unverifiable, or they use often costly methods to planarize the communication graph (e.g. CLDP~\cite{bib_cldp} and improvements~\cite{bib_cldp_lazy}). Unfortunately, the cost of planarization can defeat the whole purpose of geographic routing as a lightweight protocol for resource-constrained networks. Frequently, it is required that such preprocessing must work for any connected network where nodes are localized. Obviously, if the positions of the nodes bear no relation to the communication graph topology, geographic routing consists in making purely arbitrary decisions and its cost is bound to be high.

\begin{figure}[h]
\begin{center}
\includegraphics[width=10cm]{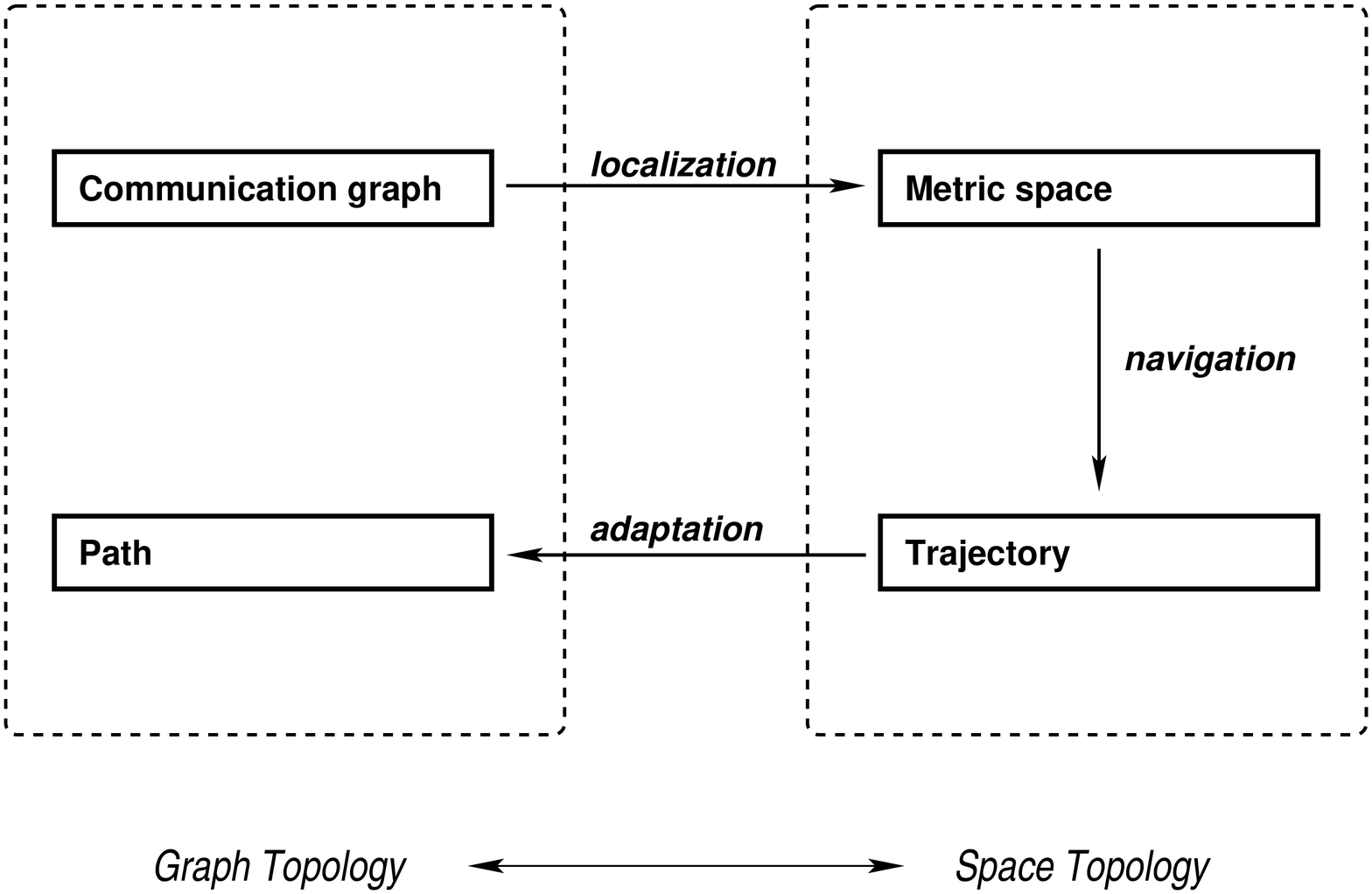}
\caption{Geographic routing framework.}
\label{fig_routing}
\end{center}
\end{figure}

Therefore, the overarching questions that drive us are the following. ``When, and how should we use geographic routing? Is there a criterion to tell whether a communication network is fit for geographic routing? When exactly does geographic routing make sense?''
In fact, geographic routing strategies can be roughly grouped into three categories.
\begin{enumerate}
\item Decrease the distance between the current position and the target position.
\item Use the right hand rule to circumvent obstacles on planar graphs.
\item Use precomputed clues from local topology discovery.
\end{enumerate}
In all cases, some navigation decisions in a continuous metric space $S$ (where nodes are localized) inform path construction in the communication graph $G$. In other words, the actual process of geographic routing uses a {\em navigation engine} that computes a {\em trajectory} (see Figure~\ref{fig_routing}). The implicit assumption is that it is reasonable to put the topology of the communication graph $G$ in relation with the topology of the continuous metric space $S$.

\subsection*{Our contribution}
The originality of our approach is that we start from a generic standpoint on figuring out what are the mechanisms of geographic routing. We are the first to provide a coherent theoretical framework for geographic routing grounded into four core principles. From these principles, we derive properties on the topology of the graph and the metric space that are necessary and sufficient to cater for geographic routing algorithms. We provide a fitness measure, the geographic eccentricity, which measures the cost of applying geographic routing algorithms. This measure is not linked in any way to any particular routing scheme, but depends instead on the four principles.
 This measure can be computed offline, but we also provide a distributed algorithm that attempts to build a geographic enabling of a network. If the algorithm succeeds, then {\bf \em any} geographic strategy can be successfully applied on the network. If it fails, it proves that the geographic eccentricity of the network is too high. Our simulations show that the geographic eccentricity has a linear dependency to localization errors.

\subsection*{Related work}
The subject of routing protocols for wireless sensor networks represents quite an intense research field. Since our work has a potentially large application scope, presenting a fair and balanced view of all the related studies constitutes a substantial task in and of itself that remains well beyond the scope of this paper.
Therefore, we refer the reader to the surveys in~\cite{bib_survey1,bib_survey2} while in this subsection we modestly review some of the approaches that can be seen as precursors to our work.

Some authors explicitly use {\em trajectories} in the continuous metric space. These trajectories are usually computed at the source node, and represent different disjoint ways to reach the destination. The messages then try to follow one of the trajectories in the network. Contrary to our approach, the concept of simulated zone is usually ignored, so the messages have to jump from node position to node position. This line of work was initiated to our knowledge in~\cite{bib_trajectory_old}, followed by a substantial amount of articles from a variety of authors. Recently, the authors in~\cite{bib_trajectory_new} used this approach to improve face routing on planar graphs.

The whole concept of geocasting has made it natural to consider tessellations of the plane such as Voronoi diagrams~\cite{bib_voronoi}, hexagons~\cite{bib_hexagon}, or squares~\cite{bib_square}, where each {\em zone} is attributed to a node.
In the field of mobile networks (including mobile wireless sensor networks), routing management can be done by dividing the network into dynamic clusters, where the clusters are formed according to the position of nodes.
 The routing is thus separated into inter and intra cluster routing.
In our formal framework (see Section~\ref{sec_simul_metric}), we could interpret that each cluster simulates a particular zone of the continuous metric space. The reader can find a description of a recent zone based clustering protocol, as well as recent developments in the field in~\cite{bib_zone} and citations therein. Whereas we do consider Voronoi diagrams, a notable difference with our conception of geographic routing is that we allow for the overlapping of zones instead of imposing a one-to-one correspondence between nodes and positions in the continuous metric space.

The authors in~\cite{bib_embedding} come close to actually formulate the {\em link embedding principle}. In an effort to adapt planar routing methods to non-planar graphs, they consider the line segments corresponding to all the communication links, and then they construct the virtual planar graph where crossings are represented by virtual nodes. The link crossings are simulated by one endpoint of the concerned links. In order to guarantee some performance for their scheme, they formulate the {\em constant intersection closed property} which is different but which nonetheless bears some similarities with the {\em constant spanning ratio principle} (see Subsection~\ref{sub_spanning}). They do not however consider simulating every type of trajectory
 nor measure the value of their constant for a given network.
 
\subsection*{Network Model and Notation}
We model a communication network by a communication graph $G$, with vertex set $V(G)$ and edge set $E(G)$. The communication graph may be directed or undirected. However, generally speaking, distributed routing algorithms do not rely exactly on the communication graph. Instead, each node in the network builds some knowledge on its local neighbors that are one or a few hops away, and how to get to them.  This defines the {\em knowledge graph} $H$, where an edge $uv$ represents the relation ``$u$ knows $v$''. We note $E(H)$ the edge set of $H$ (the vertex set of $H$ is equal to $V(G)$).
In distributed algorithms, it is desirable that the nodes need not know about other nodes that are far away.
The locality of $H$ can thus be expressed as the smallest $k$ such that $H$ is a subgraph of $G^k$, where edges in $G^k$ correspond to paths of length up to $k$ in $G$. An edge in $H$ that is not an edge of $G$ is often called a {\em virtual link} in the literature~\cite{bib_book}.

\begin{figure}[h]
\begin{center}
\includegraphics[width=6cm]{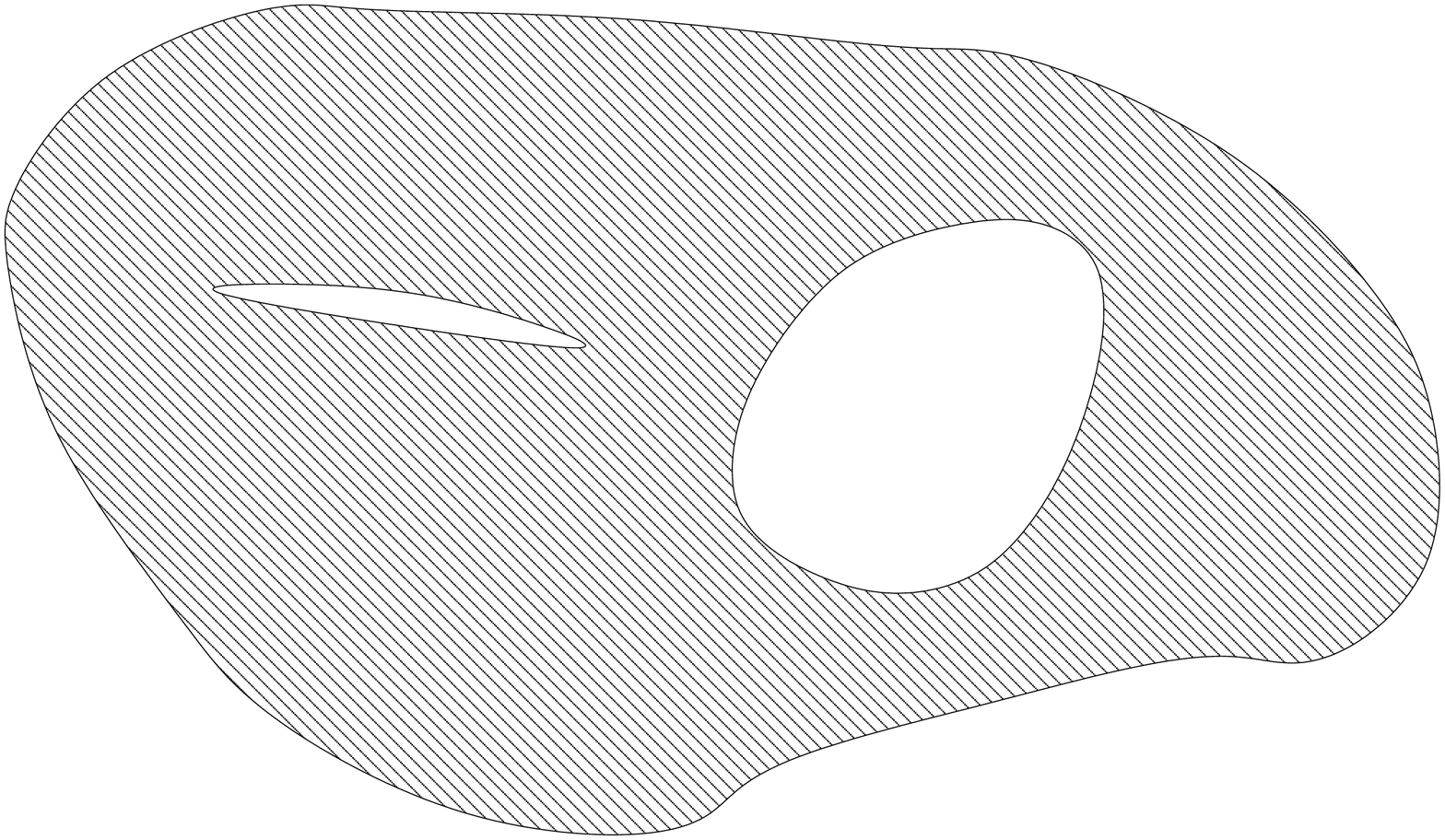}
\caption{A metric space $S$ in the Euclidean plane.}
\label{fig_space}
\end{center}
\end{figure}

In a geographic routing scenario, each node $u$ of the network has a position $p_u$ in a continuous metric space $S$, although the navigation engine doesn't usually have a direct access to the continuous metric space $S$. Instead, what is usually known is an englobing space $\englobing$, where $S\subset\englobing$. The topology of the englobing space $\englobing$ is determined by the known distance function $d:\englobing\times\englobing\rightarrow\mathbb{R}^+$.
The specific topology of $S$, and its distance function $d_S$, can then be discovered as a part of the navigation process. For instance, it is frequently the case that the space $S$ considered is an unknown bounded subset of $\englobing=\mathbb{R}^2$ with a finite number of holes, as illustrated in Figure~\ref{fig_space}.

\begin{figure}[ħtb]
\begin{center}
\includegraphics[width=7cm]{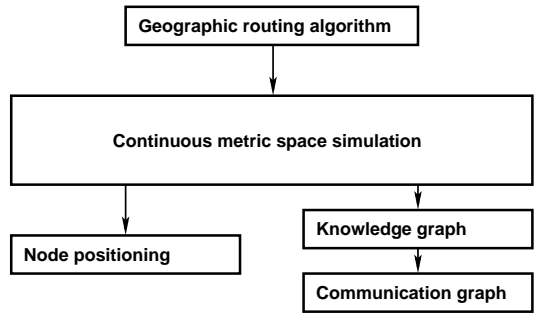}
\caption{The geographic routing stack.}
\label{fig_stack}
\end{center}
\end{figure}

Distributed geographic routing algorithms then rely on a local view of the continuous metric space $S$ to make routing decisions. Figure~\ref{fig_stack} describes the geographic routing stack.

\subsection*{Outline}

How does geographic routing work? In Section~\ref{sec_simul_metric} we postulate that the algorithmic mechanisms of geographic routing derive from two principles, from which we get to the concept of {\em canonical simulation}. From these two principles we define a generic geographic routing algorithm, described in Section~\ref{sec_navigation}, and interpret existing routing strategies. In the light of these principles, we then review in Section~\ref{sec_examples} some classical topological examples.

When does geographic routing make sense? In Section~\ref{sec_criteria} we postulate that the hitherto implicit relation between the continuous metric space and the communication graph can be expressed in the form of two additional principles, from which we get to the concept of {\em geographic eccentricity}. From the four principles we define a distributed algorithm, described in Section~\ref{sec_distributed}, that either enables geographic routing on a given localized network or reveals that its geographic eccentricity is too high. Finally, we carry out in Section~\ref{sec_experiment} geographic eccentricity measurements as well as other topological measurements on computer generated networks following several classical communication models.

\section{Simulating the Metric Space}
\label{sec_simul_metric}
As we have stated earlier, in a geographic routing scenario, the network must be able to simulate a continuous metric space $S$ in order to compute trajectories and infer routing decisions. We first postulate in Subsection~\ref{sub_def} that such a simulation is built around two principles and their fundamental properties. We then remark in Subsection~\ref{sub_voronoi} that all simulations are invariably related in the same way to Voronoi diagrams, and we derive from there the concept of {\em canonical simulation}.

\subsection{Definitions and fundamental properties}
\label{sub_def}
Since we expect distant nodes in the network to communicate via geographic routing, and since the simulation of $S$ occurs as a precondition, the simulation is invariably distributed. We postulate therefore that the metric space $S$ is divided into {\em zones}, and that each node $u$ of the network is responsible for the simulation of a zone $Z_u$. The simulated zones may overlap, and generally do, as we will see later on. Taken together they cover the continuous metric space $S$, so $S\subset(\cup_{u\in V(G)} Z_u)$. The englobing space $\englobing$ may still be larger, with $(\cup_{u\in V(G)} Z_u)\subset\englobing$.

Then, in order to simulate the trajectory $f$ of a message in its zone, each node $u$ must be able to deliver the message to its destination if the trajectory ends in $Z_u$. This is the {\em geocasting principle}. Furthermore, whenever the trajectory exits its zone $Z_u$, each node $u$ must be able to deliver the message to another node simulating the correct neighboring zone. This is the {\em handover principle}.

\paragraph{Geocasting principle.} A node $u$ follows the geocasting principle if for any position $p\in Z_u$, there is a node $v$ such that $d(p,p_v) = \min_{w\in V} \{d(p,p_w)\}$ and such that either $u=v$ or $uv$ is an edge of $H$. In other words, $u$ must be able to transmit a message to the node or to one of the nodes closest to $p$ in $\englobing$. The weaker {\bf specific geocasting principle} consists in choosing instead the node closest to $p$ in $S$, that is $d_S(p,p_v) = \min_{w\in V} \{d_S(p,p_w)\}$. A topological example which follows the specific geocasting principle is discussed in Subsection~\ref{sub_planar}.

\paragraph{Handover principle.} A node $u$ follows the handover principle if for any position $p$ on the boundary between $Z_u$ and $S\backslash Z_u$, there is a node $v$ and an open ball $B$ of $S$ such that $p\in B\subset Z_v$ and such that $uv$ is an edge of $H$.
In other words, $u$ must be able to transmit a message to a node simulating the vicinity of $p$ in $S$.

\begin{figure}[h]
\begin{center}
\includegraphics[width=8cm]{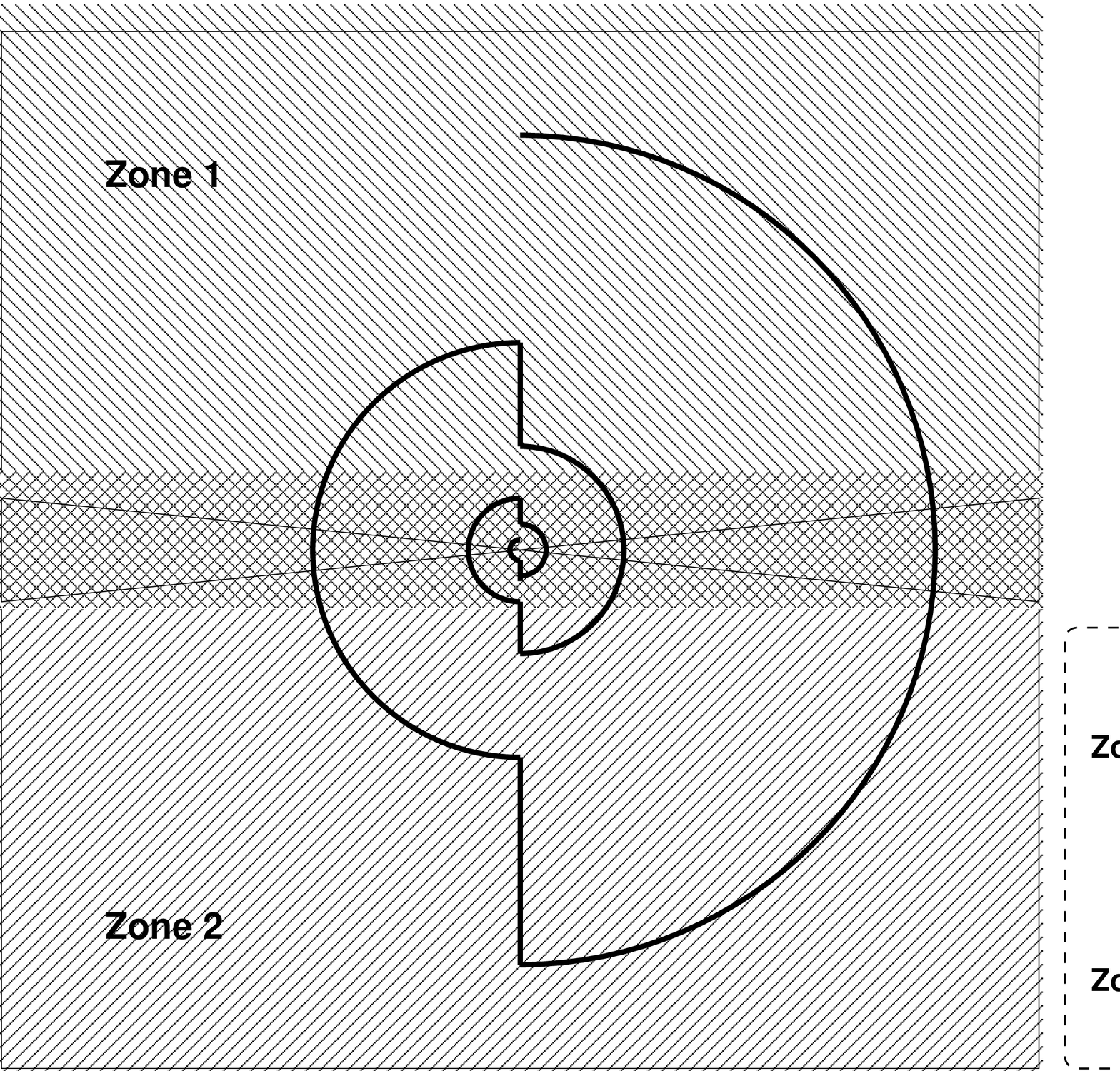}
\caption{Trajectory of length $2\pi+1$ oscillating between two zones.}
\label{fig_oscillation}
\end{center}
\end{figure}

\paragraph{}
Note that the handover principle implies that the different zones must overlap. In a general setting, the overlapping of zones is necessary for two distinct reasons. Firstly, given a message currently at position $p$, the local navigation engine reasonably needs to know what is the vicinity of $p$ to compute a trajectory. Secondly, the zones must overlap to prevent the possible oscillation of a trajectory between several zones, as expressed in Theorem~\ref{th_oscillation} and illustrated in Figure~\ref{fig_oscillation}.

\begin{theorem}
Consider a continuous metric space $S$ and a finite set $\cal Z$ of zones.
The two following propositions are equivalent.
\begin{enumerate}
\item For every position $p$ in $S$, there is an open ball $B$ of $S$ and a zone $Z\in \cal Z$ such that $p\in B\subset Z$.
\item For every continuous function $f:[0,1]\rightarrow S$ there is a finite sequence $\theta_0,\theta_1,\dots,\theta_k$ of real numbers in $[0,1]$ and a finite sequence $Z_0,\dots,Z_k$ of zones in $\cal Z$ such that $0=\theta_0<\theta_1<\dots<\theta_{k-1}<\theta_k=1$, such that for all $i$ in $\{0,\dots,k-1\}$ $f([\theta_i\theta_{i+1}))\subset Z_i$ and such that $f(\theta_k)\in Z_k$.
\end{enumerate}
\label{th_oscillation}
\end{theorem}

\begin{proof}
Suppose that Proposition 1 is true. Consider a continuous function $f:[0,1]\rightarrow S$. We call $P(\theta)$ the proposition {\em ``there is a finite sequence $\theta_0,\theta_1,\dots,\theta_k$ of real numbers in $[0,\theta]$ and a finite sequence $Z_0,\dots,Z_k$ of zones in $\cal Z$ such that $0=\theta_0<\theta_1<\dots<\theta_{k-1}<\theta_k=\theta$, such that for all $i$ in $\{0,\dots,k-1\}$ $f([\theta_i\theta_{i+1}))\subset Z_i$ and such that $f(\theta_k)\in Z_k$.''} We want to prove Proposition 2, which corresponds to $P(1)$. Note that $P(0)$ is true.
Consider a real number $\theta$ in $[0,1)$ such that $P(\theta)$ is true. From Proposition 1, there is an open ball $B$ and a zone $Z$ such that $f(\theta)\in B\subset Z$. Since $f$ is a continuous function, there is a strictly positive number $\epsilon$ such that $f([\theta,\theta+\epsilon])\subset B\subset Z$. Therefore, $P(\theta+\epsilon)$ is true.
Now consider the largest number $\overline \theta$ such that $P(\theta)$ is true for all $\theta$ in $[0,\overline\theta)$. From Proposition 1, there is an open ball $B$ and a zone $Z$ such that $f(\overline\theta)\in B\subset Z$. Since $f$ is a continuous function, there is a strictly positive number $\epsilon$ such that $f([\overline\theta-\epsilon,\overline\theta])\subset B\subset Z$. Therefore, $P(\overline\theta)$ is true. The largest number $\overline\theta$ can only be $1$, which proves Proposition 2.

Suppose that Proposition 1 is false. Let $p$ be a position such that any ball around $p$ can not be contained in a single zone. We call $Z_0, \dots,Z_{|{\cal Z}|-1}$ the zones in $\cal Z$. For any natural number $i$, let $p_i$ be a position at distance less than $\frac{1}{i}$ of $p$ that does not belong to $Z_{i[|{\cal Z}|]}$. We construct $f$ such that for each segment $[1-\frac{1}{i}, 1-\frac{1}{i+1}]$, $f$ is a continuous curve between $p_i$ and $p_{i+1}$ in the ball of center $p$ and radius $\frac{1}{i}$, and such that $f(1)=p$. The function $f$ is continuous and can not be covered by a finite sequence of zones, which disproves Proposition 2.
\end{proof}

Having said that, the trajectories computed by navigation engines are practically always constructed from a finite sequence of shortest curves or straight lines, and are thus piecewise geodesic. In that case, the surface $S$ may be partitioned into convex zones; the nodes of the network must then follow the weaker {\em directional handover principle}. Note that a geodesic curve may only enter and exit once any convex subset of $\englobing$.

\label{sub_directional}
\paragraph{Directional handover principle.} A node $u$ follows the directional handover principle if the zone $Z_u$ is a finite union of convex subsets of $\englobing$ and
if for any position $p$ on the boundary between $Z_u$ and $S\backslash Z_u$, there is an open ball $B$ of $S$ such that $p\in B\subset(Z_u\bigcup (\cup_{uv\in E(H)} (Z_v)))$.
In other words, $u$ and its neighbors must collectively simulate the vicinity of $p$ in $S$.
\paragraph{}

As a final remark, we can see that if a network $\cal N$ follows the directional handover principle, then we may consider in the same settings the network $\cal N'$ with knowledge graph $H'=H^2$ and simulated zones $Z'_u=Z_u\bigcup (\cup_{uv\in E(H)} (Z_v))$ for all $u\in V(G)$. Observe that the network $\cal N'$ follows the handover principle.

\subsection{Voronoi diagram and canonical simulation}
\label{sub_voronoi}
Consider a Voronoi diagram of the englobing space $\englobing$ by the positions $\{p_u\}_{u\in V}$. We call $C_u$ the Voronoi cell of $p_u$ for each node $u$. The geocasting principle means that any node simulating a position $p$ of $S$ in the interior of a Voronoi cell $C_u$ (i.e. $p\in C_u^\circ\cap S$) must be $u$ or one of its neighbors in $H$, as stated in Theorem~\ref{th_voronoi1}. By additionally taking the handover principle into account, we may bound the distance between the nodes of two Voronoi cells that are adjacent in $S$, as stated in Theorem~\ref{th_voronoi}.

\begin{theorem}
Consider a network $\cal N$ simulating a metric space $S$. For all nodes $u$ we have $Z_u \subset C_u\bigcup (\cup_{uv\in E(H)} (C_v))$.
\label{th_voronoi1}
\end{theorem}
\begin{proof}
The geocasting principle means that if the simulated zone $Z_u$ of a node $u$ contains a position $p$, then either $p\in C_u$ or there is an edge $uv\in H$ such that $p\in C_v$. Therefore $Z_u\subset C_u\bigcup (\cup_{uv\in E(H)} (C_v))$. 
\end{proof}

\begin{theorem}
Consider a network $\cal N$ simulating a metric space $S$. Consider two Voronoi cells $C_u$ and $C_v$ that are adjacent in $S$. Then $u$ and $v$ are at distance at most $2$ in $H$ (at distance $3$ in case of directional handover).
\label{th_voronoi}
\end{theorem}

\begin{proof}
Consider a position $p$ on the boundary of both $C_u$ and $C_v$ in $S$, so that $p\in(\overline{C_u^\circ\cap S})\cap(\overline{C_v^\circ\cap S}$). Consider an infinite sequence $\cal P$ of positions in $C_u^\circ\cap S$ that converges to $p$. Since there is a finite number of zones, there is a zone $Z_w$ that contains an infinite subsequence of $\cal P$. 
If there is an open ball $B$ of $S$ such that $p\in B\subset Z_w$ then the node $w$ simulates a position of $C_u^\circ\cap S$ and a position of $C_v^\circ\cap S$, so because of the geocasting principle $w$ is neighbor of both $u$ and $v$ in $H$. Therefore, the distance between $u$ and $v$ in $H$ is at most $2$.
Otherwise, $p$ is on the boundary between $Z_w$ and $S\backslash Z_w$.

The handover principle tells that there is an open ball $B$ of $S$ and a zone $Z_{w'}$ such that $p\in B\subset Z_{w'}$. Then, as before, the distance between $u$ and $v$ in $H$ is at most $2$.

In case of directional handover, there is an open ball $B$ of $S$ such that $p\in B\subset(Z_w\bigcup (\cup_{ww'\in E(H)} (Z_{w'})))$. This means that the node $w$ is at distance at most $1$ of $u$ and $2$ of $v$ in $H$. Therefore the distance between $u$ and $v$ in $H$ is at most $3$.
\end{proof}

It is then natural to consider the {\em canonical simulation of $S$} where the knowledge graph $H$ contains the edge $uv$ for every pair  $(C_u,C_v)$ of Voronoi cells adjacent in $S$, and where each node $u$ simulates its own Voronoi cell in addition to the Voronoi cells of all its neighbors in $H$, so $Z_u = C_u\bigcup (\cup_{uv\in E(H)} (C_v))$. 
Observe that each node $u$ follows the geocasting principle. Each node $u$ follows the handover principle by handing a message at position $p$ over to a neighbor $v$ such that $p\in C_v$.

A consequence of Theorem~\ref{th_voronoi} is that if there is any simulation of the continuous metric space $S$ with knowledge graph $H$, then we can build a canonical simulation of $S$ with knowledge graph $H^2$ (or $H^3$ in case the original simulation had directional handover). We can thus define the {\em canonical simulation with respect to H}.

\paragraph{Canonical simulation w.r.t. H} Consider a Voronoi diagram of the englobing space $\englobing$ by the positions $\{p_u\}$, where $C_u$ is the cell corresponding to $p_u$ for each node $u$. The canonical simulation with respect to $H$ consists in defining the simulated zone $Z_u$ for each node $u$ so that $Z_u = C_u\bigcup (\cup_{uv\in E(H)} (C_v))$. In this context, the simulated continuous metric space $S$ is a subset of $\englobing$ such that for each pair $u,v$ of nodes, the cells $C_u\cap S$ and $C_v\cap S$ are adjacent if and only if $uv\in E(H)$.

\begin{figure}[htb]
\begin{center}
\includegraphics[width=6cm]{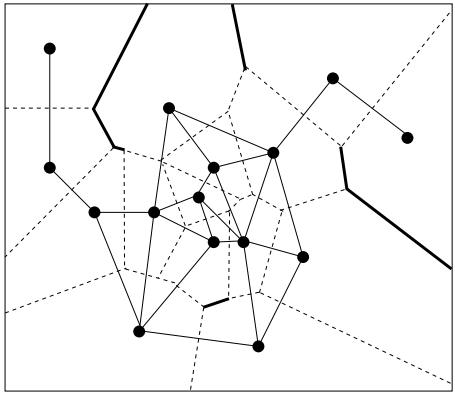}
\caption{Delaunay triangulation and Voronoi diagram of the continuous metric space. The bold boundaries must be removed from $S$ in order to comply with the handover principle.}
\label{fig_delaunay}
\end{center}
\end{figure}

\paragraph{}
Given the knowledge graph $H$, we know from Theorem~\ref{th_voronoi1} that in any simulation following the geocasting principle the simulated zones will be included in the zones of the canonical simulation. We also know that in any simulation following the handover principle, the continuous metric space $S$ may not contain any open ball around a position on the boundary of two Voronoi cells $C_u$, $C_v$ where $uv\notin E(H)$, which means that the Delaunay triangulation of $S$ is a subgraph of $H$.
 As we will see in Section~\ref{sec_navigation}, a desirable property of $S$ is that it is a closed subset of $\englobing$. Therefore, we consider that the continuous metric space $S$ in a canonical simulation is obtained from $\englobing$ by removing an open set of arbitrary small width $\epsilon$ around each boundary between two cells $C_u,C_v$ where $uv\notin E(H)$, as illustrated in Figure~\ref{fig_delaunay}.

\section{Navigation and Routing}
\label{sec_navigation}
In this paper we make a clear distinction between navigation in $S$ on the one hand, and routing in $G$ on the other hand. We postulate that the trajectory in $S$ is continuous and may be described by a continuous (and preferably derivable) function $f:[0,1]\rightarrow S$.
In this section, we first describe the generic adaptation of a navigation engine into a geographic routing algorithm in Subsection~\ref{sub_scheme}. We then examine the navigation engines that exist behind some well known geographic routing algorithms, in Subsection~\ref{sub_greedy} and Subsection~\ref{sub_perimeter}.
Note that a large variety of geographic routing algorithms currently exist; the interested reader may find a comprehensive view in one of the surveys~\cite{bib_survey1,bib_survey2}. The small number of algorithms we take as typical examples in this section should not be taken as an authoritative selection.

\subsection{Generic geographic routing scheme}
\label{sub_scheme}
The function of the navigation engine is to compute a trajectory from the starting position $p_s$ of the message towards its final destination $p_t$. Note that the position of the message is in the continuous metric space $S$, and as such is a {\em virtual position}. In the real world, messages are located in nodes and in the wireless medium. Whether the nodes themselves have real-world coordinates is irrelevant for us as long as they have a position in a continuous metric space. The position $p$ of the message doesn't have to correspond to the position of a node, and most of the time it can't, because its trajectory is a continuous curve in $S$, described by the function $f:[0,1]\rightarrow S$. If the function $f$ is derivable, the trajectory's direction at position $f(\theta)$ may be given by the derivative $\frac{\partial f}{\partial\theta}(\theta)$. This direction is necessary in case of directional handover (see Subsection~\ref{sub_directional}).

The navigation engine doesn't have to compute the whole trajectory at once, because the space $S$ is generally not globally known. Instead, it may gradually compute the trajectory from the current position $p$ using local topological information as well as some global properties of the englobing metric space $\englobing$ (see Table~\ref{tab_navigation}).

\begin{table}[h]
\begin{center}
\framebox{
  \parbox{0.9\linewidth}{
\small
~\\
\centerline{\bf Navigation Engine API}
~\\
{\bf Input:} $(p,p_t,I)$, where $p$ is the current position on the trajectory, $p_t$ is the target position,
and $I$ is some extra information optionally needed for the navigation.\\
~\\
{\bf Output:} $(p',I')$ (or $(p',\vec{\delta},I')$ in case of directional handover), where either $p'=p_t\in Z_u$ or $p'$ is on the boundary between $Z_u$ and $S\backslash Z_u$ (and $\vec{\delta}$ gives the direction of the trajectory at $p'$).\\
  }
}
\end{center}
\caption{Navigation engine API at node $u$.}
\label{tab_navigation}
\end{table}

When a network correctly simulates a continuous metric space $S$ and has a navigation engine, the actual routing algorithm can then be simply inferred from the engine as follows. Whenever a node $u$ has a message at position $p$, it uses the navigation engine to compute its trajectory until the message reaches its final destination or until it leaves the simulated zone $Z_u$. Depending on the outcome, the node then sends the message to its final recipient (geocasting), or to another node simulating an adjacent zone (handover). In any case the message is sent to a neighbor $v$ of $u$ in $H$, which means that the node $u$ uses local topological information of the communication graph $G$ to actually send the message. Table~\ref{tab_routing} formally describes the generic geographic routing algorithm.

\begin{table}[htb]
\begin{center}
\framebox{
  \parbox{0.9\linewidth}{
\small
~\\
\centerline{\bf Generic Geographic Routing Algorithm}
~\\
{\bf Input:} $(p,p_t,I,M)$, where $p$ is the current position of the message in $Z_u$, $p_t$ is the final destination of the message, $I$ is some extra information required by the navigation engine, and $M$ is the message payload.

\begin{enumerate}
\item If $p_t\in Z_u$ then {\bf send} the message $M$ to its final destination using the geocasting principle, and {\bf exit}.
\item Use the navigation engine with input $(p,p_t,I)$ to compute $(p',I')$ (resp. $(p',\vec{\delta},I')$ in case of directional handover).
\item Select the neighbor $v$ such that an open ball of $S$ containing $p'$ (resp. $p'+\epsilon\vec{\delta})$ is included in $Z_v$ by using the handover principle.
\item {\bf Send} $(p',p_t,I',M)$ (resp. $(p'+\epsilon\vec{\delta},p_t,I',M)$) to neighbor~$v$.
\end{enumerate}
}
}
\end{center}
\caption{Generic geographic routing algorithm at node $u$.}
\label{tab_routing}
\end{table}

\begin{figure}[htb]
\begin{center}
\includegraphics[width=5cm]{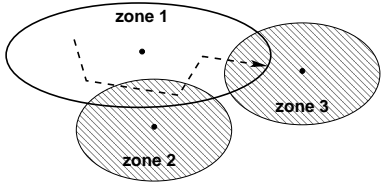}
\caption{Trajectory prone to virtual hops.}
\label{fig_virtual_hop}
\end{center}
\end{figure}

It is interesting to observe that the handover takes place only when the trajectory reaches the boundary of the current simulated zone. In many cases, this means that the trajectory may go near one or several neighbors before reaching the handover position (see Figure~\ref{fig_virtual_hop}). In the literature, where the distinction is not made between routing in $G$ and navigating in $S$, this effect can occur in the form of two consecutive hops $uv$ and $vw$, where both $v$ and $w$ are neighbors of $u$. Routing optimization demands the message to be directly sent from $u$ to $w$, and the missing hop through $v$ is called {\em virtual hop}. This situation occurs most notably in the cases of compass routing~\cite{bib_compass} and face routing~\cite{bib_face}.

\subsection{Steepest gradient and greedy navigation}
\label{sub_greedy}

\begin{figure}[h]
\begin{center}
\includegraphics[width=6cm]{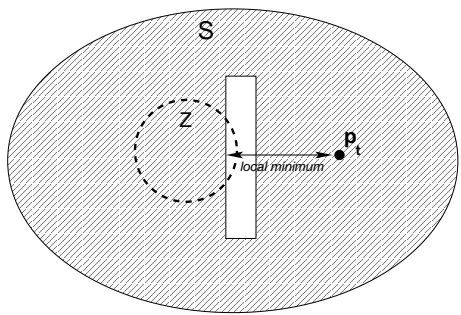}
\caption{Local minimum for the distance to $p_t$, which is reached on the boundary of $S$ in $\englobing$.}
\label{fig_lower_bound}
\end{center}
\end{figure}

In practically all applications, the englobing space $\englobing$ -- whose properties are known to the navigation engine -- is a convex subset of a Euclidean space of finite dimension $\mathbb{R}^k$ (often with $k=2$), and $S$ is a closed subset of $\englobing$. In this case the distance between the positions $p$ and $p_t$ in $\englobing$ can be immediately computed from the positions themselves.
The most obvious navigation method consists then in assuming that the continuous metric space $S$ is similar to the englobing space $\englobing$, and that reducing the distance to destination in $\englobing$ reduces it in $S$. 
Two navigation engines are built around this idea. The {\em steepest gradient navigation engine} aims at reducing the distance to destination in $\englobing$ as soon as possible, and thus produces a trajectory which is a steepest gradient curve in $S$, often along the straight line $pp_t$ in $\englobing$. The {\em greedy navigation engine} aims at reducing the distance to destination in $\englobing$ to the minimum in the simulated zone.
 Obviously, the trajectory may stop prematurely with both engines whenever there is a local minimum, which corresponds to a position on the boundary of $S$ in $\englobing$, as illustrated in Figure~\ref{fig_lower_bound}. In the literature this situation is called the {\em local minimum problem} or the {\em dead end problem}~\cite{bib_book}.
 In order to overcome this situation, other engines must be used, such as those described later on. Of the two engines discussed here, the steepest gradient navigation engine described in Table~\ref{tab_gradient} is the purest, since the chosen trajectory does not depend on the choice of the simulated zone. On the other hand, the greedy navigation engine may possibly avoid some of the local minima.

In the literature, the distinction used to hardly ever be made between routing in $G$ and navigation in $S$. Because of this, the position $p$ of the message had to jump to a node position $p_v$ at each hop $uv$. For instance, the greedy navigation engine is implemented in the greedy routing algorithm~\cite{bib_greedy}, where the distance between $p_v$ and $p_t$ in $\englobing$ is taken as a heuristic for the distance in the communication graph $G$ with no further consideration for the continuous metric space $S$.
In the more sophisticated compass routing algorithm~\cite{bib_compass}, the line segment $[p_u,p_v]$ is chosen as close as possible to the line segment $[p_u,p_t]$. This latter algorithm, if not for the requirement that the trajectory has to pass through the position of individual nodes, comes close to actually simulate a steepest gradient trajectory.

\begin{table}[h]
\begin{center}
\framebox{
  \parbox{0.9\linewidth}{\small
~\\
\centerline{\bf Steepest Gradient Navigation Engine}
~\\
{\bf Input:} $(p,p_t,\emptyset)$, where $p$ is the current position on the trajectory and $p_t$ is the target position.
\begin{enumerate}
\item Compute a steepest gradient curve in $S$ starting from $p$ for the distance to $p_t$ in $\englobing$ function. The curve is composed of straight lines and parts of the boundary of $S$ in $\englobing$.
\item If the curve ends in $p_t\in Z_u$ then {\bf output} $(p_t,\emptyset)$.
\item If the curve ends in $p$ on the boundary between $Z_u$ and $S\backslash Z_u$ then {\bf output} $(p,\emptyset)$.
\item If the curve ends in $p$ on the boundary of $S$ in $\englobing$ then {\bf fail} with output $(p,-1)$.
\end{enumerate}
}
}
\end{center}
\caption{Steepest gradient navigation engine at node $u$.}
\label{tab_gradient}
\end{table}

As stated in the first paragraph, neither the steepest gradient nor the greedy navigation engine can proceed if the trajectory leads to a local minimum on the boundary of $S$. There are three types of approach to deal with this problem, which are not necessarily mutually exclusive. The first one is to switch to another navigation engine (see for instance Subsection~\ref{sub_perimeter}). The second is to build up enough information on the continuous metric space $S$ so as to completely avoid the dead end. This type of approach has been taken for instance in the {\em Routing with Obstacle Avoidance (ROAM)} algorithms~\cite{bib_roam}. The third way is to build enough information on the communication graph $G$ to get out of the dead end by increasing $H$ and the size of the simulated zones. For instance the {\em Greedy Distributed Spanning Tree Routing (GDSTR)} algorithm~\cite{bib_gdstr} builds spanning trees in dead end zones.

\subsection{Perimeter navigation}
\label{sub_perimeter}

When the englobing space is the Euclidean plane $\mathbb{R}^2$ and when the continuous metric space $S\subset\mathbb{R}^2$ is closed, connected and bounded, then it is possible to use the right hand rule in order to overcome the local minimum problem. Imagine there is a position $p$ on the boundary of $S$ in $\mathbb{R}^2$ such that $p$ is a local minimum for the distance to a destination $p_t$, as illustrated in Figure~\ref{fig_lower_bound}. There is a point $p'$ in $\mathbb{R}^2\backslash S$ and in the vicinity of $p$ in $\mathbb{R}^2$ which is closer to the destination $p_t$. We call {\em hole} the component of $\mathbb{R}^2 \backslash S$ containing $p'$.
 Consider the line segment $[p',p_t]$. Since $p_t$ is in $S$ and since $S$ is connected, the boundary of the hole intersects $[p',p_t]$ at some position $p''$, with $d(p'',p_t)<d(p',p_t)<d(p,p_t)$. Since $S$ is bounded, following the boundary of the hole in any direction leads to a position closer to $p_t$ than the previous local lower bound.

Perimeter navigation consists in choosing a direction and in following the boundary of a hole accordingly. As soon as the boundary of $S$ is of bounded length, the combination of gradient and perimeter navigation gives a sure and efficient way to reach the destination, as described in Table~\ref{tab_perimeter}.
In the literature, perimeter navigation is mostly used along with planar graphs, and has been implemented in Face Routing. The combined Gradient/Perimeter navigation engine is embodied in the {\em Greedy Perimeter Stateless Routing (GPSR)} algorithm~\cite{bib_gpsr} (or equivalently, {\em Greedy-Face-Greedy (GFG)}~\cite{bib_gfg}), which also relies on a planarized communication graph.

\begin{table}[h]
\begin{center}
\framebox{
  \parbox{0.9\linewidth}{\small
~\\
\centerline{\bf Gradient/Perimeter Navigation Engine}
~\\
{\bf Input:} $(p,p_t,d_o)$, where $p$ is the current position on the trajectory, $p_t$ is the target position and $d_o$ is a distance value.

\begin{enumerate}
\item If $d_o$ is not defined, then replace $d_o$ with $d(p,p_t)$.
\item If $d(p,p_t)\leq d_o$, then replace $p$ with the position computed by the steepest gradient navigation engine (see Table~\ref{tab_gradient}). If the steepest gradient navigation engine has ended successfully, then {\bf output} $(p,d_o)$, otherwise replace $d_o$ with $d(p,p_t)$.
\item Compute the trajectory that follows the boundary of $S$ in $\englobing$ starting at $p$ with the hole $\englobing\backslash S$ on the right hand side.
\item If the trajectory contains a position $p'$ such that $d(p',p_t)<d_o$ then replace $p$ with $p'$ and {\bf goto} 2; otherwise the trajectory reaches the boundary between $Z_u$ and $S\backslash Z_u$ at position $p''$: {\bf output} $(p'',d_o)$.
\end{enumerate}
}
}
\end{center}
\caption{Gradient/perimeter navigation engine at node $u$.}
\label{tab_perimeter}
\end{table}

\section{Topological Examples}
\label{sec_examples}
In this section, we review some classical assumptions on the communication graph with respect to the node positions, and we interpret them in our framework.

\subsection{Planar graphs}
\label{sub_planar}
Many works on geographic routing refer to planar graphs because of the Face Routing algorithm, which has been one of the rare ways to guarantee end-to-end routing success. In the context of geographic routing, graph planarity is usually tied to the position of nodes in the plane $\mathbb{R}^2$, which means that for all pairs of distinct edges ($uv$, $u'v'$) the open line segments $(p_u,p_v)$ and $(p_{u'},p_{v'})$ do not intersect.

\begin{figure}[h]
\begin{center}
\includegraphics[width=5cm]{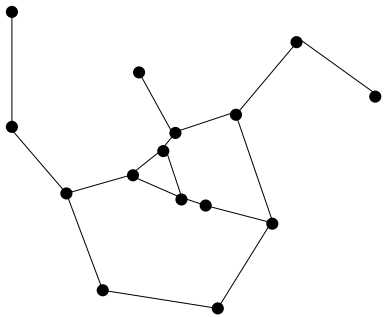}
\caption{Metric space simulated by a planar graph.}
\label{fig_planar_graph}
\end{center}
\end{figure}

In this setup, we consider the knowledge graph $H$ to be equal to the planar communication graph $G$, the simulated zones $Z_u=\{p_u\} \bigcup \cup_{uv\in E(G)}[p_u,p_v]$ for each node $u$ and $S=\cup_{u\in V(G)}(Z_u)$. In effect, $S$ is the union of the line segments corresponding to the edges of $G$ (plus some isolated points in case of isolated vertices), as illustrated in Figure~\ref{fig_planar_graph}. If we consider a position $p$ on the line segment $[p_u,p_v]$ in a simulated zone $Z_u$, the closest node position in $S$ is either $p_u$ or $p_v$, which proves that node $u$ follows the specific geocasting principle. The boundary of $Z_u$ in $S$ is the union of the positions $\{p_v\}$ where each vertex $v$ is a neighbor of $u$, so $u$ also follows the handover principle.

Note however that the stronger geocasting principle is in general not respected. In order to make it valid, a position on a line segment $[p_u,p_v]$ must be closest to either $p_u$, $p_v$ or the position $p_w$ of a common neighbor of $u$ and $v$. For instance, Gabriel graphs~\cite{bib_gabriel}, where no node $w$ may have its position $p_w$ in the interior of a disc of diameter $[p_u,p_v]$ if $uv\in E(G)$, follow the geocasting principle.

\subsection{Unit disc graphs}

The unit disc graphs have been introduced as an idealized model of a radio network~\cite{bib_udg}, where each node has the same constant range $r$. In this model, two nodes $u$ and $v$ may communicate if and only if their distance in the plane is less than $r$. In other words, the edge $uv$ exists if and only if $d(p_u,p_v)\leq r$.

\begin{figure}[h]
\begin{center}
\includegraphics[width=6cm]{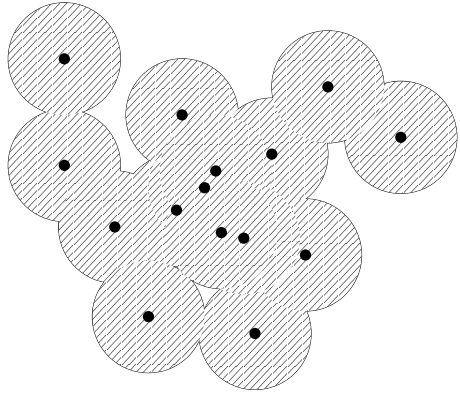}
\caption{Metric space simulated by a unit disc graph.}
\label{fig_disc_graph}
\end{center}
\end{figure}

In this setup, we consider the knowledge graph $H$ to be equal to the unit disc graph $G$, and the simulated zones $Z_u$ to be equal to the closed ball
 $\overline{B}_{p_u}(\frac{r}{2})$ centered in $p_u$ of radius $\frac{r}{2}$ for each node $u$ of the network. The continuous metric space $S\subset\mathbb{R}^2$ is the union of all these closed balls, as illustrated in Figure~\ref{fig_disc_graph}.
Consider a node $u$ and a position $p$ in $\overline{B}_{p_u}(\frac{r}{2})$. If there is a vertex $v$ such that $d(p,p_v) < d(p,p_u)$, then $d(p,p_v) < \frac{r}{2}$ so $d(p_u,p_v) < r$, which means that $v$ is a neighbor of $u$ in $H$. Therefore, $u$ follows the geocasting principle. Likewise, a position $p$ on the boundary of $\overline{B}_{p_u}(\frac{r}{2})$ in $S$ must belong to another closed ball $\overline{B}_{p_v}(\frac{r}{2})$, where $v$ is a neighbor of $u$ in $H$. This means that node $u$ follows at least the directional handover principle. It also follows the stronger handover principle if there is no node $v$ such that $d_\englobing(p_u,p_v)=\frac{r}{2}$.

Unit disc graphs have become popular in the literature because they could be planarized in a distributed manner with Gabriel subgraphs~\cite{bib_gabriel}, which enables the use of planar routing strategies (see for instance~\cite{bib_gfg,bib_gpsr}). The planarization comes with a spanning ratio of $\sqrt{|V(G)|}$, which means that an edge in $G$ might correspond to a path of length $\sqrt{|V(G)|}$ in the subgraph. Of course, we have just seen that it is quite unnecessary to planarize a unit disc graph in order to carry out geographic routing, and one may advantageously use the gradient/perimeter navigation engine directly on the surface $S$.

It has also been argued~\cite{bib_sensing} that in a model where the communication radius $r$ was more than the double of a sensing radius $\rho$, if sensors were deployed in a convex area $\englobing$ so as to sense the entire area -- the area would be called sensing-covered -- then greedy routing would always succeed. In our setup, we can easily see that the sensing-covered condition means that $S = \englobing$, and immediately infer that any trajectory in $\englobing$ can be simulated.

\subsection{Quasi unit disc graphs}
\label{sub_quasi}
In an effort to have a model that is somewhat more realistic than unit disc graphs, the authors in~\cite{bib_quasi_udg} have come up with the quasi unit disc graph model. Here, there are two communications radii $r_{min}$ and $r_{max}$. Two nodes will communicate if their distance is smaller than $r_{min}$, will not communicate if their distance is greater than $r_{max}$, and may or may not communicate if their distance is between $r_{min}$ and $r_{max}$. The ratio $\frac{r_{max}}{r_{min}}$ is classically bounded by $\sqrt{2}$.

\begin{figure}[h]
\begin{center}
\includegraphics[width=4cm]{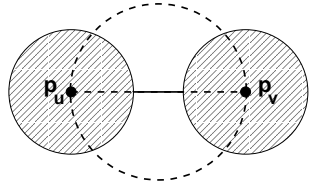}
\caption{The set of points that are closer to a position in $[p_u,p_v]$ than both $p_u$ and $p_v$ is the open disc of diameter $(p_u,p_v)$.}
\label{fig_disc}
\end{center}
\end{figure}

In this setup we consider the knowledge graph $H$ to be equal to $G^2$ (the two hop neighborhood graph). For each node $u$, the simulated zone $Z_u$ is the closed ball $\overline{B}_{p_u}(\frac{r_{min}}{2})$ plus the line segments $[p_u,p_v]$ for all the neighbors $v$ of $u$ in $G$, so $Z_u=\overline{B}_{p_u}(\frac{r_{min}}{2})\bigcup(\cup_{uv\in E(G)}[p_u,p_v])$ and $S=\cup_{u\in V(G)}(Z_u)$.
Consider a node $u$ and a position $p$ in $Z_u$. If $p$ is in $\overline{B}_{p_u}(\frac{r_{min}}{2})$ then the node closest to $p$ is a neighbor of $u$ for the same reasons as we have seen in the unit disc graph example. If $p$ is on a line segment $[p_u,p_v]$ then the closest node $w$ is either $u$,$v$, or a node at position $p_w$ in the open disc of diameter $(p_u,p_v)$, as illustrated in Figure~\ref{fig_disc}. The radius of this disc is smaller than $\frac{r_{max}}{2}$ which means that $d(p_u,p_w)<\frac{r_{max}}{\sqrt{2}}$ or $d(p_v,p_w)<\frac{r_{max}}{\sqrt{2}}$. If $r_{max}\leq\sqrt{2}r_{min}$ then $w$ must be in the neighborhood of either $u$ or $v$, so $uw$ is an edge of $H$. Therefore $u$ follows both the geocasting and the handover principles.

In the literature, quasi unit disc graphs used to be planarized with the help of virtual links corresponding to two existing links, which is consistent with $H=G^2$. Unfortunately, virtual links could also be constituted of a physical and another virtual link, which led to a possibly large spanning ratio. Again, we have shown that these planarization techniques are no longer necessary, nor desirable.

\section{Geographic Routing Criteria}
\label{sec_criteria}
It is not enough to be able to simulate a continuous metric space for geographic routing to make sense. As we have seen in the introduction, a fundamental assumption of geographic routing is to assume that the node coordinates carry relevant information for the routing in $G$. In contrast, even if the {\em Cross Link Detection Protocol (CLDP)} algorithm~\cite{bib_cldp} may planarize a graph where nodes have any random two-dimensional coordinates -- so they carry no relevant topological information whatsoever -- it inevitably has a costly overhead.
In this section, we address the two following questions.
``What criteria must the continuous metric space verify to make geographic routing a sensible choice? Given a communication graph and some node coordinates, can we tell whether there is a continuous metric space that makes geographic routing a sensible choice, and if so can we build it efficiently?''

In Subsection~\ref{sub_simul_com} we consider connectivity issues and formulate the {\em link embedding principle}. We prove that the three first principles put strong constraints on the topology of the the continuous metric space $S$.
In Subsection~\ref{sub_spanning} we consider efficiency issues and formulate the {\em constant spanning ratio principle}. We prove that this fourth principle can be interpreted as a condition on the size of the knowledge graph $H$, which gives us a measure on the appropriateness of geographic routing, which we call {\em geographic eccentricity}.

\subsection{Connectivity issues}
\label{sub_simul_com}
\begin{figure}[h]
\begin{center}
\subfigure[Communication graph.]{\includegraphics[width = 5cm]{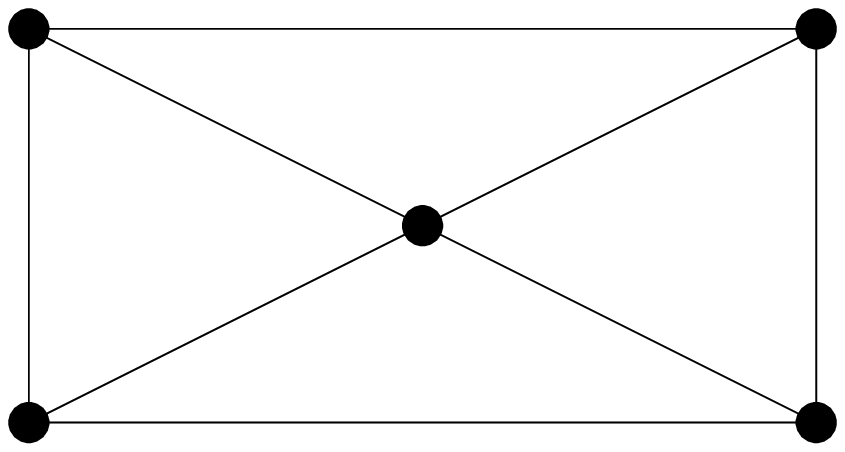}}
\hspace{1cm}
\subfigure[Topology of the Gabriel subgraph.]{\includegraphics[width=5cm]{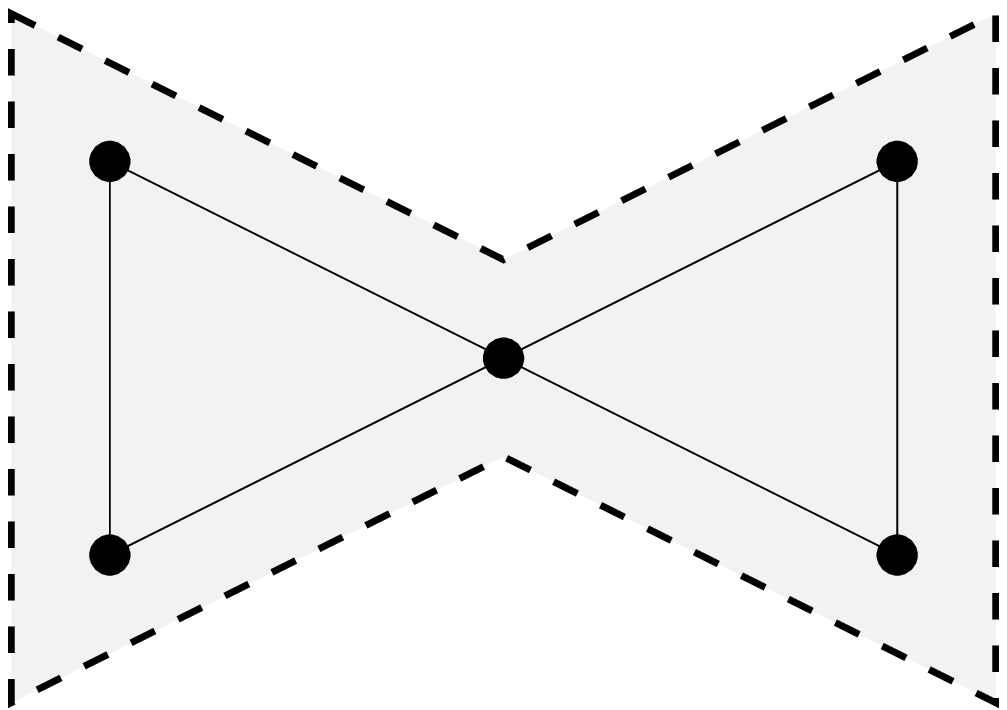}}
\\
\subfigure[Communication graph.]{\includegraphics[width = 5cm]{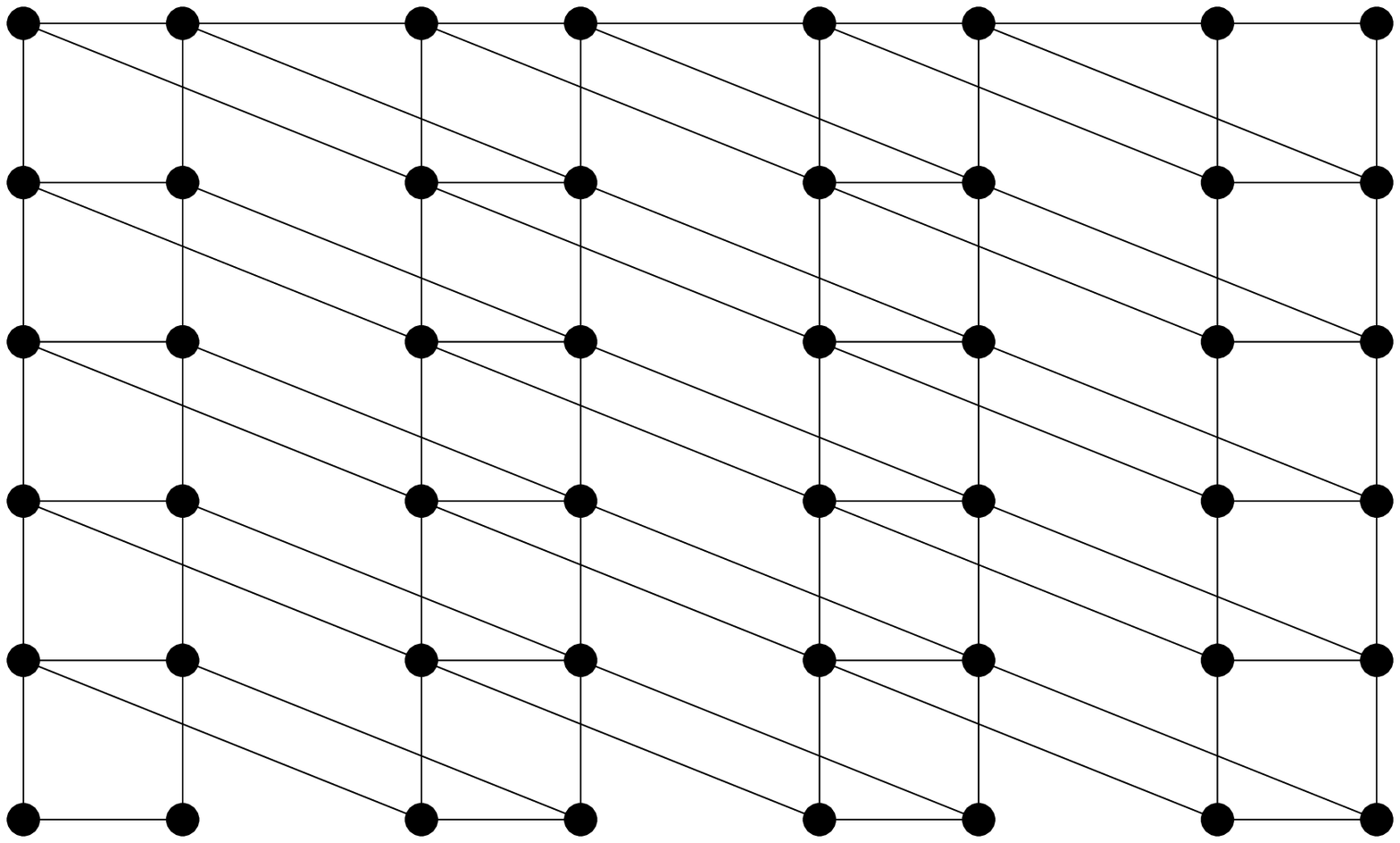}}
\hspace{1cm}
\subfigure[Topology after crossing link removal.]{\includegraphics[width=5cm]{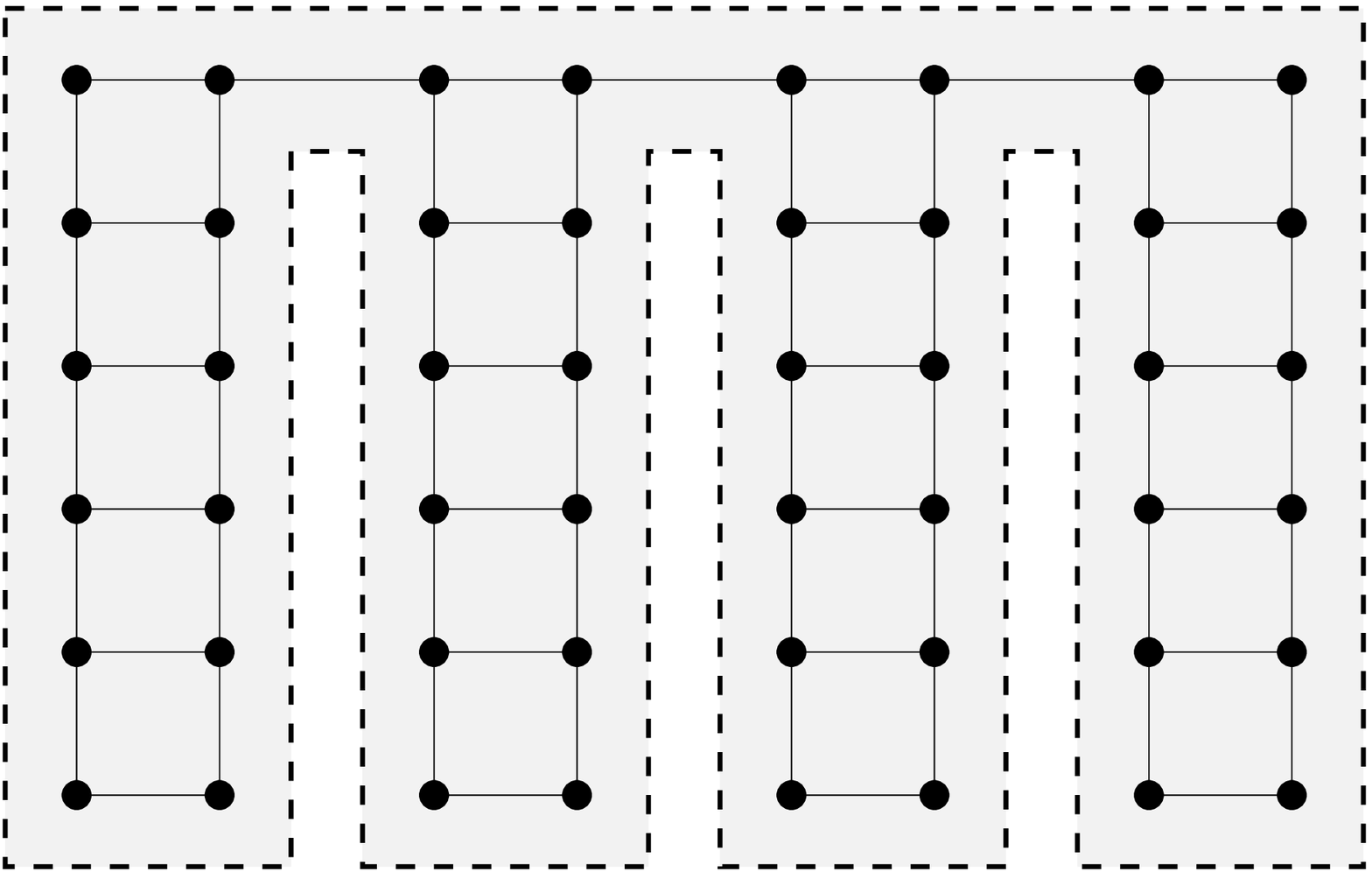}}
\caption{Removing links leads to loss of connectivity.}
\label{fig_cldp}
\end{center}
\end{figure}

We have previously focused in Section~\ref{sec_simul_metric} on how to simulate a continuous metric space. Now we take the opposite view: ``given a communication graph $G$ with node positions, what would be the appropriate space to consider?'' As we have seen in Subsection~\ref{sub_planar}, any planar subgraph of the knowledge graph $H$ could be used to build a continuous metric space $S$. However, planarizing a graph consists in deliberately ignoring some links and may very well lead to a severe loss of connectivity, as illustrated in Figure~\ref{fig_cldp}. If the only relevant criterion is to enforce simple connectivity, even a spanning tree with a planar embedding could provide a connected metric space. In that case, deliberately building a pure spanning tree structure would probably be more efficient than any geographic routing scheme. Instead, we postulate that an implied expectation of geographic routing is that to any path in $G$ corresponds a trajectory in $S$ consisting in piecewise geodesic curves of $\englobing$. We call this expected property the {\em link embedding principle}.

\paragraph{Link embedding principle.} A continuous metric space $S$ follows the link embedding principle if for all edges $uv$ of $G$ there is a geodesic curve $L$ from $p_u$ to $p_v$ in $\englobing$ such that $L\subset S$. If $\englobing$ is a Euclidean space then for all $uv\in E(G)$ the line segment $[p_u,p_v]$ is included in $S$.

\paragraph{}
The link embedding principle has very strong implications on the topology of $S$. Consider a Voronoi diagram of the englobing space $\englobing$ by the positions $\{p_u\}$, where $C_u$ is the cell corresponding to $p_u$ for each node $u$. If a line segment $[p_{u_1},p_{u_2}]$ traverses the boundary of two adjacent Voronoi cells $C_{v_1},C_{v_2}$ of $\englobing$, then the edge $v_1v_2$ must be in the knowledge graph $H$ (see Subsection~\ref{sub_voronoi}). 
Therefore, the canonical simulation with respect to $H$ follows the link embedding principle if and only if $H$ contains all such edges $v_1v_2$.
This induces a lower bound on the size of the smallest $k$ such that $H$ is a subgraph of $G^k$, which can be measured offline (see Section~\ref{sec_experiment} and the Annex). 

\subsection{Efficiency issues}
\label{sub_spanning}

\begin{figure}[h]
\begin{center}
\includegraphics[width=6cm]{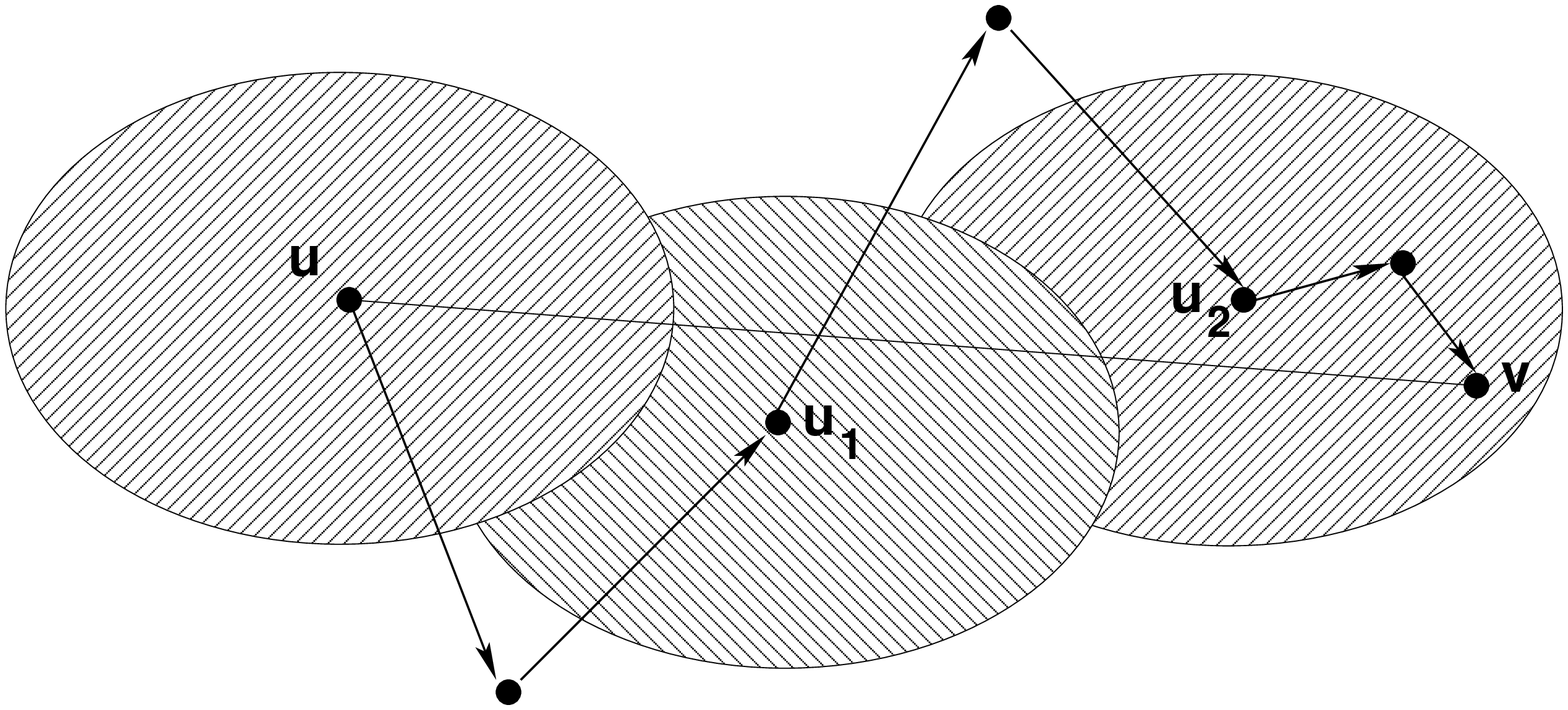}
\caption{The line segment $[p_u,p_v]$ is covered by 3 zones in a simulation with knowledge graph $H=G^2$. Following the trajectory $[p_u,p_v]$ in $S$ takes 6 hops from $u$ to $v$.}
\label{fig_spanning}
\end{center}
\end{figure}

Apart from connectivity factors, the efficiency of geographic routing can be measured by comparing the length of the paths yielded by the navigation/simulation framework to the length of the shortest paths in the communication graph $G$. This spanning ratio can be bounded by the length of paths yielded by following a geodesic curve for each edge of $G$, which should be a small number (see Figure~\ref{fig_spanning}). We call this number {\em spanning ratio} and postulate that geographic routing schemes must follow the {\em constant spanning ratio principle}. We then prove that the constant spanning ratio can be exclusively linked to the size of the knowledge graph, as stated in Theorem~\ref{th_constant}. Finally we define the {\em geographic eccentricity} of a localized network, and its {\em canonical simulation}.

\paragraph{Constant spanning ratio principle.} The simulation of a continuous metric space $S$ by a network $\cal N$ has spanning ratio $k$ if the knowledge graph $H$ is a subgraph of $G^{k_1}$ and if for all edges $uv$ of $G$ there is a sequence $L_1\subset\dots\subset L_{k_2}$ of continuous geodesic curves is $S$ and a sequence $u_0,\dots,u_{k_2}$ of nodes such that $k_1\times k_2\leq k$, such that $u_0=u$, $u_{k_2}=v$, such that $L_{k_2}$ connects $p_u$ to $p_v$, and such that for all $i\in\{1,\dots,k\}$ $u_{i-1}u_i\in H$ and $L_i\subset(\cup_{0\leq j\leq i-1} Z_{u_j})$.

\newpage
\begin{theorem}
Consider a communication graph $G$ and a continuous metric space $S$ following the link embedding principle. If there is a simulation of $S$ with spanning ratio $k$ and if there is a canonical simulation of $S$ with knowledge graph $G^k$, then for each edge $uv\in E(G)$ there is a geodesic curve $L$ of $\englobing$ from $p_u$ to $p_v$ included in the zone simulated by $u$ in the canonical simulation.
\label{th_constant}
\end{theorem}
\begin{proof}
Consider a simulation of $S$ with knowledge graph $H$ and spanning ratio $k_1\times k_2$, such that $H$ is a subgraph of $G^{k_1}$ (but not a subgraph of $G^{k_1-1}$).
Let $Z_u$ be the simulated zone of $S$ for each node $u$. We construct a new simulation with knowledge graph $G^k$ and the simulated zone $Z_u' = Z_u\bigcup(\cup_{uv\in E(H^{k_2-1})} Z_v)$ for each node $u$. Note that for each node $v$ such that $uv\in H^{k_2-1}$, the neighbors of $v$ in $H$ are neighbors of $u$ in $H^{k_2}$, so $Z_u'$ follows the geocasting and handover principles as does $Z_u$. Since the original simulation has spanning ratio $k$, for every edge $uv$ of $G$ there is a geodesic curve $L$ connecting $p_u$ to $p_v$ and a sequence $u_0,\dots,u_{k_2}$ of nodes such that $u_0=u$, $u_{k_2}=v$, such that for all $i\in\{0,\dots,k_2\}$ $u_{i-1}u_i\in H$, and such that $L\subset(\cup_{1\leq i\leq k_2-1} Z_{u_i})$. Observe that all the nodes $u_1,\dots,u_{k_2-1}$ are neighbors of $u$ in $H^{k_2}$ so $L\subset Z_u'$. According to Theorem~\ref{th_voronoi1}, $Z_u'$ is included in the zone simulated by $u$ in the canonical simulation of $S$ with knowledge graph $G^k$.
\end{proof}

\paragraph{Geographic eccentricity} The geographic eccentricity of a localized network is the smallest number $k$ such that the canonical simulation with respect to $H=G^k$ follows the link embedding principle and has constant spanning ratio $k$.

\paragraph{Canonical simulation}
Let $k$ be the geographic eccentricity of a network with localized nodes. We call canonical simulation of the network the canonical simulation with respect to $H=G^k$.
\paragraph{}

An interesting consequence of the constant spanning ratio principle is that in the context of a Euclidean englobing space, a node $u$ is neighbor in a canonical simulation with all the nodes $w$ such that the line segment $[p_u,p_v]$ traverses the Voronoi cell $C_w$ for all edges $uv$ of $G$. This property enables the distributed computation of the Delaunay triangulation, as we will see in the next section.

\section{Distributed space simulation}
\label{sec_distributed}

In this section, we assume that we have a connected communication graph with coordinates in $\mathbb{R}^2$. Given a small number $k$, we propose a distributed algorithm that either constructs a continuous metric space $S\subset\mathbb{R}^2$ and a canonical simulation of it with knowledge graph $G^k$, or reveals that the geographic eccentricity of the network is strictly greater than $k$.

\subsection{Delaunay triangulation}
\label{sub_dist_delaunay}

We know from Subsection~\ref{sub_voronoi} that however chosen the continuous metric space $S$, the Delaunay triangulation of $S$ by the positions $\{p_u\}$ must be a subgraph of the knowledge graph $G^k$. Furthermore, the link embedding and the constant spanning ratio principles tell that for all edges $uv$ of $G$, the centers of the Voronoi cells that are traversed by a line segment $[p_up_v]$ are neighbors of $u$ in the knowledge graph $G^k$ (see Subsection~\ref{sub_spanning}). This means that the node $u$ may locally compute a Voronoi diagram of $\englobing$ by the set of the positions of its neighbors in the knowledge graph $G^k$ and consider the one traversed by the line segments $[p_up_v]$ in order to compute some edges of the Delaunay triangulation. This is done in the algorithm described in Table~\ref{tab_delaunay}. Conversely, if two Voronoi cells $C_u$, $C_v$ that are adjacent in $\englobing$ are not traversed by a same line segment corresponding to an edge of $G$, then the edge $uv$ may or may not belong to the Delaunay triangulation of $S$, depending on $S$. Also note that for any edge $uv$ of $G$, the computed triangulation will contain a path from $u$ to $v$; the computed triangulation is therefore connected.

\begin{figure}[htbp]
\begin{center}
\includegraphics[width=5cm]{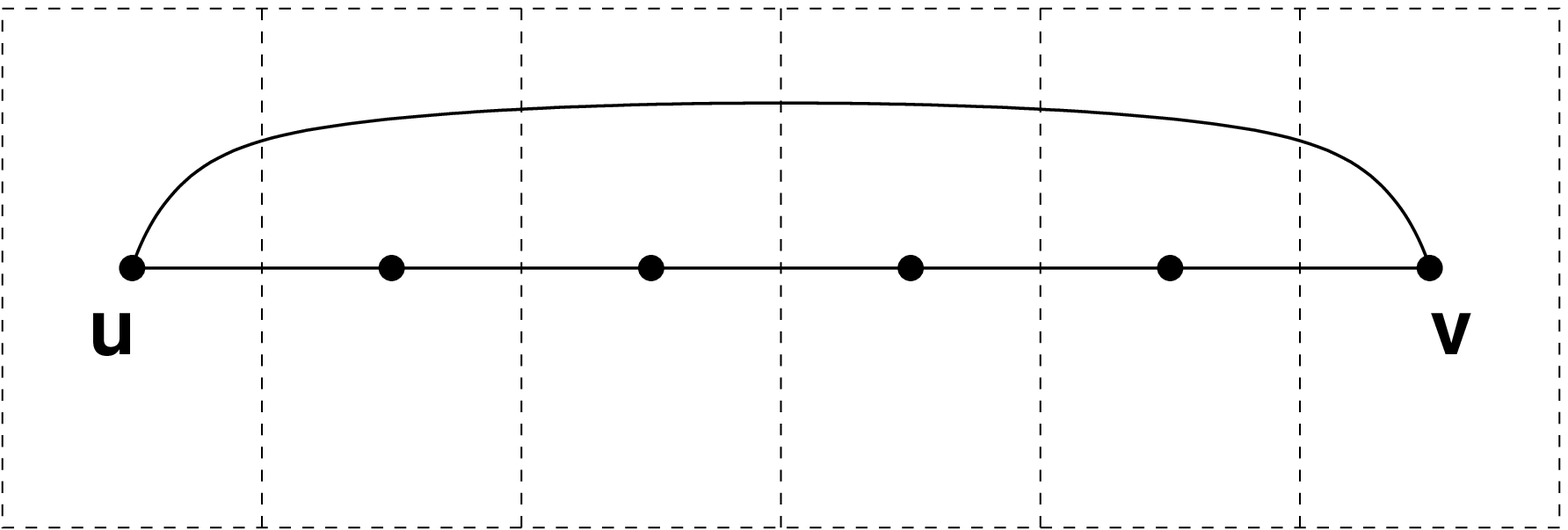}
\caption{Localized network with constant spanning ratio 5.}
\label{fig_delaunay_fail}
\end{center}
\end{figure}

\begin{table}[htbp]
\begin{center}
\framebox{
  \parbox{0.9\linewidth}{\small
~\\
\centerline{\bf Distributed Delaunay Triangulation}
~\\
{\bf Input:} $(p_u,k)$, where $p_u$ is the position of node $u$ and $k$ is a small number.
\begin{enumerate}
\item {\bf Broadcast} the position $p_u$ at $k$ hops.
\item {\bf Receive} the position information of neighbors in $G^k$ as well as a path to them in $G$.
\item Compute the Voronoi diagram of $\englobing$ by the set of points $\{p_v$ such that $v=u$ or $uv\in G^k\}$.\label{adr_diagram}
\item For each edge $uv\in E(G)$, compute the Voronoi cells $C_{u_1},C_{u_2},\dots$ that $[p_u,p_v]$ traverses and store the pairs ($u_i,u_{i+1}$) of nodes with adjacent Voronoi cells on the line segment $[p_u,p_v]$. The pairs ($u_i,u_{i+1}$) constitute edges of the Delaunay triangulation.\label{adr_link}
\item {\bf Broadcast} the computed Delaunay edges at $k$ hops.
\item {\bf Receive} the Delaunay edges $uv$ computed by neighbors in $G^k$.
\item If a Delaunay edge $uv$ has been received such that the Voronoi cells $C_u,C_v$ locally computed at Instruction~\ref{adr_diagram} are not adjacent, then {\bf assert global failure}.\label{adr_fail}
\end{enumerate}
{\bf Output:} the neighborhood $\Gamma_T$ of $u$ in the Delaunay triangulation.\\
}
}
\end{center}
\caption{Distributed Delaunay triangulation at node $u$.}
\label{tab_delaunay}
\end{table}

The communication cost of the distributed Delaunay triangulation (described in Table~\ref{tab_delaunay}) consists in two broadcasts at $k$ hops for each node. These broadcasts are needed to exchange local topological information. The nodes first compute a Voronoi diagram of $\englobing$ of the positions they locally know. This diagram can be computed, for instance, with Fortune's algorithm~\cite{bib_fortune}. As we have seen in the preceding paragraph, Instruction~\ref{adr_link} correctly computes adjacent nodes in the Delaunay triangulation of any continuous metric space $S$ if the geographic eccentricity is smaller than or equal to $k$. Contrariwise, if the 3 first principles are respected but the spanning ratio is greater than $k$, then there is a line segment $[p_up_v]$ corresponding to an edge of $G$ that traverses a Voronoi cell $C_w$ such that $u$ and $w$ are not neighbors in $G^k$ (see Figure~\ref{fig_delaunay_fail}). In that case, Instruction~\ref{adr_fail} will cause a global failure.

\subsection{Distributed zone computation}
\label{sub_dist_voronoi}

\begin{figure}[htbp]
\begin{center}
\subfigure[Cycle with crossing links.\label{fig_crossing}]{\includegraphics[width = 5.5cm]{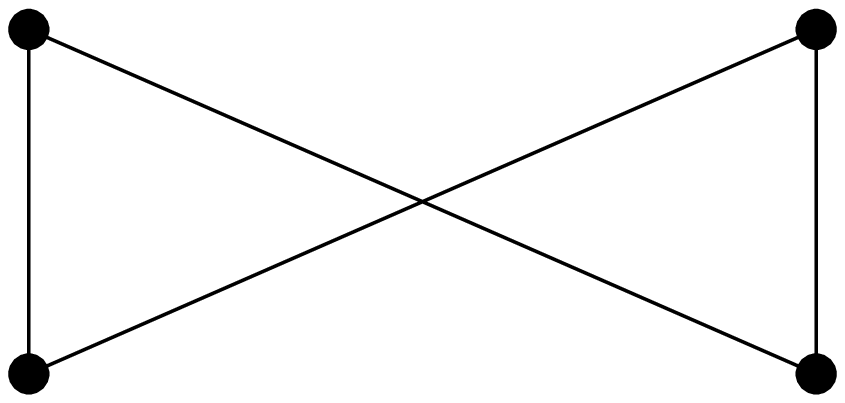}}
\hspace{0.5cm}
\subfigure[Faces with non-adjacent Voronoi cells.\label{fig_face_fail}]{\includegraphics[width=5.5cm]{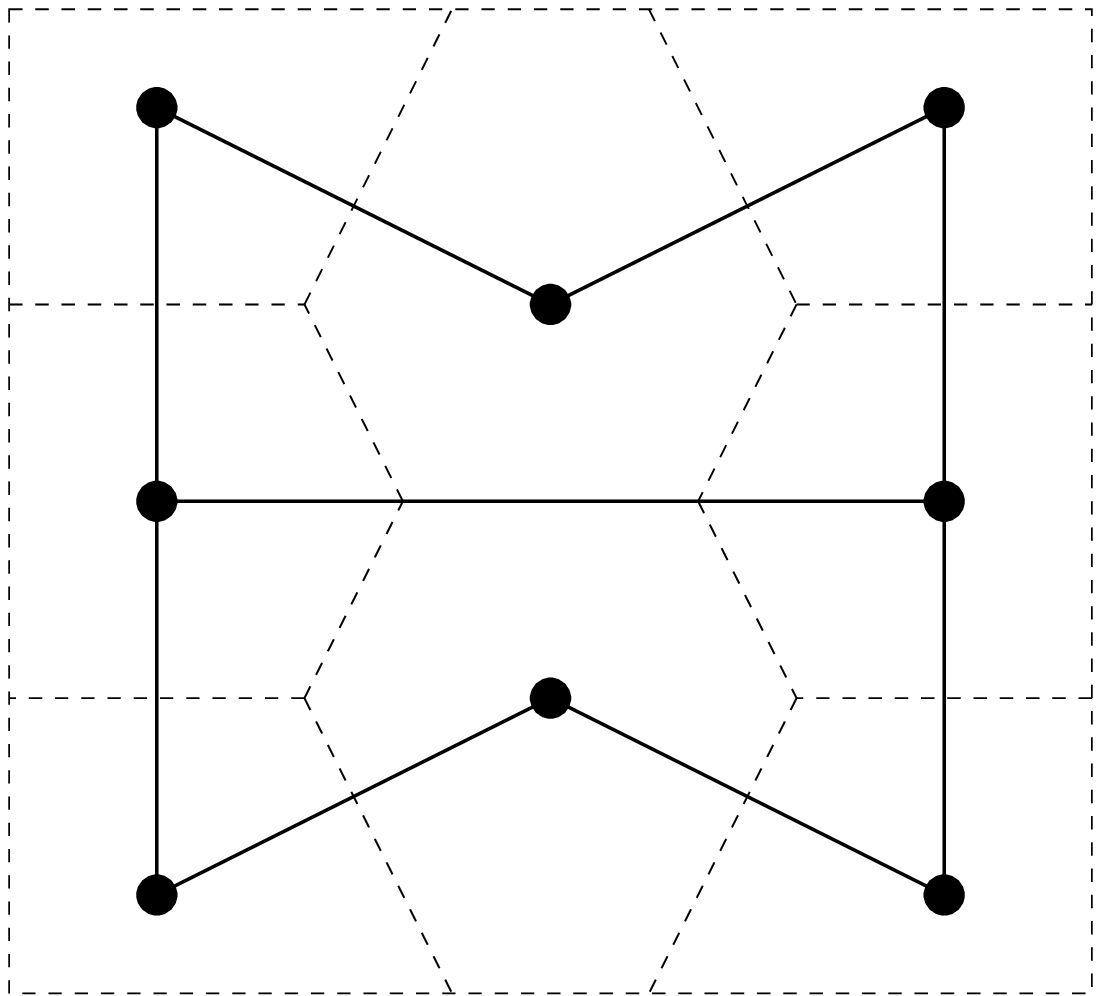}}
\caption{Diagram computation failure cases.}
\label{fig_sinr}
\end{center}
\end{figure}

Once the Delaunay triangulation has been locally computed, verifying that the result is effectively a planar graph and computing Voronoi cells is simply a matter of probing faces, as described in Table~\ref{tab_probe}. The probe sending mechanism is similar to the one used for CLDP. If a probe detects two crossing links then the computed triangulation was not a planar graph (see Figure~\ref{fig_crossing}). Conversely, if all the probes do not detect any crossing links, then the computed triangulation is a planar subgraph of $G^k$~\cite{bib_cldp}.

\begin{table}[htbp]
\begin{center}
\framebox{
  \parbox{0.9\linewidth}{\small
~\\
\centerline{\bf Face Probe}
~\\
{\bf Input:} $(p_u,k,\Gamma_T)$, where $p_u$ is the position of node $u$, $k$ is a small number and $\Gamma_T$ is the neighborhood of $u$ in the computed Delaunay triangulation.
\begin{enumerate}
\item {\bf Send a probe} along an edge $uv$ where $v\in\Gamma_T$.
  \begin{enumerate}
  \item The probe is forwarded according to the right-hand rule in the computed Delaunay triangulation through $u_1,u_2,\dots$ and stops when returning at $u=u_0$.
  \item The probe stores the positions $p_{u_i}$ of the nodes encountered on the Delaunay face along the way.
  \item If the probe detects two crossing links $[p_{u_i}p_{u_{i+1}}]$, $[p_{u_j}p_{u_{j+1}}]$ then {\bf assert global failure}.\label{adr_crossing}
  \end{enumerate}
\item {\bf Receive} the returning probe.
  \begin{enumerate}
  \item Compute the Voronoi diagram $\{C_{u_0}^F,C_{u_1}^F,\dots\}$ of $\englobing$ by the set of points $\{p_{u_0},p_{u_1},\dots\}$.
  \item For each pair of adjacent cells $(C_{u_i}^F,C_{u_j}^F)$ such that $u_i$ and $u_j$ are not neighbors in $G^k$, compute an open subset $B_{u_iu_j}$ of small width $\epsilon$ covering the boundary between $C_{u_i}^F$ and $C_{u_j}^F$. The subsets $B_{u_iu_j}$ represent holes in $\englobing\backslash S$.
  \end{enumerate}
\item {\bf Send a message} containing the computed cells $\{C_{u_i}^F\}$ and the holes $\{B_{u_iu_j}\}$ to the other nodes $u_1,u_2,\dots$ of the face $F$.
\end{enumerate}
}
}
\end{center}
\caption{Face probe sent from node $u$.}
\label{tab_probe}
\end{table}

The communication cost caused by the face probing consists in sending two messages around each face of the Delaunay triangulation.
Sending a message on a edge of a face in $G^k$ can take up to $k$ hops, so the communication cost consists in an average of at most $6$ messages sent at $k$ hops for each node\footnote{The average degree of a planar graph is upper bounded by 6.}. Finally, the individual nodes can compute their own Voronoi cell and their simulation zone as described in Table~\ref{tab_face}. This last step may only fail if the computed triangulation was not a subgraph of the Delaunay triangulation of $\englobing$, and this is detected when two nodes adjacent in the computed triangulation do not correspond to adjacent cells (see Figure~\ref{fig_face_fail}).

\begin{table}[htp]
\begin{center}
\framebox{
  \parbox{0.9\linewidth}{\small
~\\
\centerline{\bf Distributed Zone Computation}
~\\
{\bf Input:} $(k,\Gamma_T)$, where $k$ is a small number and $\Gamma_T$ is the neighborhood of $u$ in the computed Delaunay triangulation.
\begin{enumerate}
\item {\bf Receive} the computed cells $C_u^F$ and possibly the holes $B_{uv}$ for every Delaunay face $u$ is part of.
\item Compute the Voronoi cell $C_u=\cap (C_u^F)$.
\item Compute the hole set $B_u=C_u\cap(\cup (B_{uv}))$. The hole set $B_u$ can be empty.
\item {\bf Broadcast} the cell $C_u$ and the hole set $B_u$ at $k$ hops in $G$.
\item {\bf Receive} the cell $C_v$ and the hole set $B_v$ for each neighbor $v$ in $G^k$.
 \item If there is a neighbor $v$ in $\Gamma_T$ such that the cells $C_u$ and $C_v$ are not adjacent, then {\bf assert global failure}.
\item Compute the simulated zone $Z_u = C_u\bigcup(\cup_{uv\in E(G^k)} C_v)$.
\item Compute the local hole set $Z_u\backslash S = B_u\bigcup(\cup_{uv\in E(G^k)} B_v)$.
\end{enumerate}
{\bf Output:} the Voronoi cells $C_u$, $C_v$ for all $uv\in E(G^k)$ and the simulated zone $Z_u = C_u\bigcup(\cup_{uv\in E(G^k)} C_v)$. The simulated space $S$ is equal to $\englobing\backslash\cup_{u\in V(G)} B_u$.\\
}
}
\end{center}
\caption{Distributed zone computation at node $u$.}
\label{tab_face}
\end{table}

The communication cost of the distributed zone computation algorithm (described in Table~\ref{tab_face}) consists in a broadcast at $k$ hops for each node. Note that if any of the algorithm of this section ends in global failure, then the geographic eccentricity of the localized network must be strictly greater than $k$. According to Theorem~\ref{th_voronoi} and Theorem~\ref{th_constant}, a global failure means that there can't be any simulation of the localized network with knowledge graph $H$ and spanning ratio $k'$ that follows the four principles such that $H$ is a subgraph of $G^{\lfloor\frac{k}{2}\rfloor}$ and such that $k'\leq k$.

\section{Topological Measurements}
\label{sec_experiment}
In this section we present the topological measurements we made on several types of localized networks obtained through computer simulations.
These measurements were made in order to address the following questions.
With which communication models is geographic routing a relevant choice?
What are the effects of localization errors?
How costly is the constant spanning ratio principle? What are the consequences of removing particularly long links?
How do the metrics vary when we change the scale of the network?

\begin{figure}[h]
\begin{center}
\includegraphics[width=7cm]{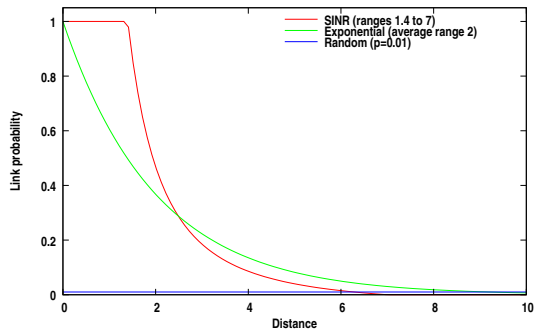}
\caption{Link probability as a function of distance in the three communication models: SINR, exponential and random.}
\label{fig_model}
\end{center}
\end{figure}

The networks were generated by scattering $4L^2$ nodes (usually $2500$) in a rectangle of size $L\times 4L$ according to a uniform distribution law, so the average density is one node per unit square. The communication graphs were built according to one of the following communication models: random, SINR, or exponential (see Figure~\ref{fig_model} and Subsection~\ref{sub_model}).

For each type of network and each set of parameters we have run enough simulations so as to obtain a hundred instances of connected communication graphs. For each simulated network, we compared its geographic eccentricity $k_g$ to its diameter $D$, and to two other measures. The first measure is the largest distance $k_T$ in the communication graph $G$ between two neighbors in the Delaunay triangulation $T$. This latter distance, which we call the {\em Delaunay locality}, is also the smallest number such that the whole plane (without holes) is simulated by the canonical simulation with respect to $H=G^{k_T}$. The second measure is the minimum number $k_e$ such that the canonical simulation with respect to $H=G^{k_e}$ follows the link embedding principle. We call $k_e$ the {\em embedding locality}, and we are mostly interested by the difference $(k_g-k_e)$ which can be viewed as the additional cost carried by the constant spanning ratio principle. 

\subsection{Communication models}
\label{sub_model}
\begin{figure}[h]
\begin{center}
\includegraphics[width=7cm]{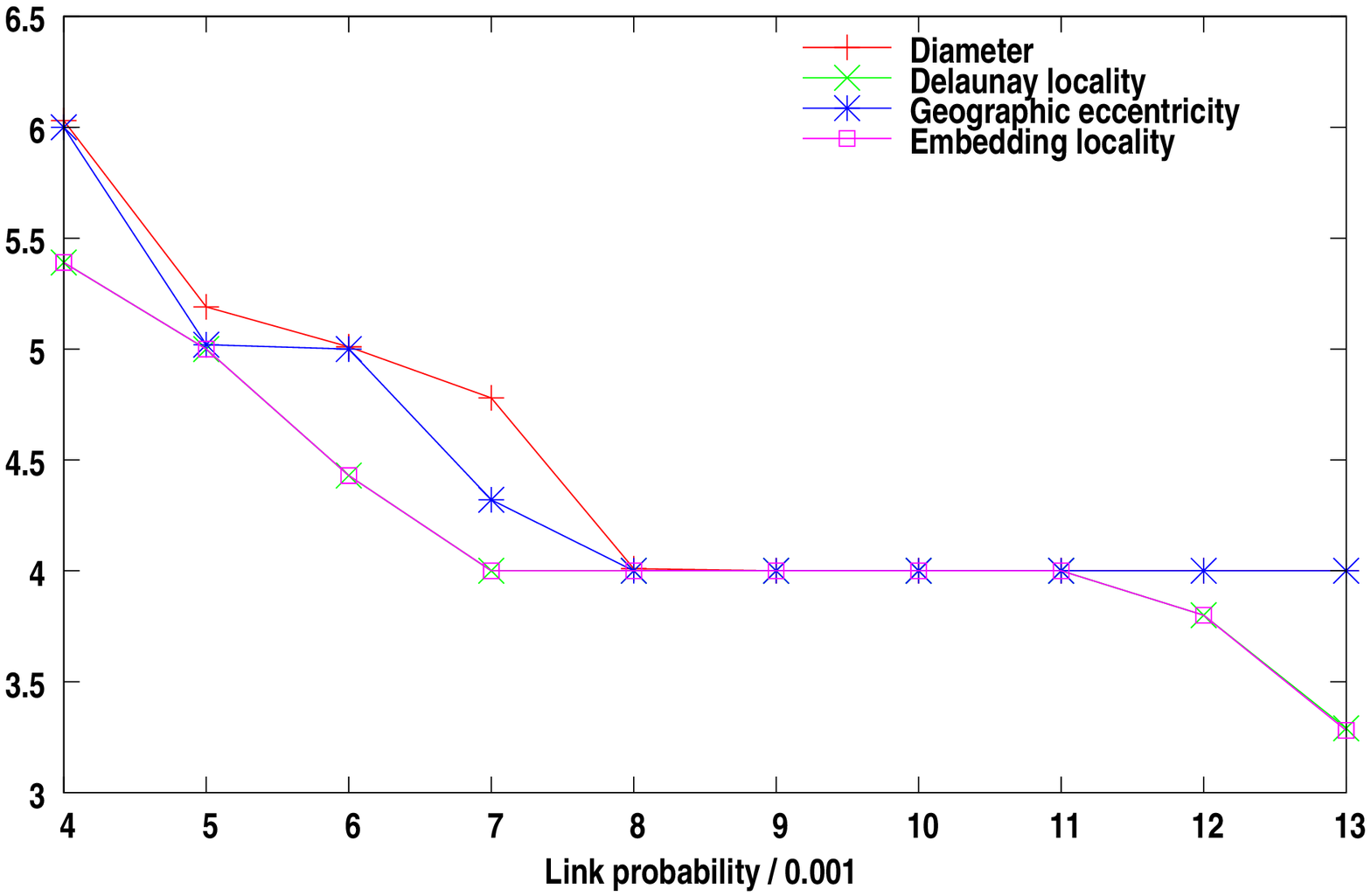}
\caption{Random graph metrics with varying link probability.}
\label{fig_rand}
\end {center}
\end{figure}

The first communication model we have taken as a control is the random graph model, where there is a constant probability $p$ that a link exists for any given pair of nodes. Random graphs have notoriously low diameters, and this can be seen in our simulation results, where the diameter of the generated random graphs decreases from 6 to 4 when the link probability increases from $0.004$ to $0.013$. With values of $p$ under $0.004$ the communication graphs are generally disconnected. The simulation results show that the embedding locality $k_e$ remains close to the Delaunay locality $k_T$, whereas the geographic eccentricity $k_g$ is close or equal to the diameter $D$, which means that the knowledge graph $H=G^{k_g}$ is close to the complete graph $K_{2500}$ (see Figure~\ref{fig_rand}).

We then considered the SINR communication model, where the link probability is commensurate with the difference between the signal strength and the ambient noise. We parametrized a minimum range $r$ under which the link probability is one, and a maximum range $R$ over which the link probability is zero. Between the two ranges, the link probability is proportional to $\frac{1}{d^2(u,v)}-\frac{1}{R^2}$, where $d(u,v)$ is the link length. Quasi unit disc graphs are a special case where $\frac{R}{r}\leq \sqrt{2}$ and their geographic eccentricity is theoretically upper bounded by 3 (this is left as an exercise for the reader). However, even with a ratio $\frac{R}{r}=5$, our measurements show that this communication model is highly amenable to geographic routing with a low geographic eccentricity (consistently 3), as can be seen in Figure~\ref{fig_sinr}.
\begin{figure}[h]
\begin{center}
\subfigure[Quasi unit disc graph with $R = 1.4r$.]{\includegraphics[width = 6cm]{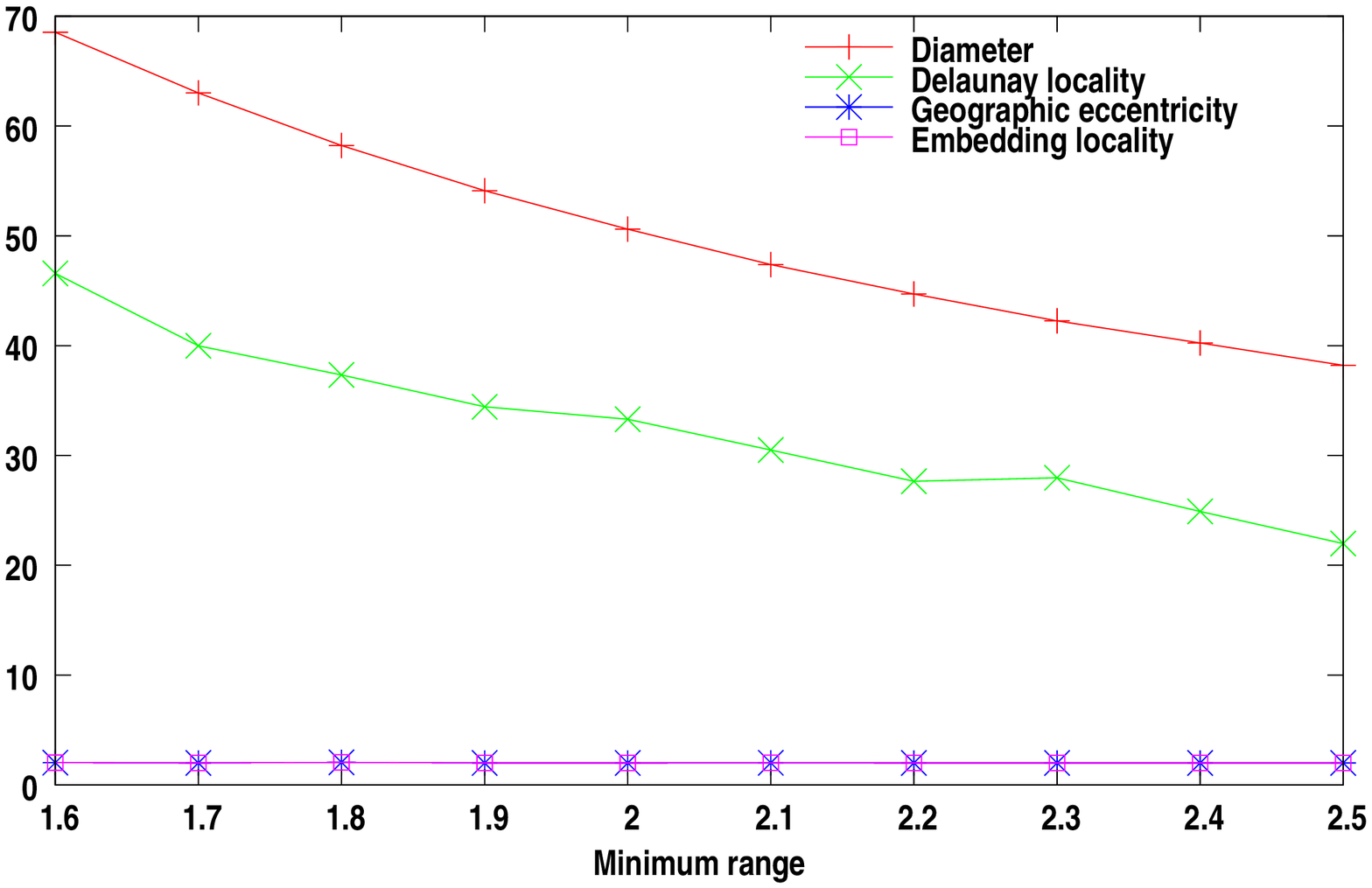}}
\subfigure[Communication graph with $R = 5r$.]{\includegraphics[width=6cm]{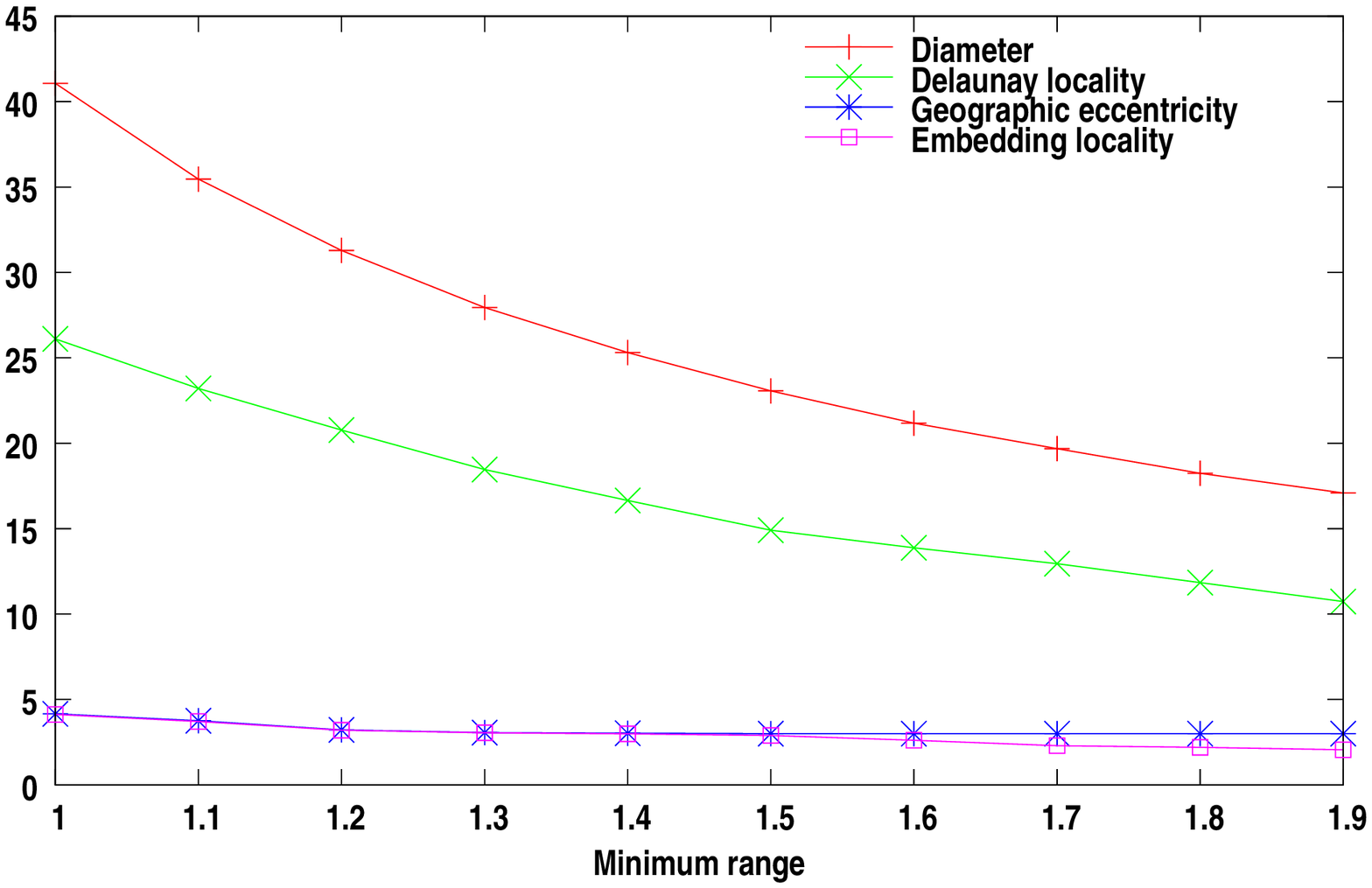}}
\caption{Network metrics based on the SINR communication model with varying minimum range $r$ and maximum range $R$.}
\label{fig_sinr}
\end{center}
\end{figure}

\begin{figure}[h]
\begin{center}
\includegraphics[width = 7cm]{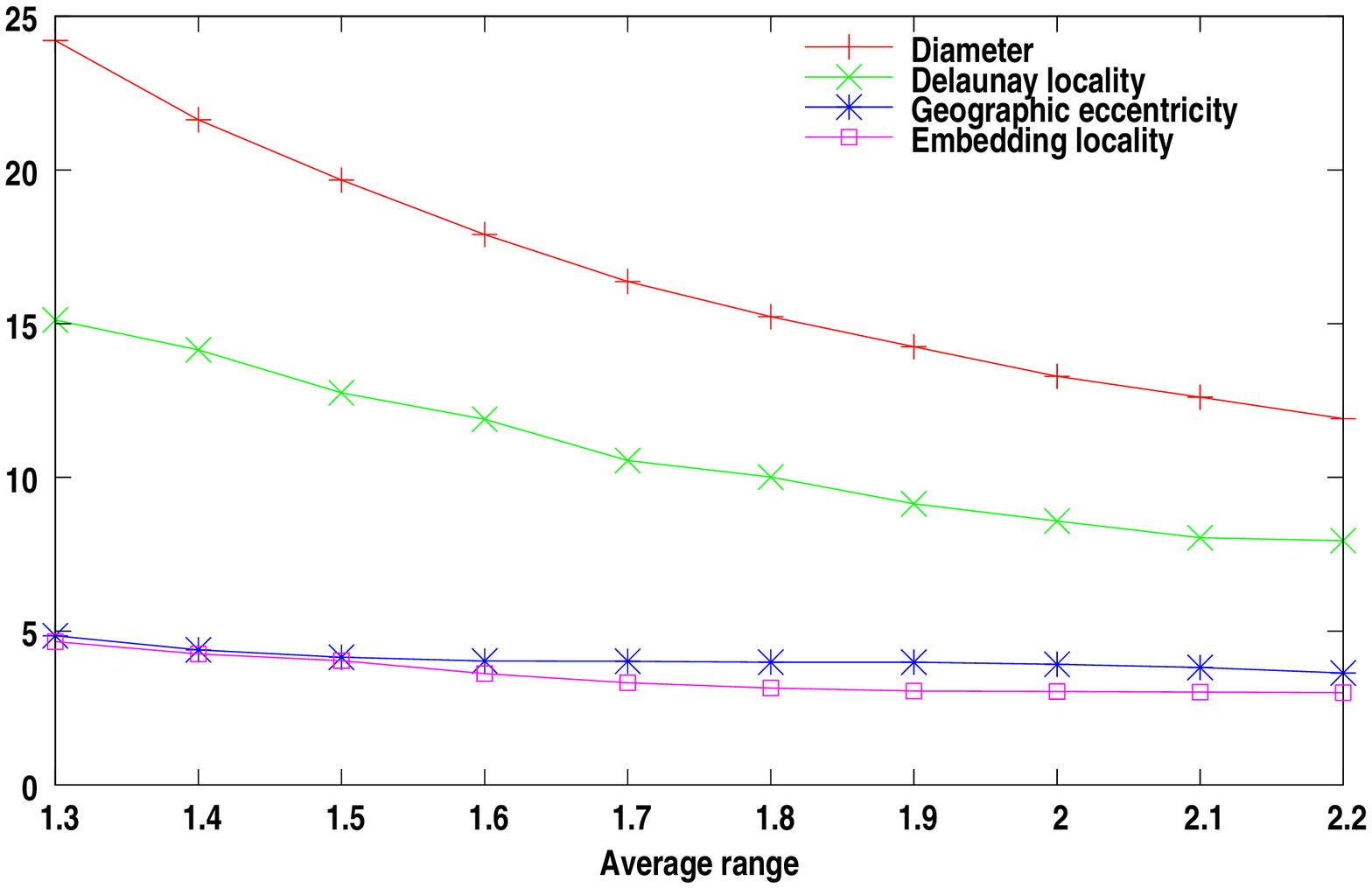}
\caption{Network metrics based on the exponential link probability model.}
\label{fig_exp}
\end{center}
\end{figure}

We finally considered the exponential communication model, where the link probability is proportional to $\exp(-\frac{d(u,v)}{r_{avg}})$, where $d(u,v)$ is the link length and $r_{avg}$ is the average communication range. We can see in Figure~\ref{fig_exp} that the geographic eccentricity of these networks is low (4-5), but the diameter is also relatively low. This means that a large part of the network (up to 40\%) is reached within $k_g$ hops, which makes geographic routing a poor choice. We could interpret these networks as the union of a SINR network for the local links, plus a random graph representing the randomly selected long links (see Subsection~\ref{sub_trunc}).

\subsection{Localization errors}
\begin{figure}[h]
\begin{center}
\subfigure[Quasi u.d.g. with $(r,R)=(1.6,2.24)$.]{\includegraphics[width = 6cm]{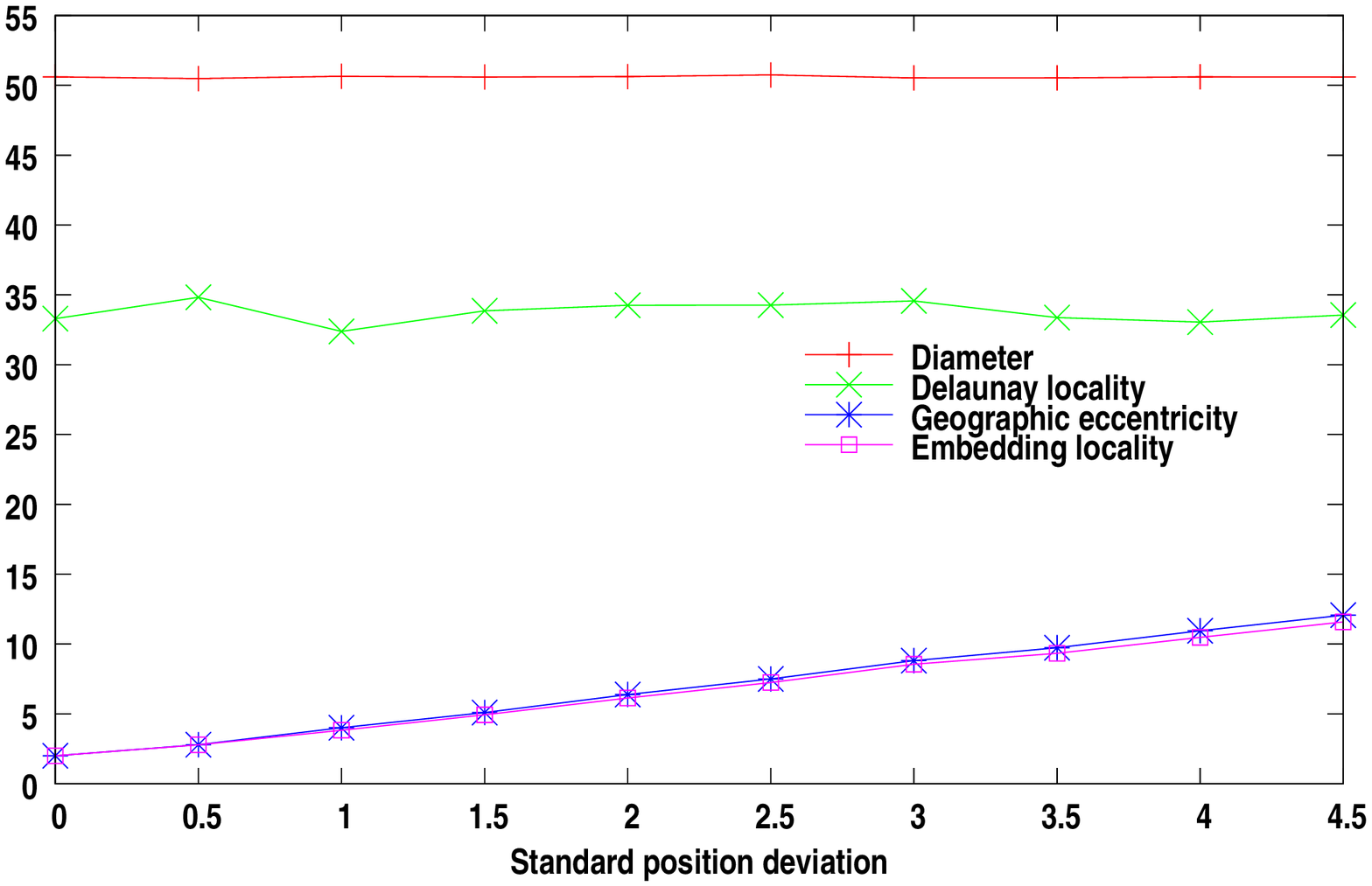}}
\subfigure[Comm. graph with $(r,R)=(1,5)$.]{\includegraphics[width=6cm]{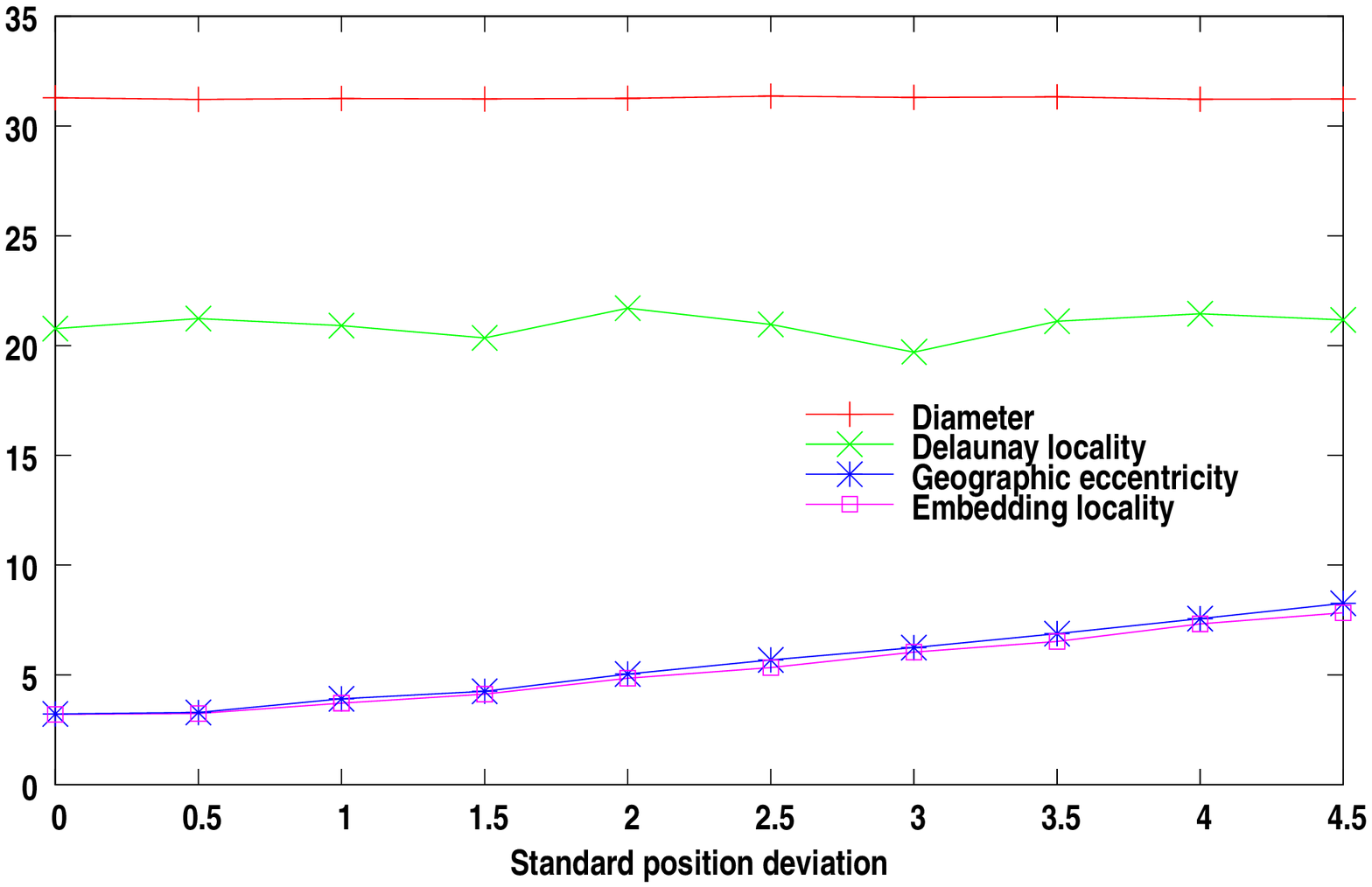}}
\caption{Networks metrics based on the SINR communication model with minimum range $r$ and maximum range $R$. A Gaussian localization error has been added with varying standard deviation $\sigma_{err}$.}
\label{fig_sinr_e}
\end{center}
\end{figure}

In order to measure the effect of localization errors on the geographic eccentricity, we have a focused on communications graph generated with the SINR model and have added a Gaussian localization error.
 This has been done by selecting for each node a direction in $[0,\pi)$ with uniform distribution, and a relative radius $r$ subject to a Gaussian distribution with mean value zero and standard deviation $\sigma_{err}$. The simulation results (see Figure~\ref{fig_sinr_e}) clearly show that there is a near perfect linear dependency between the geographic eccentricity and the localization error.

\subsection{Removing long links}
\label{sub_trunc}

\begin{figure}[h]
\begin{center}
\subfigure[Initial average range $1.7$ and varying maximum range.]{\includegraphics[width=6cm]{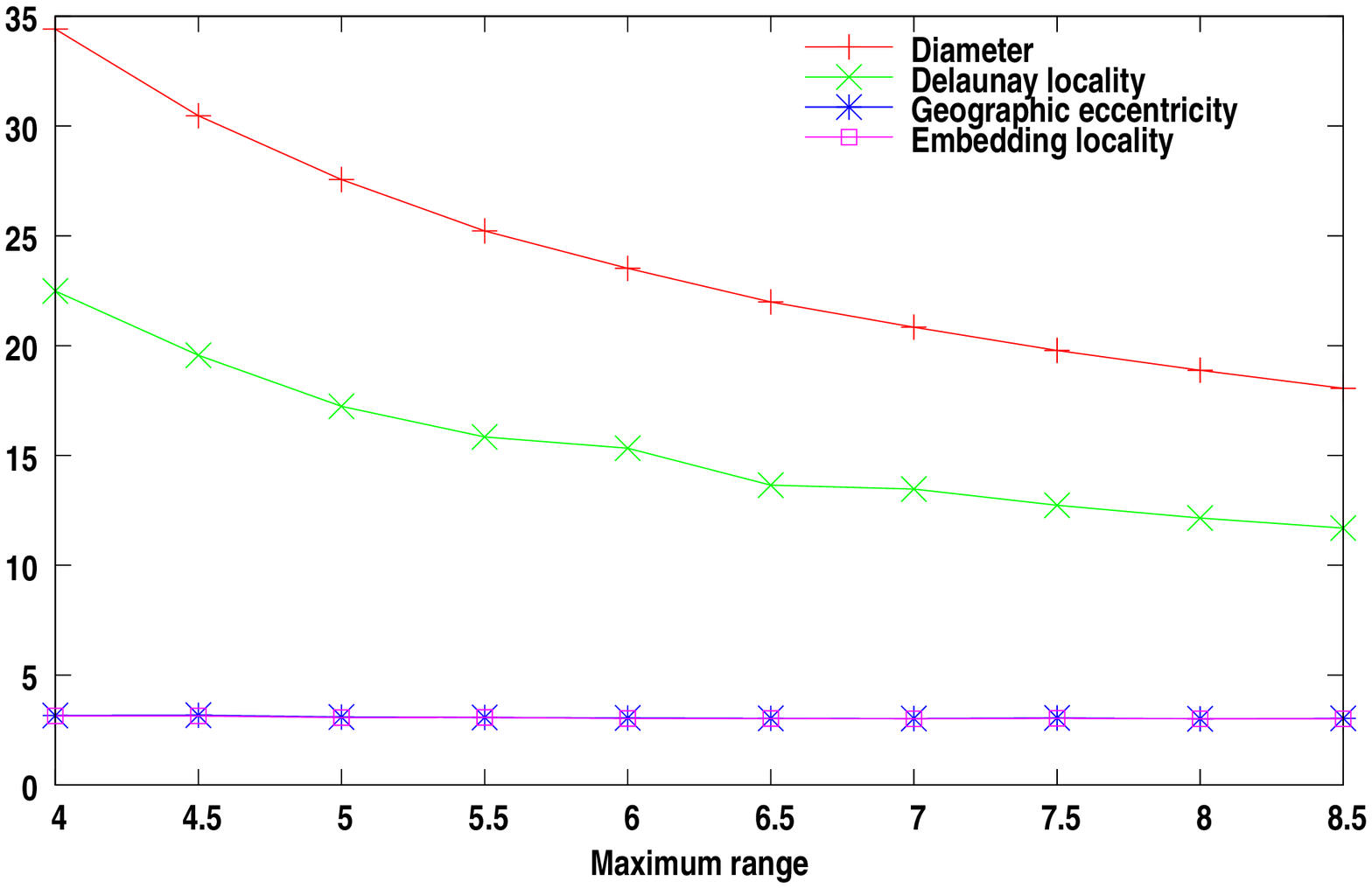}}
\subfigure[Initial average range $2$ and maximum range $6$. Varying Gaussian localization error.]{\includegraphics[width = 6cm]{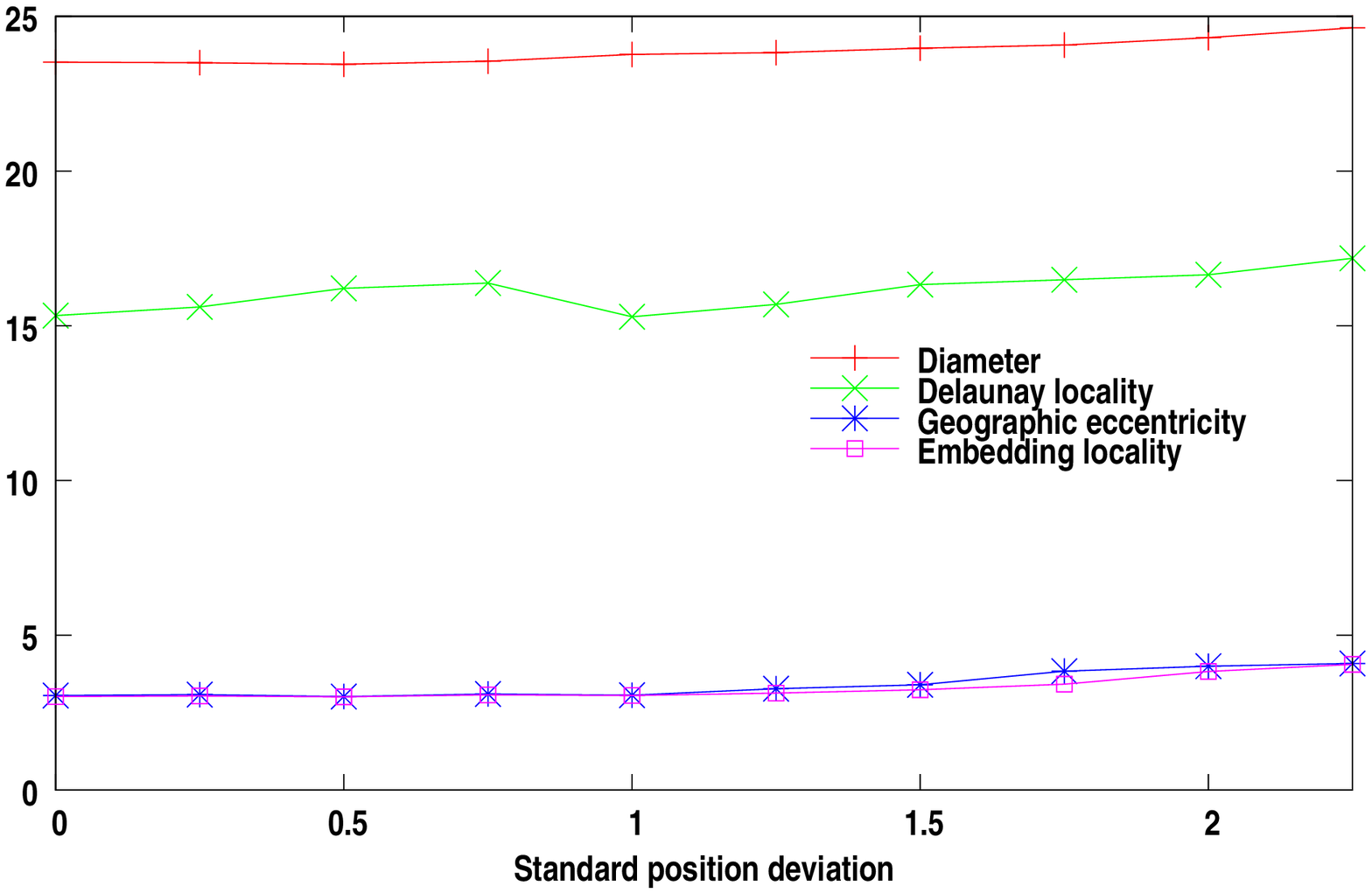}}
\caption{Network metrics based on the exponential link probability model with fixed initial average range. Long links are removed.}
\label{fig_exp_tr}
\end{center}
\end{figure}

Conventional wisdom tells that long links are the most likely to cause problems, since their embedding in the continuous metric space is the most constraining (see Subsection~\ref{sub_simul_com}). On a similar note, planar subgraphs used to typically favor short links~\cite{bib_book}. Therefore, we investigated the consequences of deliberately violating the link embedding principle by discarding long links in the exponential communication model (see Figure~\ref{fig_exp_tr}).
In the first set of experiments we varied the length at which links were discarded. The two main effects of decreasing the maximum admissible length is to artificially increase the communication graph diameter, and to decrease the size of the neighborhood at $k_g$ hops (from \~500 to \~200 nodes). In the second set of experiments, we introduced localization errors and discarded links with long apparent length (the length is computed after the error is factored in). The first effect is to apparently nullify the effect of increasing localization error, but what happens in reality is that the networks become increasingly disconnected when the error grows; most networks were disconnected when the standard error became greater than $2.5$ units in this set of experiments.

The simulation results show that removing long links may make geographic routing easier but comes at a steep price, either in greatly increasing path lengths, or in sheer loss of connectedness.

\subsection{Scalability}
\begin{figure}[h]
\begin{center}
\includegraphics[width=7cm]{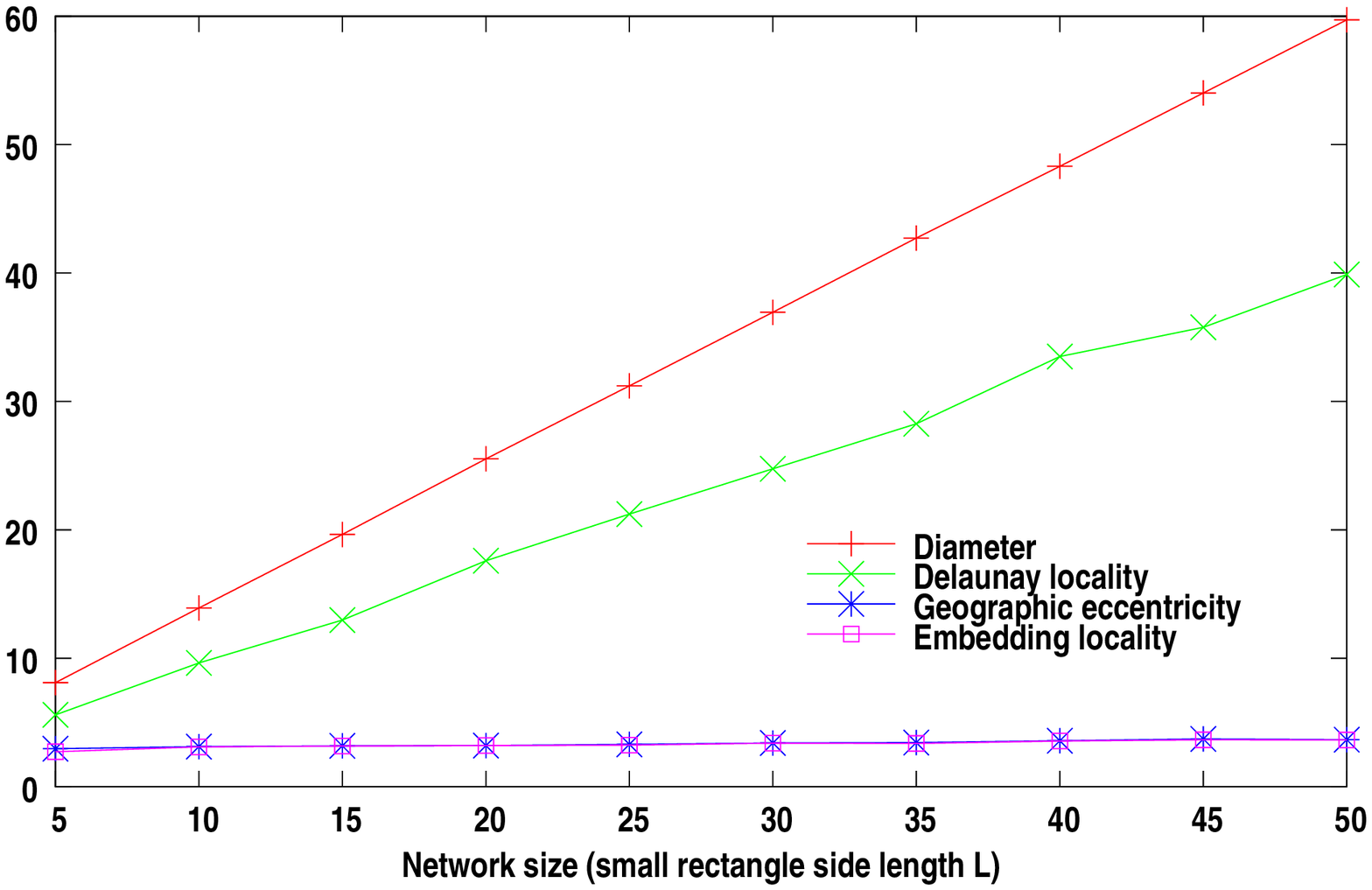}
\caption{Network metrics based on the SINR communication model with minimum range 1.2 and maximum range 6. A Gaussian localization error has been added with standard deviation 0.5. Each network is composed of n nodes
scattered in a $L\times 4L$ rectangle.}
\label{fig_scal}
\end {center}
\end{figure}

Finally, we have tested the scalability of the geographic eccentricity metric by simulating networks following the SINR communication model that ranged from 100 to 10,000 nodes (see Figure~\ref{fig_scal}). The simulation results show that multiplying the number of nodes by 100 hardly affects the geographic eccentricity measure; its average increases only slightly ($+0.71$).

\subsection{Observation summary}

The measurements that we have carried out show that the geographic eccentricity is very low for the networks that intuitively seemed suited to geographic routing. The network size has practically no influence on the geographic eccentricity, but the localization errors do influence it linearly.
 Furthermore, we witness in all the measurements
a negligible difference between the two constants $k_e$ and $k_g$, which means that the spanning ratio principle hardly affects the network metrics. 
A complete account of the simulations can be found in the Appendix.

\section{Conclusion}

We have studied the concepts surrounding geographic routing by taking the view that navigation hints are relevant for routing if and only if there is a relation between the topology of the communication graph on the one hand, and the topology of a continuous metric space on the other hand. We have postulated that this relation is built around four principles: {\bf geocasting}, {\bf handover}, {\bf link embedding} and {\bf constant spanning ratio}, and have proven that these four principles have strong algorithmic and topological implications.

\begin{figure}[ħtb]
\begin{center}
\includegraphics[width=7cm]{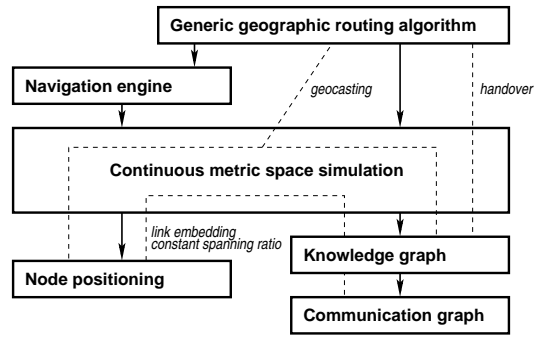}
\caption{Geographic routing schema.}
\label{fig_outline}
\end{center}
\end{figure}

Secondly, we have precisely defined the mechanisms of geographic routing (see Figure~\ref{fig_outline}) and interpreted some classical scenarios in our framework, which revealed that graph planarization techniques are unnecessary and may even be counterproductive.

Thirdly, we have derived from the four principles the concepts of {\bf canonical simulation} and {\bf geographic eccentricity}. We have proposed a distributed and lightweight algorithm for localized networks in the two-dimensional space. Our algorithm either enables geographic routing, or reveals that the geographic eccentricity of the network is too high. We have also measured the geographic eccentricity of various computer generated networks. Our measurements show that the geographic eccentricity depends chiefly on the communication model, remains low on models intuitively suited for geographic routing and depends linearly on localization errors.

\section*{Acknowledgments}
This work was greatly improved following the constructive feedback of Nawfel Jarry Naciri, Eugenio Noto, Christoforos Raptopoulos and Ricardo Wehbe.

\newpage
\bibliographystyle{plain}
\bibliography{georouting}

\newpage
\appendix
\section*{Appendix -- Simulation Runs}
We have run computer simulations in order to carry out measurements on various types of localized networks. For each type of network and each set of parameters we have run enough simulations so as to obtain a hundred instances of connected communication graphs.
All the simulations runs were done in the following manner.
\begin{enumerate}
\item A rectangle of size $L\times 4L$ is filled with $4L^2$ nodes, randomly scattered according to a uniform distribution law. This corresponds either to 2500 nodes (Figure~\ref{exp_random} to~\ref{exp_exp_r_e}) or to a number of nodes ranging from 100 to 10,000 (Figure~\ref{exp_sinrScal}).
\item A communication graph is built according to one of the communication models below.
  \begin{enumerate}
  \item Random -- there is a constant link probability $p$ between any pair of nodes (Figure~\ref{exp_random}).
  \item SINR -- the link probability is proportional to $\frac{1}{d^2(u,v)}-v_{min}$, which represents the value of signal strength that is above a minimum threshold; there is a minimum range $r$ under which the link probability is one, and a maximum range $R$ over which the link probability is zero (Figure~\ref{exp_quasi},\ref{exp_sinr},\ref{exp_quasi_e},\ref{exp_sinr_e} and~\ref{exp_sinrScal}).
  \item Exponential -- the link probability is proportional to $\exp(-\frac{d(u,v)}{r_{avg}})$ (Figure~\ref{exp_exp},\ref{exp_exp_r} and~\ref{exp_exp_r_e}).
  \end{enumerate}
\item A localization error is optionally added to each position. This is done by selecting a direction in $[0,\pi)$ with uniform distribution, and a relative radius $r$ subject to a Gaussian distribution with mean value zero and standard deviation $\sigma_{err}$ (Figure~\ref{exp_quasi_e} to~\ref{exp_sinrScal}).
\item Links with apparent distance greater than a maximum range $R$ are optionally removed. The apparent distance is subject to the localization error (Figure~\ref{exp_exp_r} and~\ref{exp_exp_r_e}).
\item Simulations where the communication graph is not connected are discarded. Otherwise measurements are normally carried out.
\end{enumerate}
The terms used in the simulation reports are explained in Table~\ref{tab_key}. The source code of the program that was used to produce the network samples and measurements can be found at https://github.com/ajarry/geographic-eccentricity.

\begin{table}[h]
\begin{center}
\small
\renewcommand{\arraystretch}{1.1}
\begin{tabular}{l p{0.7\linewidth}}
\hline
Net. & network characteristics\\
$D$ & graph diameter\\
$N_i$ & average number of neighbors at $i$ hops\\
$k_T$ & largest distance in $G$ between two neighbors of the Delaunay triangulation\\
$k_e$ & embedding locality\\
$k_g$ & geographic eccentricity\\
$dk$ & $k_g-k_e$\\
$dN$ & $N_{k_g}-N_{k_e}$\\
\hline
$\Sigma$ & total number of runs\\
$\Delta$ & number of discarded runs because of a disconnected network\\
$\overline{x}$ & average value over 100 connected networks\\
$\sigma$ & standard deviation over 100 connected networks\\
\hline
\end{tabular}
\end{center}
\caption{Simulation report key.}
\label{tab_key}
\end{table}

\input{experiments.incl}

\end{document}

%% file: experiments.incl
\renewcommand{\arraystretch}{1}
\small

\begin{figure}[h]
\begin{center}
\includegraphics[width=9cm]{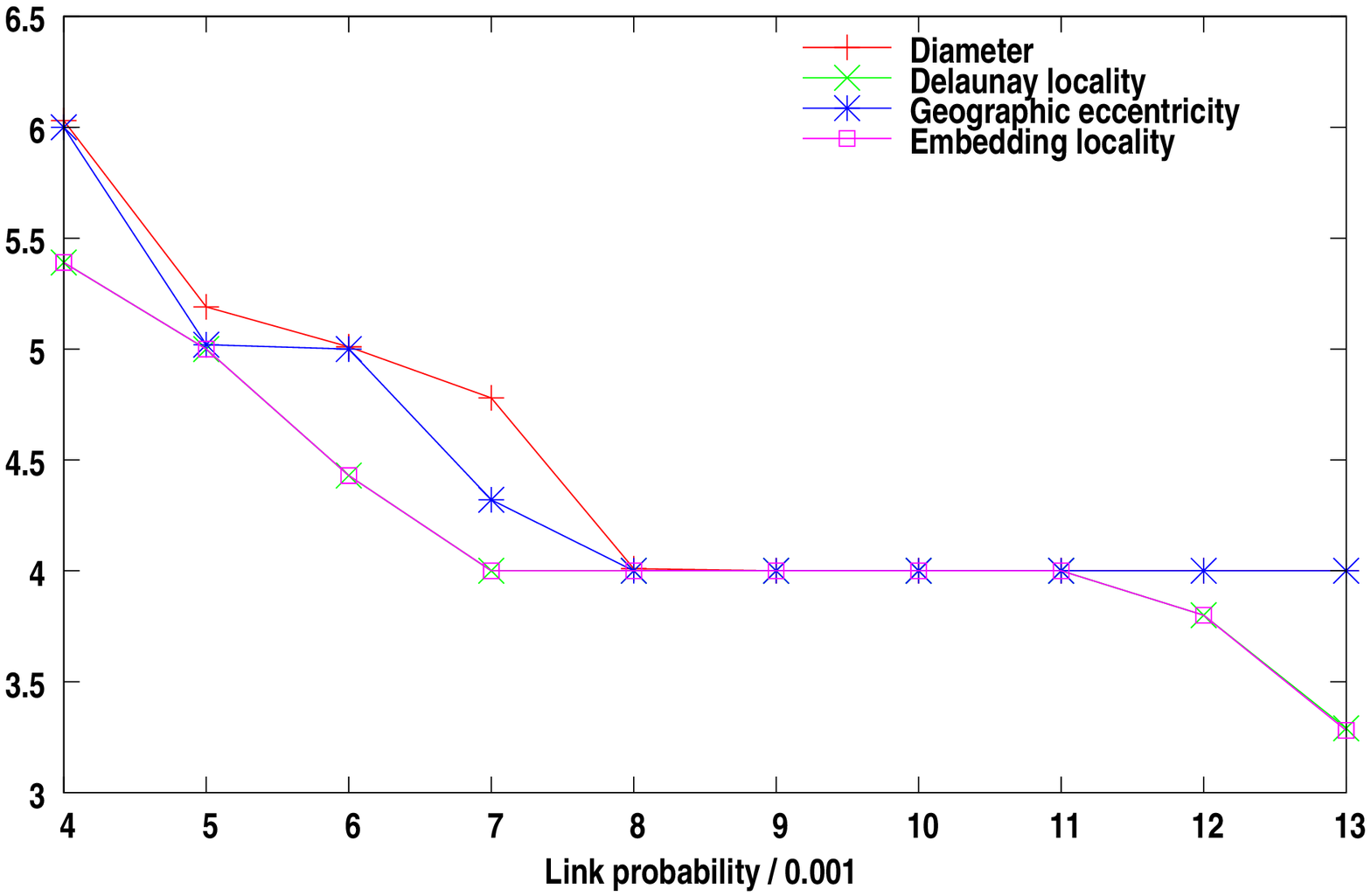}
~\\~\\
\begin{tabular}{|r|rr|rr|rr|rr|}
\hline
\multicolumn{3}{|c|}{Simulation} & \multicolumn{6}{|c|}{Communication graph}\\
\multicolumn{1}{|c}{Net.} & \multicolumn{2}{c|}{Runs} & \multicolumn{2}{|c}{$D$} & \multicolumn{2}{c}{$N_1$} & \multicolumn{2}{c|}{$k_T$}\\
\hline
\multicolumn{1}{|c|}{$p$} & \multicolumn{1}{|c}{$\Sigma$} & \multicolumn{1}{c|}{$\Delta$} & \multicolumn{1}{|c}{$\overline{x}$} & \multicolumn{1}{c|}{$\sigma$} & \multicolumn{1}{|c}{$\overline{x}$} & \multicolumn{1}{c|}{$\sigma$} & \multicolumn{1}{|c}{$\overline{x}$} & \multicolumn{1}{c|}{$\sigma$}\\
0.004 & 119 & 19 & 6.03 & 0.17 & 10.98 & 0.08 & 5.39 & 0.49\\
0.005 & 100 &  0 & 5.19 & 0.39 & 13.50 & 0.11 & 5.00 & 0.00\\
0.006 & 100 &  0 & 5.01 & 0.10 & 15.98 & 0.10 & 4.43 & 0.50\\
0.007 & 100 &  0 & 4.78 & 0.41 & 18.49 & 0.12 & 4.00 & 0.00\\
0.008 & 100 &  0 & 4.01 & 0.10 & 20.98 & 0.11 & 4.00 & 0.00\\
0.009 & 100 &  0 & 4.00 & 0.00 & 23.48 & 0.13 & 4.00 & 0.00\\
0.010 & 100 &  0 & 4.00 & 0.00 & 26.01 & 0.14 & 4.00 & 0.00\\
0.011 & 100 &  0 & 4.00 & 0.00 & 28.50 & 0.15 & 4.00 & 0.00\\
0.012 & 100 &  0 & 4.00 & 0.00 & 31.01 & 0.17 & 3.80 & 0.40\\
0.013 & 100 &  0 & 4.00 & 0.00 & 33.46 & 0.15 & 3.29 & 0.45\\
\hline
\end{tabular}\\
~\\~\\
\begin{tabular}{|r|rr|rr|rr|rr|rr|}
\hline
& \multicolumn{4}{|c|}{Embedding locality} & \multicolumn{4}{|c|}{Geographic eccentricity} & \multicolumn{2}{|c|}{Difference}\\
Net. & \multicolumn{2}{c}{$k_e$} & \multicolumn{2}{c|}{$N_{k_e}$} & \multicolumn{2}{c}{$k_g$} & \multicolumn{2}{c|}{$N_{k_g}$}& $dk$ & $dN$\\
\hline
\multicolumn{1}{|c|}{$p$} & \multicolumn{1}{|c}{$\overline{x}$} & \multicolumn{1}{c|}{$\sigma$} & \multicolumn{1}{|c}{$\overline{x}$} & \multicolumn{1}{c|}{$\sigma$} & \multicolumn{1}{|c}{$\overline{x}$} & \multicolumn{1}{c|}{$\sigma$} & \multicolumn{1}{|c}{$\overline{x}$} & \multicolumn{1}{c|}{$\sigma$} & \multicolumn{1}{|c}{$\sigma$} & \multicolumn{1}{c|}{$\sigma$}\\
0.004 & 3.28 & 0.45 & 2499.93 & 0.05 & 4.00 & 0.00 & 2500.00 & 0.00 & 0.45 & 0.05\\
0.005 & 5.00 & 0.00 & 2500.00 & 0.00 & 5.02 & 0.14 & 2500.00 & 0.00 & 0.14 & 0.00\\
0.006 & 4.43 & 0.50 & 2499.90 & 0.10 & 5.00 & 0.00 & 2500.00 & 0.00 & 0.50 & 0.10\\
0.007 & 4.00 & 0.00 & 2500.00 & 0.02 & 4.32 & 0.47 & 2500.00 & 0.00 & 0.47 & 0.02\\
0.008 & 4.00 & 0.00 & 2500.00 & 0.00 & 4.00 & 0.00 & 2500.00 & 0.00 & 0.00 & 0.00\\
0.009 & 4.00 & 0.00 & 2500.00 & 0.00 & 4.00 & 0.00 & 2500.00 & 0.00 & 0.00 & 0.00\\
0.010 & 4.00 & 0.00 & 2500.00 & 0.00 & 4.00 & 0.00 & 2500.00 & 0.00 & 0.00 & 0.00\\
0.011 & 4.00 & 0.00 & 2500.00 & 0.00 & 4.00 & 0.00 & 2500.00 & 0.00 & 0.00 & 0.00\\
0.012 & 3.80 & 0.40 & 2499.89 & 0.22 & 4.00 & 0.00 & 2500.00 & 0.00 & 0.40 & 0.22\\
0.013 & 3.28 & 0.45 & 2499.93 & 0.05 & 4.00 & 0.00 & 2500.00 & 0.00 & 0.45 & 0.05\\
\hline
\end{tabular}
\caption{Random graph metrics with link probability $p$.}
\label{exp_random}
\end{center}
\end{figure}

\begin{figure}[h]
\begin{center}
\includegraphics[width=9cm]{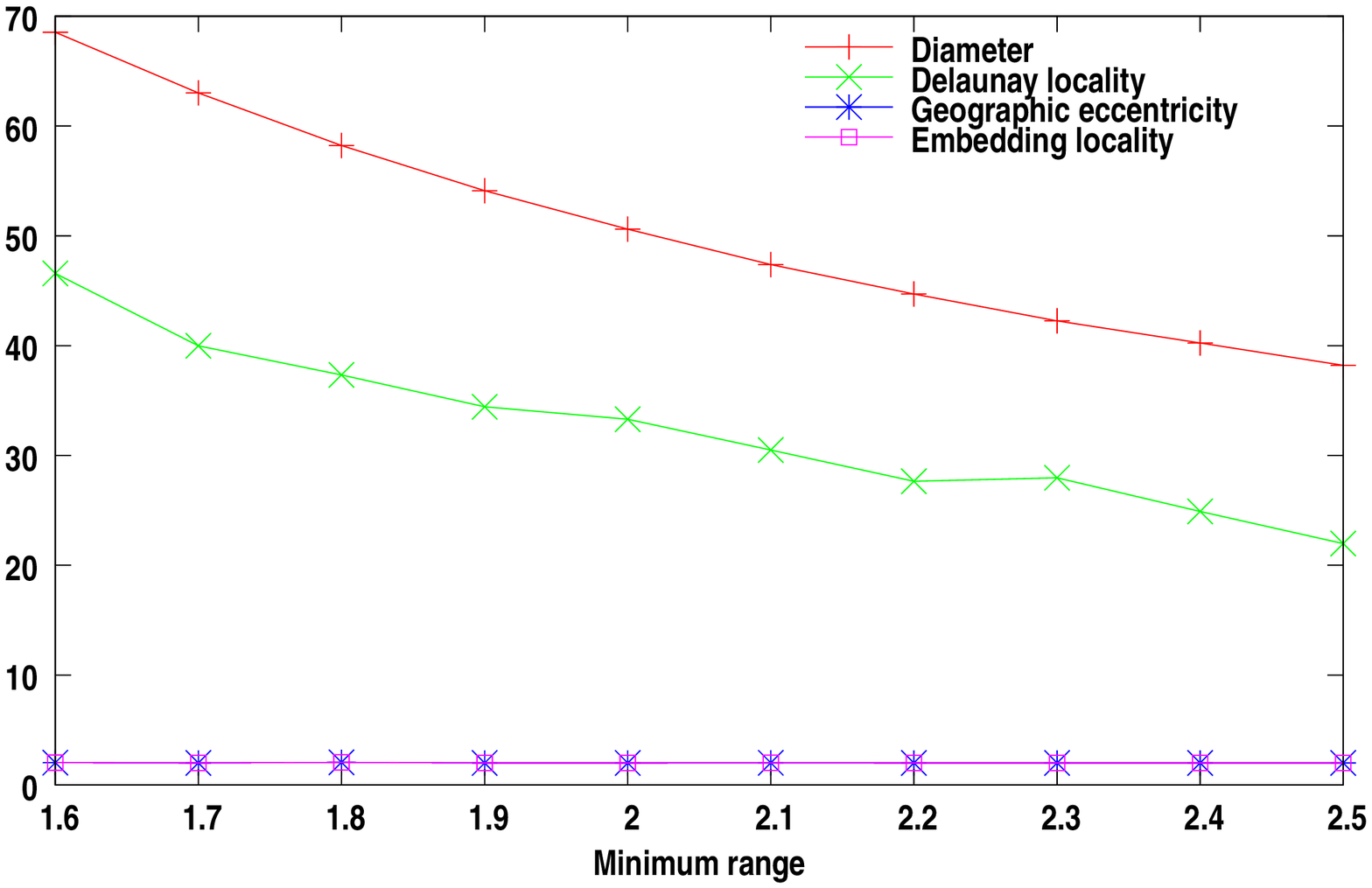}
~\\~\\
\begin{tabular}{|rr|rr|rr|rr|rr|}
\hline
\multicolumn{4}{|c|}{Simulation} & \multicolumn{6}{|c|}{Communication graph}\\
\multicolumn{2}{|c}{Net.} & \multicolumn{2}{c|}{Runs} & \multicolumn{2}{|c}{$D$} & \multicolumn{2}{c}{$N_1$} & \multicolumn{2}{c|}{$k_T$}\\
\hline
\multicolumn{1}{|c}{$r$} & \multicolumn{1}{c|}{$R$} & \multicolumn{1}{|c}{$\Sigma$} & \multicolumn{1}{c|}{$\Delta$} & \multicolumn{1}{|c}{$\overline{x}$} & \multicolumn{1}{c|}{$\sigma$} & \multicolumn{1}{|c}{$\overline{x}$} & \multicolumn{1}{c|}{$\sigma$} & \multicolumn{1}{|c}{$\overline{x}$} & \multicolumn{1}{c|}{$\sigma$}\\
1.60 & 2.24 & 160 & 60 & 68.54 & 1.00 & 11.59 & 0.11 & 46.58 & 9.79\\
1.70 & 2.38 & 131 & 31 & 63.01 & 0.99 & 12.94 & 0.12 & 39.99 & 8.74\\
1.80 & 2.52 & 116 & 16 & 58.23 & 0.85 & 14.35 & 0.12 & 37.33 & 7.64\\
1.90 & 2.66 & 105 &  5 & 54.11 & 0.82 & 15.82 & 0.13 & 34.43 & 7.85\\
2.00 & 2.80 & 103 &  3 & 50.61 & 0.77 & 17.38 & 0.15 & 33.30 & 7.83\\
2.10 & 2.94 & 102 &  2 & 47.38 & 0.58 & 19.00 & 0.16 & 30.49 & 7.06\\
2.20 & 3.08 & 101 &  1 & 44.70 & 0.61 & 20.71 & 0.17 & 27.65 & 6.40\\
2.30 & 3.22 & 100 &  0 & 42.26 & 0.56 & 22.53 & 0.17 & 27.96 & 6.75\\
2.40 & 3.36 & 100 &  0 & 40.25 & 0.48 & 24.38 & 0.17 & 24.90 & 5.47\\
2.50 & 3.50 & 100 &  0 & 38.21 & 0.45 & 26.26 & 0.21 & 21.97 & 5.05\\
\hline
\end{tabular}\\
~\\~\\
\begin{tabular}{|r|rr|rr|rr|rr|rr|}
\hline
& \multicolumn{4}{|c|}{Embedding locality} & \multicolumn{4}{|c|}{Geographic eccentricity} & \multicolumn{2}{|c|}{Difference}\\
Net. & \multicolumn{2}{c}{$k_e$} & \multicolumn{2}{c|}{$N_{k_e}$} & \multicolumn{2}{c}{$k_g$} & \multicolumn{2}{c|}{$N_{k_g}$}& $dk$ & $dN$\\
\hline
\multicolumn{1}{|c|}{$r$} & \multicolumn{1}{|c}{$\overline{x}$} & \multicolumn{1}{c|}{$\sigma$} & \multicolumn{1}{|c}{$\overline{x}$} & \multicolumn{1}{c|}{$\sigma$} & \multicolumn{1}{|c}{$\overline{x}$} & \multicolumn{1}{c|}{$\sigma$} & \multicolumn{1}{|c}{$\overline{x}$} & \multicolumn{1}{c|}{$\sigma$} & \multicolumn{1}{|c}{$\sigma$} & \multicolumn{1}{c|}{$\sigma$}\\
1.60 & 2.02 & 0.14 & 31.82 & 4.07 & 2.02 & 0.14 & 31.82 & 4.07 & 0.00 & 0.00\\
1.70 & 2.00 & 0.00 & 35.99 & 0.39 & 2.00 & 0.00 & 35.99 & 0.39 & 0.00 & 0.00\\
1.80 & 2.05 & 0.22 & 43.09 & 8.98 & 2.05 & 0.22 & 43.09 & 8.98 & 0.00 & 0.00\\
1.90 & 2.00 & 0.00 & 46.42 & 0.49 & 2.00 & 0.00 & 46.42 & 0.49 & 0.00 & 0.00\\
2.00 & 2.00 & 0.00 & 52.10 & 0.58 & 2.00 & 0.00 & 52.10 & 0.58 & 0.00 & 0.00\\
2.10 & 2.01 & 0.10 & 58.63 & 5.84 & 2.01 & 0.10 & 58.63 & 5.84 & 0.00 & 0.00\\
2.20 & 2.00 & 0.00 & 64.37 & 0.65 & 2.00 & 0.00 & 64.37 & 0.65 & 0.00 & 0.00\\
2.30 & 2.00 & 0.00 & 71.03 & 0.70 & 2.00 & 0.00 & 71.03 & 0.70 & 0.00 & 0.00\\
2.40 & 2.00 & 0.00 & 77.95 & 0.68 & 2.00 & 0.00 & 77.95 & 0.68 & 0.00 & 0.00\\
2.50 & 2.00 & 0.00 & 84.93 & 0.77 & 2.00 & 0.00 & 84.93 & 0.77 & 0.00 & 0.00\\
\hline
\end{tabular}
\caption{Quasi unit disc graph metrics based on the SINR probability model with minimum range $r$ and maximum range $R = 1.4r$.}
\label{exp_quasi}
\end{center}
\end{figure}

\begin{figure}[h]
\begin{center}
\includegraphics[width=9cm]{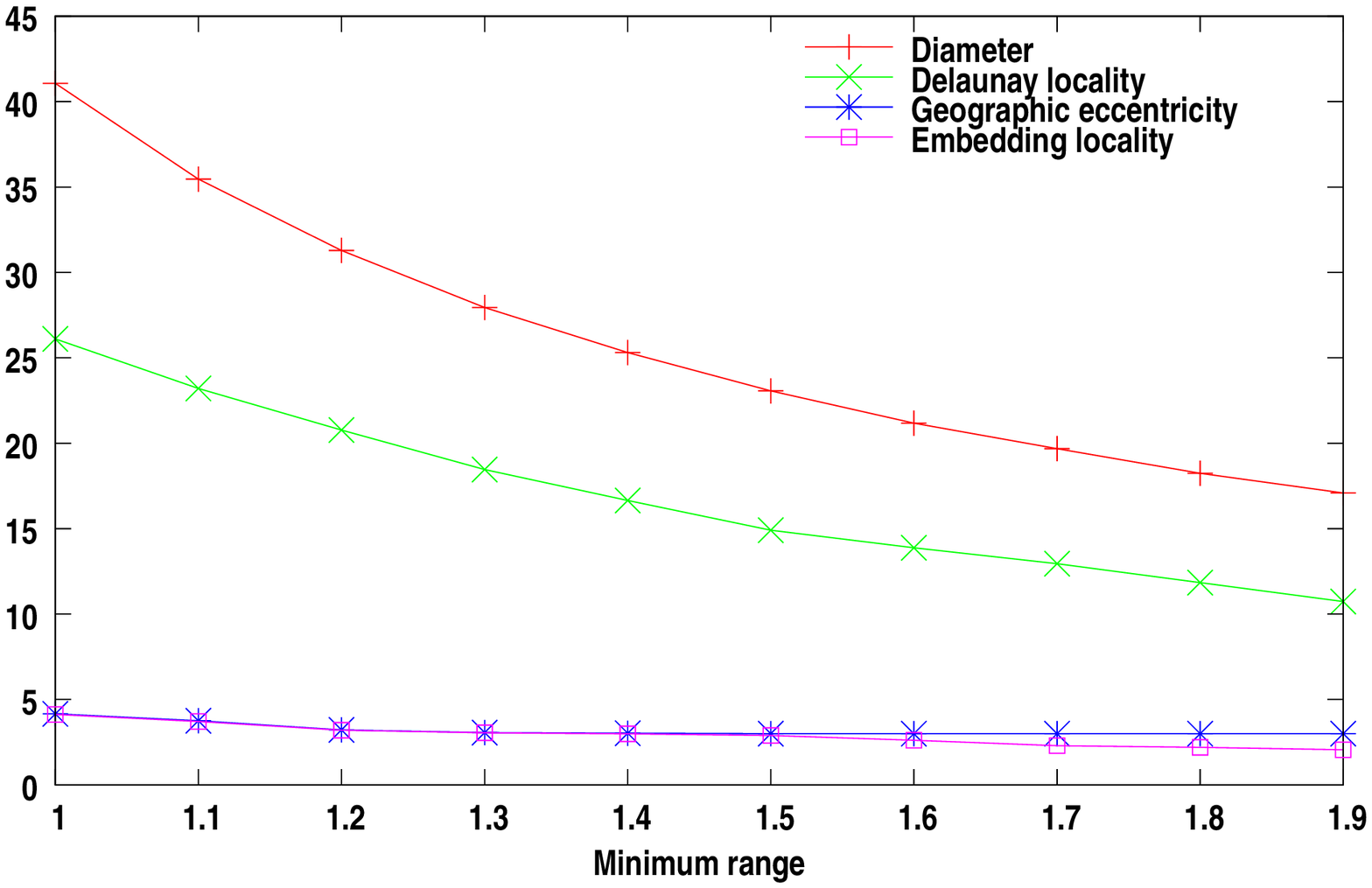}
~\\~\\
\begin{tabular}{|rr|rr|rr|rr|rr|}
\hline
\multicolumn{4}{|c|}{Simulation} & \multicolumn{6}{|c|}{Communication graph}\\
\multicolumn{2}{|c}{Net.} & \multicolumn{2}{c|}{Runs} & \multicolumn{2}{|c}{$D$} & \multicolumn{2}{c}{$N_1$} & \multicolumn{2}{c|}{$k_T$}\\
\hline
\multicolumn{1}{|c}{$r$} & \multicolumn{1}{c|}{$R$} & \multicolumn{1}{|c}{$\Sigma$} & \multicolumn{1}{c|}{$\Delta$} & \multicolumn{1}{|c}{$\overline{x}$} & \multicolumn{1}{c|}{$\sigma$} & \multicolumn{1}{|c}{$\overline{x}$} & \multicolumn{1}{c|}{$\sigma$} & \multicolumn{1}{|c}{$\overline{x}$} & \multicolumn{1}{c|}{$\sigma$}\\
1.00 & 5.00 & 164 & 64 & 41.07 & 0.91 & 10.98 & 0.09 & 26.11 & 6.03\\
1.10 & 5.50 & 117 & 17 & 35.46 & 0.93 & 13.02 & 0.11 & 23.21 & 4.32\\
1.20 & 6.00 & 111 & 11 & 31.29 & 0.71 & 15.21 & 0.13 & 20.77 & 4.23\\
1.30 & 6.50 & 104 &  4 & 27.95 & 0.54 & 17.58 & 0.15 & 18.46 & 3.56\\
1.40 & 7.00 & 100 &  0 & 25.31 & 0.52 & 20.13 & 0.17 & 16.65 & 3.34\\
1.50 & 7.50 & 100 &  0 & 23.07 & 0.47 & 22.85 & 0.18 & 14.91 & 2.80\\
1.60 & 8.00 & 100 &  0 & 21.18 & 0.41 & 25.74 & 0.21 & 13.88 & 3.14\\
1.70 & 8.50 & 100 &  0 & 19.68 & 0.49 & 28.78 & 0.24 & 12.95 & 2.57\\
1.80 & 9.00 & 100 &  0 & 18.24 & 0.43 & 31.94 & 0.25 & 11.84 & 2.34\\
1.90 & 9.50 & 100 &  0 & 17.09 & 0.29 & 35.29 & 0.29 & 10.73 & 2.37\\
\hline
\end{tabular}\\
~\\~\\
\begin{tabular}{|r|rr|rr|rr|rr|rr|}
\hline
& \multicolumn{4}{|c|}{Embedding locality} & \multicolumn{4}{|c|}{Geographic eccentricity} & \multicolumn{2}{|c|}{Difference}\\
Net. & \multicolumn{2}{c}{$k_e$} & \multicolumn{2}{c|}{$N_{k_e}$} & \multicolumn{2}{c}{$k_g$} & \multicolumn{2}{c|}{$N_{k_g}$}& $dk$ & $dN$\\
\hline
\multicolumn{1}{|c|}{$r$} & \multicolumn{1}{|c}{$\overline{x}$} & \multicolumn{1}{c|}{$\sigma$} & \multicolumn{1}{|c}{$\overline{x}$} & \multicolumn{1}{c|}{$\sigma$} & \multicolumn{1}{|c}{$\overline{x}$} & \multicolumn{1}{c|}{$\sigma$} & \multicolumn{1}{|c}{$\overline{x}$} & \multicolumn{1}{c|}{$\sigma$} & \multicolumn{1}{|c}{$\sigma$} & \multicolumn{1}{c|}{$\sigma$}\\
1.00 & 4.12 & 0.47 & 216.82 & 51.93 & 4.16 & 0.42 & 220.45 & 47.40 & 0.20 & 17.78\\
1.10 & 3.71 & 0.50 & 230.75 & 59.80 & 3.75 & 0.48 & 235.40 & 57.44 & 0.20 & 22.77\\
1.20 & 3.20 & 0.42 & 215.42 & 63.58 & 3.22 & 0.44 & 218.37 & 65.53 & 0.14 & 20.63\\
1.30 & 3.05 & 0.22 & 238.10 & 38.95 & 3.06 & 0.24 & 239.88 & 42.43 & 0.10 & 17.74\\
1.40 & 2.99 & 0.17 & 275.32 & 30.96 & 3.02 & 0.14 & 280.82 & 30.28 & 0.17 & 31.56\\
1.50 & 2.89 & 0.31 & 306.36 & 60.68 & 3.00 & 0.00 & 327.89 &  4.84 & 0.31 & 61.25\\
1.60 & 2.61 & 0.49 & 293.91 & 109.70 & 3.00 & 0.00 & 381.81 & 6.18 & 0.49 & 109.96\\
1.70 & 2.29 & 0.45 & 255.99 & 117.33 & 3.00 & 0.00 & 438.12 & 6.57 & 0.45 & 116.45\\
1.80 & 2.20 & 0.40 & 264.33 & 114.93 & 3.00 & 0.00 & 495.10 & 6.08 & 0.40 & 115.43\\
1.90 & 2.06 & 0.24 & 253.87 &  76.20 & 3.00 & 0.00 & 555.47 & 6.68 & 0.24 &  76.30\\
\hline
\end{tabular}
\caption{Network metrics based on the SINR communication model with minimum range $r$ and maximum range $R = 5r$.}
\label{exp_sinr}
\end{center}
\end{figure}

\begin{figure}[h]
\begin{center}
\includegraphics[width=9cm]{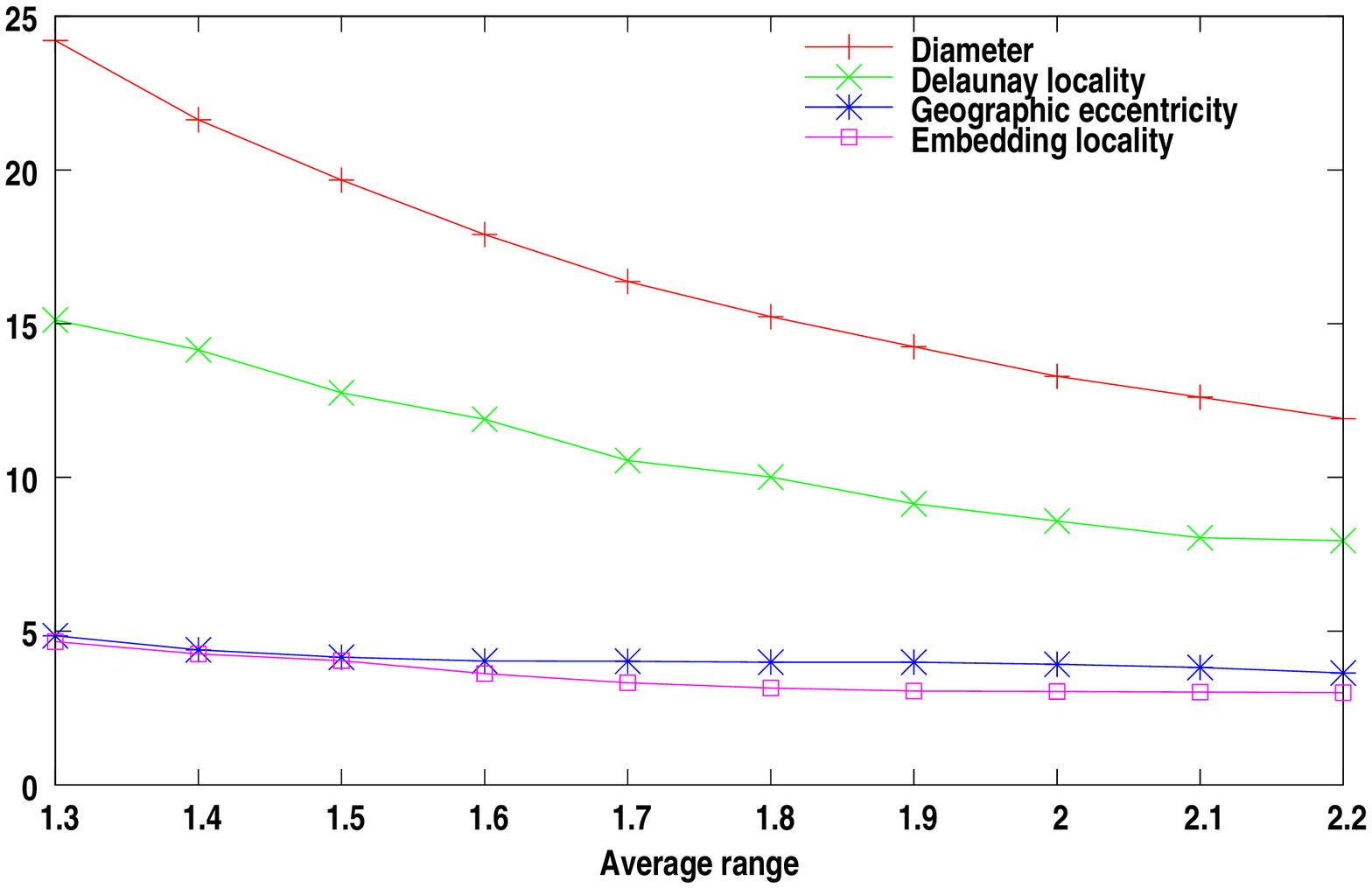}
~\\~\\
\begin{tabular}{|r|rr|rr|rr|rr|}
\hline
\multicolumn{3}{|c|}{Simulation} & \multicolumn{6}{|c|}{Communication graph}\\
\multicolumn{1}{|c}{Net.} & \multicolumn{2}{c|}{Runs} & \multicolumn{2}{|c}{$D$} & \multicolumn{2}{c}{$N_1$} & \multicolumn{2}{c|}{$k_T$}\\
\hline
\multicolumn{1}{|c|}{$r_{avg}$} & \multicolumn{1}{|c}{$\Sigma$} & \multicolumn{1}{c|}{$\Delta$} & \multicolumn{1}{|c}{$\overline{x}$} & \multicolumn{1}{c|}{$\sigma$} & \multicolumn{1}{|c}{$\overline{x}$} & \multicolumn{1}{c|}{$\sigma$} & \multicolumn{1}{|c}{$\overline{x}$} & \multicolumn{1}{c|}{$\sigma$}\\
1.30 & 188 & 88 & 24.21 & 0.90 & 10.75 & 0.10 & 15.12 & 3.01\\
1.40 & 135 & 35 & 21.63 & 0.70 & 12.23 & 0.11 & 14.15 & 2.82\\
1.50 & 117 & 17 & 19.67 & 0.69 & 13.80 & 0.10 & 12.75 & 2.56\\
1.60 & 108 &  8 & 17.90 & 0.69 & 15.48 & 0.13 & 11.89 & 2.34\\
1.70 & 103 &  3 & 16.37 & 0.50 & 17.21 & 0.15 & 10.55 & 1.94\\
1.80 & 103 &  3 & 15.23 & 0.55 & 19.05 & 0.17 & 10.01 & 1.92\\
1.90 & 102 &  2 & 14.25 & 0.46 & 21.01 & 0.18 &  9.14 & 1.76\\
2.00 & 100 &  0 & 13.29 & 0.48 & 23.00 & 0.22 &  8.58 & 1.55\\
2.10 & 100 &  0 & 12.61 & 0.51 & 25.11 & 0.25 &  8.04 & 1.74\\ 
2.20 & 100 &  0 & 11.91 & 0.35 & 27.24 & 0.22 &  7.94 & 1.34\\
\hline
\end{tabular}\\
~\\~\\
\begin{tabular}{|r|rr|rr|rr|rr|rr|}
\hline
& \multicolumn{4}{|c|}{Embedding locality} & \multicolumn{4}{|c|}{Geographic eccentricity} & \multicolumn{2}{|c|}{Difference}\\
Net. & \multicolumn{2}{c}{$k_e$} & \multicolumn{2}{c|}{$N_{k_e}$} & \multicolumn{2}{c}{$k_g$} & \multicolumn{2}{c|}{$N_{k_g}$}& $dk$ & $dN$\\
\hline
\multicolumn{1}{|c|}{$r_{avg}$} & \multicolumn{1}{|c}{$\overline{x}$} & \multicolumn{1}{c|}{$\sigma$} & \multicolumn{1}{|c}{$\overline{x}$} & \multicolumn{1}{c|}{$\sigma$} & \multicolumn{1}{|c}{$\overline{x}$} & \multicolumn{1}{c|}{$\sigma$} & \multicolumn{1}{|c}{$\overline{x}$} & \multicolumn{1}{c|}{$\sigma$} & \multicolumn{1}{|c}{$\sigma$} & \multicolumn{1}{c|}{$\sigma$}\\
1.30 & 4.66 & 0.68 & 585.33 & 164.99 & 4.85 & 0.62 &  632.48 & 151.29 & 0.39 & 97.39\\
1.40 & 4.26 & 0.46 & 592.01 & 128.25 & 4.39 & 0.53 &  628.11 & 146.15 & 0.34 & 93.44\\
1.50 & 4.04 & 0.40 & 634.33 & 121.40 & 4.15 & 0.36 &  667.75 & 110.53 & 0.31 & 95.09\\
1.60 & 3.62 & 0.49 & 596.62 & 166.53 & 4.03 & 0.17 &  736.38 &  58.44 & 0.53 & 181.04\\
1.70 & 3.32 & 0.49 & 569.88 & 182.03 & 4.02 & 0.14 &  832.04 &  49.43 & 0.48 & 179.16\\
1.80 & 3.15 & 0.36 & 581.01 & 144.73 & 3.99 & 0.10 &  921.96 &  42.10 & 0.37 & 148.97\\
1.90 & 3.05 & 0.22 & 614.67 &  96.39 & 3.99 & 0.10 & 1018.92 &  44.92 & 0.24 & 102.31\\
2.00 & 3.04 & 0.20 & 682.03 &  90.49 & 3.92 & 0.27 & 1076.31 & 120.89 & 0.32 & 145.80\\
2.10 & 3.02 & 0.14 & 747.01 &  65.91 & 3.82 & 0.38 & 1117.96 & 178.45 & 0.40 & 185.65\\
2.20 & 3.00 & 0.00 & 809.71 &  11.28 & 3.64 & 0.48 & 1112.66 & 228.36 & 0.48 & 227.35\\
\hline
\end{tabular}
\caption{Network metrics based on the exponential link probability model with average range $r_{avg}$.}
\label{exp_exp}
\end{center}
\end{figure}

\begin{figure}[h]
\begin{center}
\includegraphics[width=9cm]{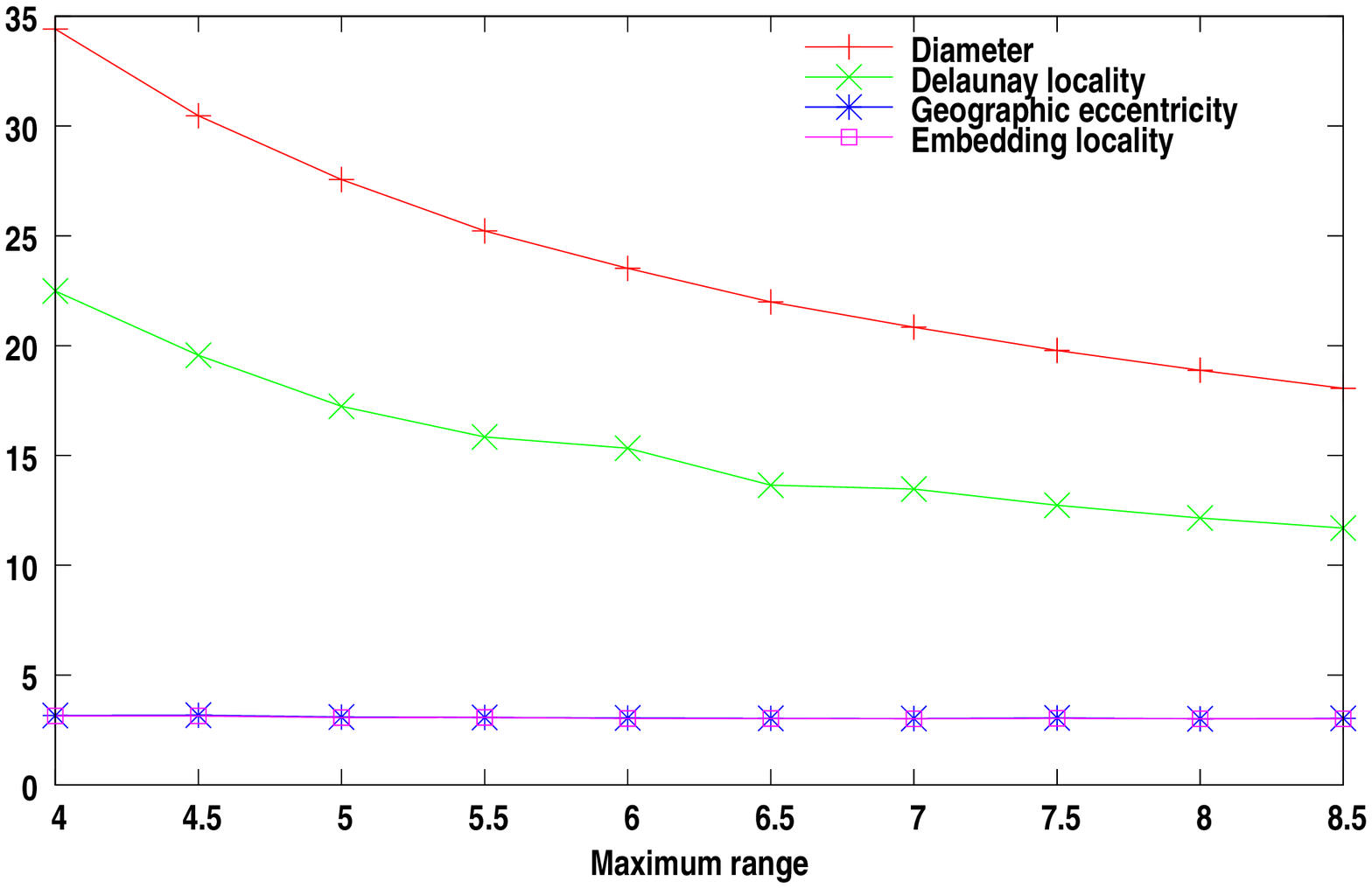}
~\\~\\
\begin{tabular}{|r|rr|rr|rr|rr|}
\hline
\multicolumn{3}{|c|}{Simulation} & \multicolumn{6}{|c|}{Communication graph}\\
\multicolumn{1}{|c}{Net.} & \multicolumn{2}{c|}{Runs} & \multicolumn{2}{|c}{$D$} & \multicolumn{2}{c}{$N_1$} & \multicolumn{2}{c|}{$k_T$}\\
\hline
\multicolumn{1}{|c|}{$R$} & \multicolumn{1}{|c}{$\Sigma$} & \multicolumn{1}{c|}{$\Delta$} & \multicolumn{1}{|c}{$\overline{x}$} & \multicolumn{1}{c|}{$\sigma$} & \multicolumn{1}{|c}{$\overline{x}$} & \multicolumn{1}{c|}{$\sigma$} & \multicolumn{1}{|c}{$\overline{x}$} & \multicolumn{1}{c|}{$\sigma$}\\
4.00 & 108 & 8 & 34.41 & 0.76 & 14.90 & 0.14 & 22.49 & 4.69\\
4.50 & 105 & 5 & 30.47 & 0.56 & 16.28 & 0.12 & 19.56 & 3.75\\
5.00 & 103 & 3 & 27.56 & 0.55 & 17.47 & 0.13 & 17.24 & 3.69\\
5.50 & 102 & 2 & 25.22 & 0.48 & 18.46 & 0.15 & 15.84 & 3.49\\
6.00 & 101 & 1 & 23.52 & 0.50 & 19.32 & 0.17 & 15.33 & 2.91\\
6.50 & 101 & 1 & 21.99 & 0.36 & 20.02 & 0.18 & 13.65 & 3.00\\
7.00 & 101 & 1 & 20.84 & 0.42 & 20.57 & 0.18 & 13.47 & 2.71\\
7.50 & 100 & 0 & 19.78 & 0.41 & 21.01 & 0.18 & 12.73 & 2.93\\
8.00 & 100 & 0 & 18.88 & 0.38 & 21.46 & 0.19 & 12.15 & 2.59\\
8.50 & 101 & 1 & 18.06 & 0.37 & 21.74 & 0.19 & 11.69 & 2.39\\
\hline
\end{tabular}\\
~\\~\\
\begin{tabular}{|r|rr|rr|rr|rr|rr|}
\hline
& \multicolumn{4}{|c|}{Embedding locality} & \multicolumn{4}{|c|}{Geographic eccentricity} & \multicolumn{2}{|c|}{Difference}\\
Net. & \multicolumn{2}{c}{$k_e$} & \multicolumn{2}{c|}{$N_{k_e}$} & \multicolumn{2}{c}{$k_g$} & \multicolumn{2}{c|}{$N_{k_g}$}& $dk$ & $dN$\\
\hline
\multicolumn{1}{|c|}{$R$} & \multicolumn{1}{|c}{$\overline{x}$} & \multicolumn{1}{c|}{$\sigma$} & \multicolumn{1}{|c}{$\overline{x}$} & \multicolumn{1}{c|}{$\sigma$} & \multicolumn{1}{|c}{$\overline{x}$} & \multicolumn{1}{c|}{$\sigma$} & \multicolumn{1}{|c}{$\overline{x}$} & \multicolumn{1}{c|}{$\sigma$} & \multicolumn{1}{|c}{$\sigma$} & \multicolumn{1}{c|}{$\sigma$}\\
4.00 & 3.14 & 0.35 & 200.85 & 45.89 & 3.16 & 0.37 & 203.54 & 48.58 & 0.14 & 18.81\\
4.50 & 3.14 & 0.35 & 243.97 & 56.24 & 3.17 & 0.38 & 248.82 & 60.75 & 0.17 & 27.54\\
5.00 & 3.06 & 0.24 & 272.36 & 45.81 & 3.09 & 0.29 & 278.24 & 56.11 & 0.17 & 33.45\\
5.50 & 3.07 & 0.26 & 313.49 & 55.87 & 3.07 & 0.26 & 313.49 & 55.87 & 0.00 & 0.00\\
6.00 & 3.02 & 0.14 & 340.48 & 34.58 & 3.05 & 0.22 & 347.69 & 52.95 & 0.17 & 41.00\\
6.50 & 3.02 & 0.14 & 375.36 & 36.80 & 3.03 & 0.17 & 377.95 & 44.59 & 0.10 & 25.79\\
7.00 & 3.01 & 0.10 & 405.85 & 29.00 & 3.02 & 0.14 & 408.71 & 40.13 & 0.10 & 28.38\\
7.50 & 3.03 & 0.17 & 440.25 & 53.51 & 3.05 & 0.22 & 446.31 & 67.21 & 0.14 & 42.41\\
8.00 & 3.01 & 0.10 & 463.18 & 29.89 & 3.01 & 0.10 & 463.18 & 29.89 & 0.00 & 0.00\\
8.50 & 3.01 & 0.10 & 487.87 & 32.45 & 3.03 & 0.17 & 494.59 & 57.68 & 0.14 & 47.05\\
\hline
\end{tabular}
\caption{Network metrics based on the exponential link probability model with average range $2$. Links longer than a maximum range $R$ have then been removed.}
\label{exp_exp_r}
\end{center}
\end{figure}

\begin{figure}[h]
\begin{center}
\includegraphics[width=9cm]{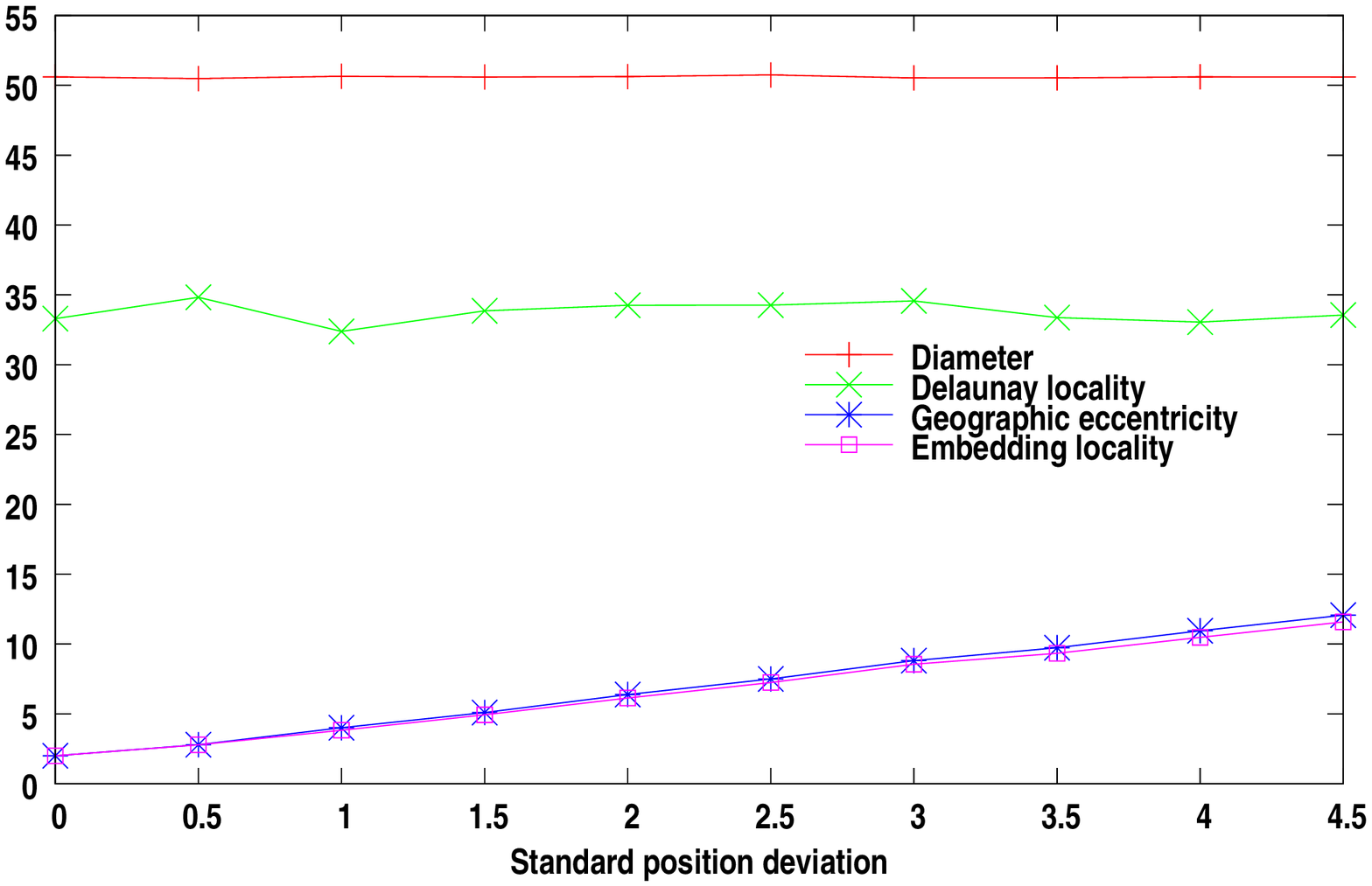}
~\\~\\
\begin{tabular}{|r|rr|rr|rr|rr|}
\hline
\multicolumn{3}{|c|}{Simulation} & \multicolumn{6}{|c|}{Communication graph}\\
\multicolumn{1}{|c}{Net.} & \multicolumn{2}{c|}{Runs} & \multicolumn{2}{|c}{$D$} & \multicolumn{2}{c}{$N_1$} & \multicolumn{2}{c|}{$k_T$}\\
\hline
\multicolumn{1}{|c|}{$\sigma_{err}$} & \multicolumn{1}{|c}{$\Sigma$} & \multicolumn{1}{c|}{$\Delta$} & \multicolumn{1}{|c}{$\overline{x}$} & \multicolumn{1}{c|}{$\sigma$} & \multicolumn{1}{|c}{$\overline{x}$} & \multicolumn{1}{c|}{$\sigma$} & \multicolumn{1}{|c}{$\overline{x}$} & \multicolumn{1}{c|}{$\sigma$}\\
0.00 & 103 & 3 & 50.61 & 0.77 & 17.38 & 0.15 & 33.30 & 7.83\\
0.50 & 100 & 0 & 50.49 & 0.70 & 17.39 & 0.15 & 34.82 & 6.27\\
1.00 & 100 & 0 & 50.65 & 0.65 & 17.38 & 0.14 & 32.38 & 7.19\\
1.50 & 101 & 1 & 50.59 & 0.63 & 17.40 & 0.15 & 33.86 & 7.32\\
2.00 & 101 & 1 & 50.63 & 0.72 & 17.40 & 0.12 & 34.25 & 6.63\\
2.50 & 101 & 1 & 50.74 & 0.66 & 17.38 & 0.16 & 34.26 & 6.33\\
3.00 & 103 & 3 & 50.53 & 0.67 & 17.39 & 0.15 & 34.56 & 6.72\\
3.50 & 100 & 0 & 50.53 & 0.75 & 17.39 & 0.14 & 33.37 & 6.67\\
4.00 & 101 & 1 & 50.61 & 0.68 & 17.37 & 0.14 & 33.05 & 6.55\\
4.50 & 105 & 5 & 50.59 & 0.72 & 17.40 & 0.14 & 33.55 & 6.19\\
\hline
\end{tabular}\\
~\\~\\
\begin{tabular}{|r|rr|rr|rr|rr|rr|}
\hline
& \multicolumn{4}{|c|}{Embedding locality} & \multicolumn{4}{|c|}{Geographic eccentricity} & \multicolumn{2}{|c|}{Difference}\\
Net. & \multicolumn{2}{c}{$k_e$} & \multicolumn{2}{c|}{$N_{k_e}$} & \multicolumn{2}{c}{$k_g$} & \multicolumn{2}{c|}{$N_{k_g}$}& $dk$ & $dN$\\
\hline
\multicolumn{1}{|c|}{$\sigma_{err}$} & \multicolumn{1}{|c}{$\overline{x}$} & \multicolumn{1}{c|}{$\sigma$} & \multicolumn{1}{|c}{$\overline{x}$} & \multicolumn{1}{c|}{$\sigma$} & \multicolumn{1}{|c}{$\overline{x}$} & \multicolumn{1}{c|}{$\sigma$} & \multicolumn{1}{|c}{$\overline{x}$} & \multicolumn{1}{c|}{$\sigma$} & \multicolumn{1}{|c}{$\sigma$} & \multicolumn{1}{c|}{$\sigma$}\\
0.00 &  2.00 & 0.00 &  52.10 &  0.58 &  2.00 & 0.00 &  52.10 &  0.58 & 0.00 & 0.00\\
0.50 &  2.78 & 0.46 &  93.71 & 25.22 &  2.79 & 0.45 &  94.24 & 24.89 & 0.10 & 5.25\\
1.00 &  3.83 & 0.53 & 163.76 & 38.71 &  4.01 & 0.41 & 176.50 & 31.32 & 0.38 & 27.26\\
1.50 &  4.93 & 0.43 & 251.91 & 37.39 &  5.09 & 0.40 & 265.94 & 38.12 & 0.37 & 32.27\\
2.00 &  6.13 & 0.83 & 364.99 & 84.54 &  6.38 & 0.75 & 389.29 & 77.11 & 0.46 & 44.49\\
2.50 &  7.24 & 0.67 & 477.17 & 70.20 &  7.50 & 0.62 & 504.37 & 66.70 & 0.50 & 52.74\\
3.00 &  8.55 & 0.75 & 616.00 & 80.75 &  8.81 & 0.70 & 644.07 & 75.27 & 0.48 & 52.08\\
3.50 &  9.34 & 0.79 & 701.74 & 84.24 &  9.73 & 0.77 & 743.27 & 80.48 & 0.55 & 58.06\\
4.00 & 10.48 & 0.79 & 819.90 & 80.12 & 10.95 & 0.75 & 868.16 & 74.79 & 0.62 & 64.03\\
4.50 & 11.58 & 0.94 & 932.29 & 87.32 & 12.07 & 0.95 & 979.29 & 86.60 & 0.77 & 73.08\\
\hline
\end{tabular}
\caption{Quasi unit disc graph metrics based on the SINR probability model with minimum range $2$ and maximum range $2.8$. A Gaussian localization error has been added with standard deviation $\sigma_{err}$.}
\label{exp_quasi_e}
\end{center}
\end{figure}

\begin{figure}[h]
\begin{center}
\includegraphics[width=9cm]{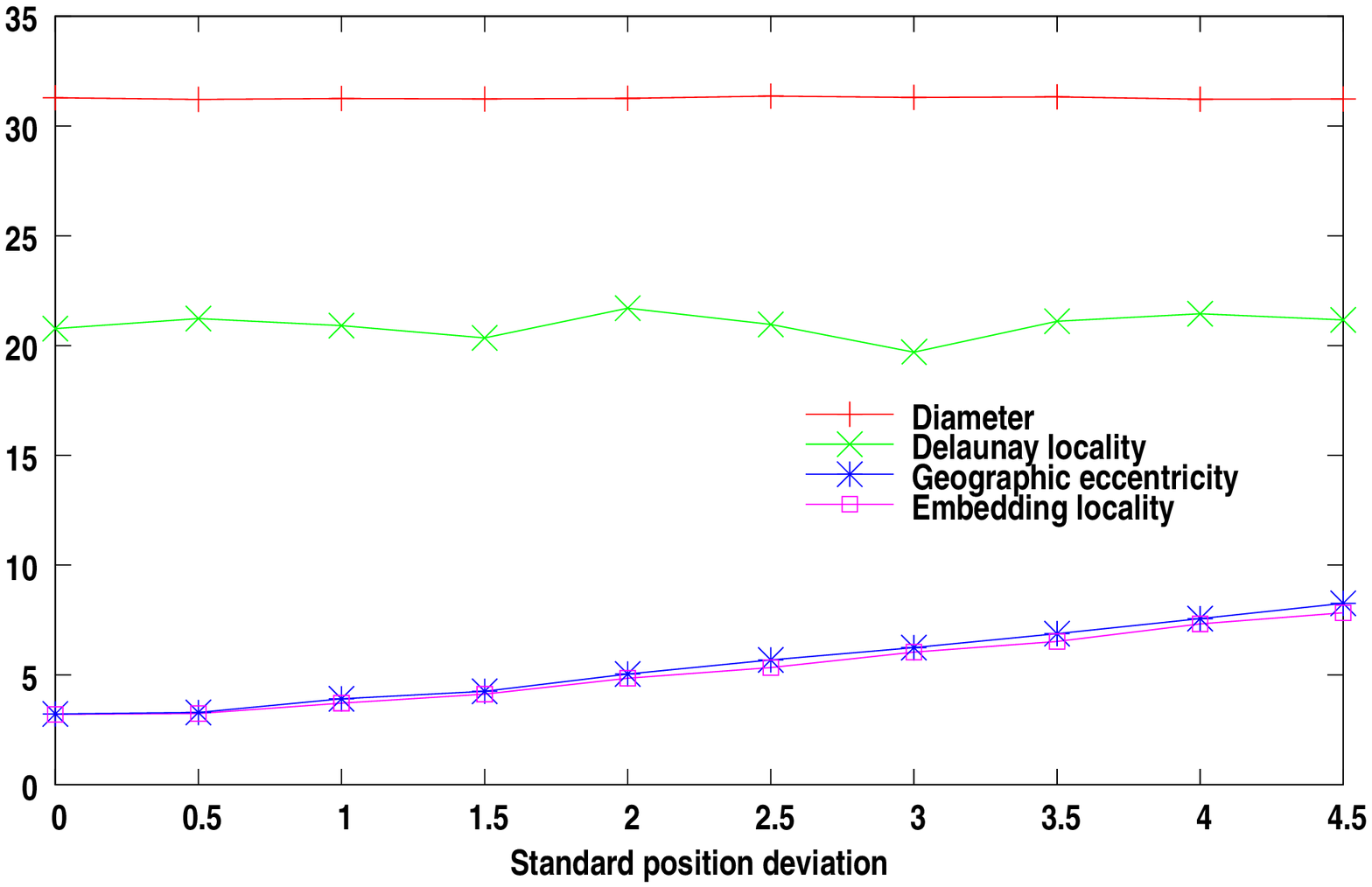}
~\\~\\
\begin{tabular}{|r|rr|rr|rr|rr|}
\hline
\multicolumn{3}{|c|}{Simulation} & \multicolumn{6}{|c|}{Communication graph}\\
\multicolumn{1}{|c}{Net.} & \multicolumn{2}{c|}{Runs} & \multicolumn{2}{|c}{$D$} & \multicolumn{2}{c}{$N_1$} & \multicolumn{2}{c|}{$k_T$}\\
\hline
\multicolumn{1}{|c|}{$\sigma_{err}$} & \multicolumn{1}{|c}{$\Sigma$} & \multicolumn{1}{c|}{$\Delta$} & \multicolumn{1}{|c}{$\overline{x}$} & \multicolumn{1}{c|}{$\sigma$} & \multicolumn{1}{|c}{$\overline{x}$} & \multicolumn{1}{c|}{$\sigma$} & \multicolumn{1}{|c}{$\overline{x}$} & \multicolumn{1}{c|}{$\sigma$}\\
0.00 & 111 & 11 & 31.29 & 0.71 & 15.21 & 0.13 & 20.77 & 4.23\\
0.50 & 105 &  5 & 31.21 & 0.71 & 15.23 & 0.14 & 21.23 & 4.13\\
1.00 & 107 &  7 & 31.25 & 0.67 & 15.21 & 0.13 & 20.91 & 4.76\\
1.50 & 100 &  0 & 31.23 & 0.66 & 15.20 & 0.12 & 20.35 & 4.08\\
2.00 & 105 &  5 & 31.26 & 0.72 & 15.23 & 0.13 & 21.70 & 4.48\\
2.50 & 103 &  3 & 31.36 & 0.74 & 15.22 & 0.12 & 20.96 & 3.80\\
3.00 & 103 &  3 & 31.30 & 0.71 & 15.22 & 0.13 & 19.70 & 3.75\\
3.50 & 108 &  8 & 31.33 & 0.65 & 15.20 & 0.14 & 21.11 & 4.31\\
4.00 & 105 &  5 & 31.22 & 0.58 & 15.21 & 0.12 & 21.45 & 3.88\\
4.50 & 109 &  9 & 31.23 & 0.71 & 15.21 & 0.14 & 21.17 & 4.25\\
\hline
\end{tabular}\\
~\\~\\
\begin{tabular}{|r|rr|rr|rr|rr|rr|}
\hline
& \multicolumn{4}{|c|}{Embedding locality} & \multicolumn{4}{|c|}{Geographic eccentricity} & \multicolumn{2}{|c|}{Difference}\\
Net. & \multicolumn{2}{c}{$k_e$} & \multicolumn{2}{c|}{$N_{k_e}$} & \multicolumn{2}{c}{$k_g$} & \multicolumn{2}{c|}{$N_{k_g}$}& $dk$ & $dN$\\
\hline
\multicolumn{1}{|c|}{$\sigma_{err}$} & \multicolumn{1}{|c}{$\overline{x}$} & \multicolumn{1}{c|}{$\sigma$} & \multicolumn{1}{|c}{$\overline{x}$} & \multicolumn{1}{c|}{$\sigma$} & \multicolumn{1}{|c}{$\overline{x}$} & \multicolumn{1}{c|}{$\sigma$} & \multicolumn{1}{|c}{$\overline{x}$} & \multicolumn{1}{c|}{$\sigma$} & \multicolumn{1}{|c}{$\sigma$} & \multicolumn{1}{c|}{$\sigma$}\\
0.00 & 3.20 & 0.42 & 215.42 & 63.58 & 3.22 & 0.44 & 218.37 & 65.53 & 0.14 & 20.63\\
0.50 & 3.24 & 0.43 & 221.40 & 62.29 & 3.29 & 0.45 & 228.90 & 66.79 & 0.22 & 32.73\\
1.00 & 3.71 & 0.52 & 291.31 & 78.46 & 3.91 & 0.38 & 320.84 & 58.40 & 0.40 & 59.06\\
1.50 & 4.12 & 0.35 & 354.02 & 61.37 & 4.25 & 0.46 & 376.42 & 78.87 & 0.34 & 57.95\\
2.00 & 4.85 & 0.46 & 481.09 & 81.16 & 5.04 & 0.28 & 513.63 & 51.17 & 0.39 & 67.21\\
2.50 & 5.33 & 0.51 & 565.92 & 93.80 & 5.68 & 0.56 & 629.42 & 101.86 & 0.50 & 89.99\\
3.00 & 6.04 & 0.47 & 692.52 & 81.22 & 6.24 & 0.47 & 728.40 & 80.48 & 0.40 & 71.77\\
3.50 & 6.52 & 0.59 & 778.40 & 104.66 & 6.88 & 0.57 & 841.43 & 99.71 & 0.50 & 87.54\\
4.00 & 7.32 & 0.53 & 917.19 & 83.70 & 7.56 & 0.57 & 955.88 & 88.74 & 0.43 & 68.91\\
4.50 & 7.83 & 0.63 & 997.90 & 99.47 & 8.26 & 0.63 & 1064.39 & 93.71 & 0.57 & 86.87\\
\hline
\end{tabular}
\caption{Network metrics based on the SINR communication model with minimum range $1.2$ and maximum range $6$. A Gaussian localization error has been added with standard deviation $\sigma_{err}$.}
\label{exp_sinr_e}
\end{center}
\end{figure}

\begin{figure}[h]
\begin{center}
\includegraphics[width=9cm]{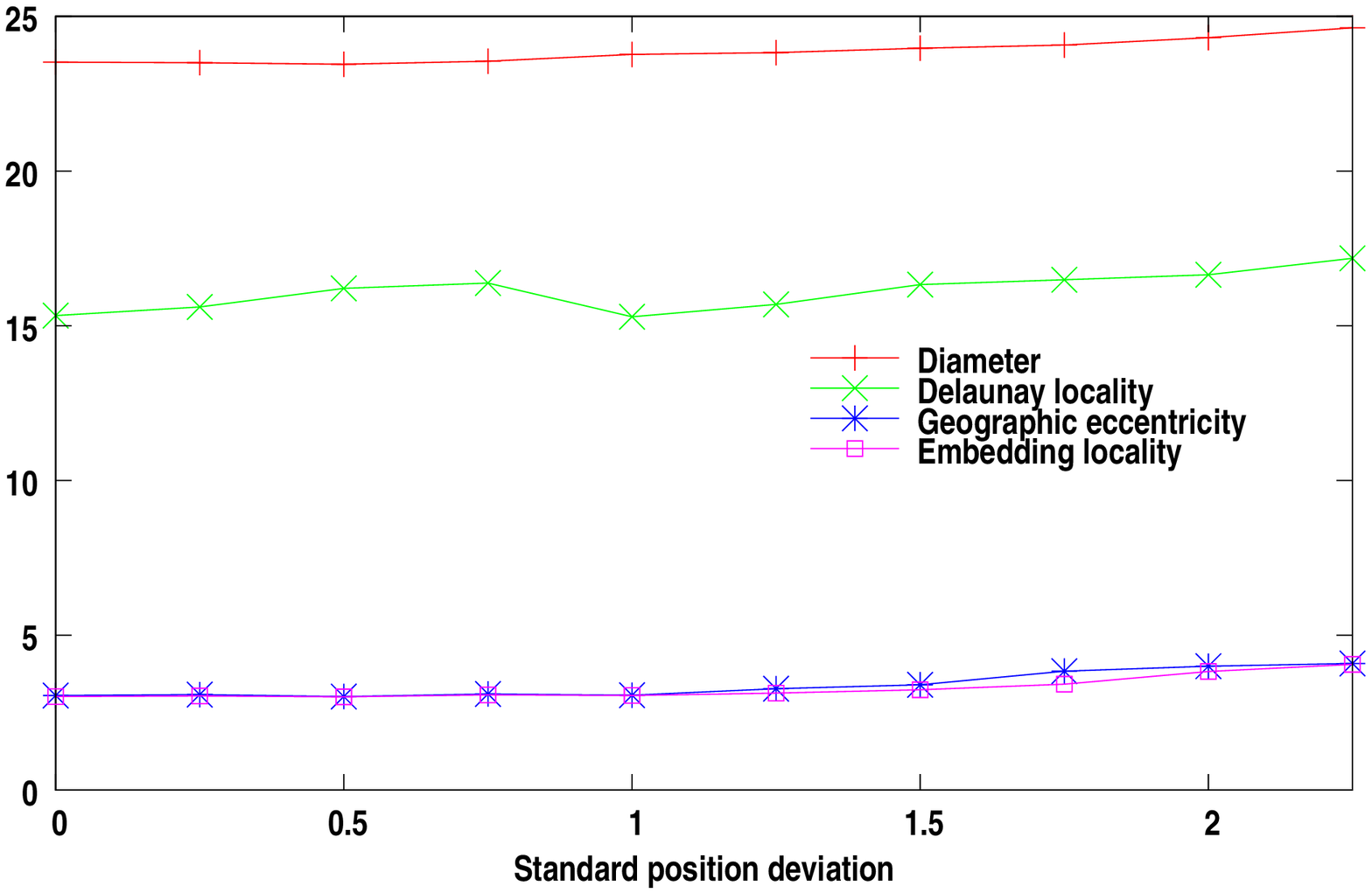}
~\\~\\
\begin{tabular}{|r|rr|rr|rr|rr|}
\hline
\multicolumn{3}{|c|}{Simulation} & \multicolumn{6}{|c|}{Communication graph}\\
\multicolumn{1}{|c}{Net.} & \multicolumn{2}{c|}{Runs} & \multicolumn{2}{|c}{$D$} & \multicolumn{2}{c}{$N_1$} & \multicolumn{2}{c|}{$k_T$}\\
\hline
\multicolumn{1}{|c|}{$\sigma_{err}$} & \multicolumn{1}{|c}{$\Sigma$} & \multicolumn{1}{c|}{$\Delta$} & \multicolumn{1}{|c}{$\overline{x}$} & \multicolumn{1}{c|}{$\sigma$} & \multicolumn{1}{|c}{$\overline{x}$} & \multicolumn{1}{c|}{$\sigma$} & \multicolumn{1}{|c}{$\overline{x}$} & \multicolumn{1}{c|}{$\sigma$}\\
0.00 & 101 &  1 & 23.52 & 0.50 & 19.32 & 0.17 & 15.33 & 2.91\\
0.25 & 101 &  1 & 23.50 & 0.54 & 19.28 & 0.17 & 15.61 & 3.09\\
0.50 & 100 &  0 & 23.45 & 0.52 & 19.21 & 0.15 & 16.21 & 3.02\\
0.75 & 102 &  2 & 23.55 & 0.54 & 19.05 & 0.17 & 16.38 & 3.10\\
1.00 & 103 &  3 & 23.77 & 0.55 & 18.87 & 0.14 & 15.29 & 3.20\\
1.25 & 104 &  4 & 23.83 & 0.51 & 18.60 & 0.17 & 15.69 & 2.80\\
1.50 & 103 &  3 & 23.97 & 0.43 & 18.30 & 0.16 & 16.33 & 3.30\\
1.75 & 108 &  8 & 24.07 & 0.45 & 17.89 & 0.14 & 16.49 & 3.09\\
2.00 & 132 & 32 & 24.31 & 0.48 & 17.41 & 0.16 & 16.65 & 3.11\\
2.25 & 174 & 74 & 24.63 & 0.56 & 16.86 & 0.17 & 17.18 & 3.09\\
\hline
\end{tabular}\\
~\\~\\
\begin{tabular}{|r|rr|rr|rr|rr|rr|}
\hline
& \multicolumn{4}{|c|}{Embedding locality} & \multicolumn{4}{|c|}{Geographic eccentricity} & \multicolumn{2}{|c|}{Difference}\\
Net. & \multicolumn{2}{c}{$k_e$} & \multicolumn{2}{c|}{$N_{k_e}$} & \multicolumn{2}{c}{$k_g$} & \multicolumn{2}{c|}{$N_{k_g}$}& $dk$ & $dN$\\
\hline
\multicolumn{1}{|c|}{$\sigma_{err}$} & \multicolumn{1}{|c}{$\overline{x}$} & \multicolumn{1}{c|}{$\sigma$} & \multicolumn{1}{|c}{$\overline{x}$} & \multicolumn{1}{c|}{$\sigma$} & \multicolumn{1}{|c}{$\overline{x}$} & \multicolumn{1}{c|}{$\sigma$} & \multicolumn{1}{|c}{$\overline{x}$} & \multicolumn{1}{c|}{$\sigma$} & \multicolumn{1}{|c}{$\sigma$} & \multicolumn{1}{c|}{$\sigma$}\\
0.00 & 3.02 & 0.14 & 340.48 &  34.58 & 3.05 & 0.22 & 347.69 &  52.95 & 0.17 & 41.00\\
0.25 & 3.04 & 0.20 & 344.28 &  47.85 & 3.08 & 0.27 & 353.82 &  65.45 & 0.20 & 46.76\\
0.50 & 3.01 & 0.10 & 335.83 &  24.89 & 3.02 & 0.14 & 338.23 &  34.17 & 0.10 & 23.84\\
0.75 & 3.07 & 0.26 & 347.58 &  60.58 & 3.10 & 0.30 & 354.70 &  71.18 & 0.17 & 40.51\\
1.00 & 3.05 & 0.22 & 339.63 &  52.14 & 3.06 & 0.28 & 342.03 &  66.33 & 0.10 & 23.90\\
1.25 & 3.13 & 0.34 & 354.75 &  79.28 & 3.27 & 0.44 & 387.71 & 104.58 & 0.35 & 81.69\\
1.50 & 3.24 & 0.43 & 376.16 & 100.81 & 3.40 & 0.49 & 413.47 & 114.74 & 0.37 & 85.49\\
1.75 & 3.42 & 0.49 & 412.64 & 114.71 & 3.84 & 0.37 & 510.51 &  85.57 & 0.49 & 115.02\\
2.00 & 3.83 & 0.38 & 498.63 &  86.04 & 4.00 & 0.00 & 537.77 &   7.49 & 0.38 & 86.49\\
2.25 & 4.06 & 0.24 & 541.32 &  58.57 & 4.09 & 0.29 & 548.31 &  68.49 & 0.17 & 39.73\\
\hline
\end{tabular}
\caption{Network metrics based on the exponential link probability model with average range $2$. A Gaussian localization error has been added with standard deviation $\sigma_{err}$. Links longer than a maximum range $6$ have then been removed.}
\label{exp_exp_r_e}
\end{center}
\end{figure}

\begin{figure}[h]
\begin{center}
\includegraphics[width=9cm]{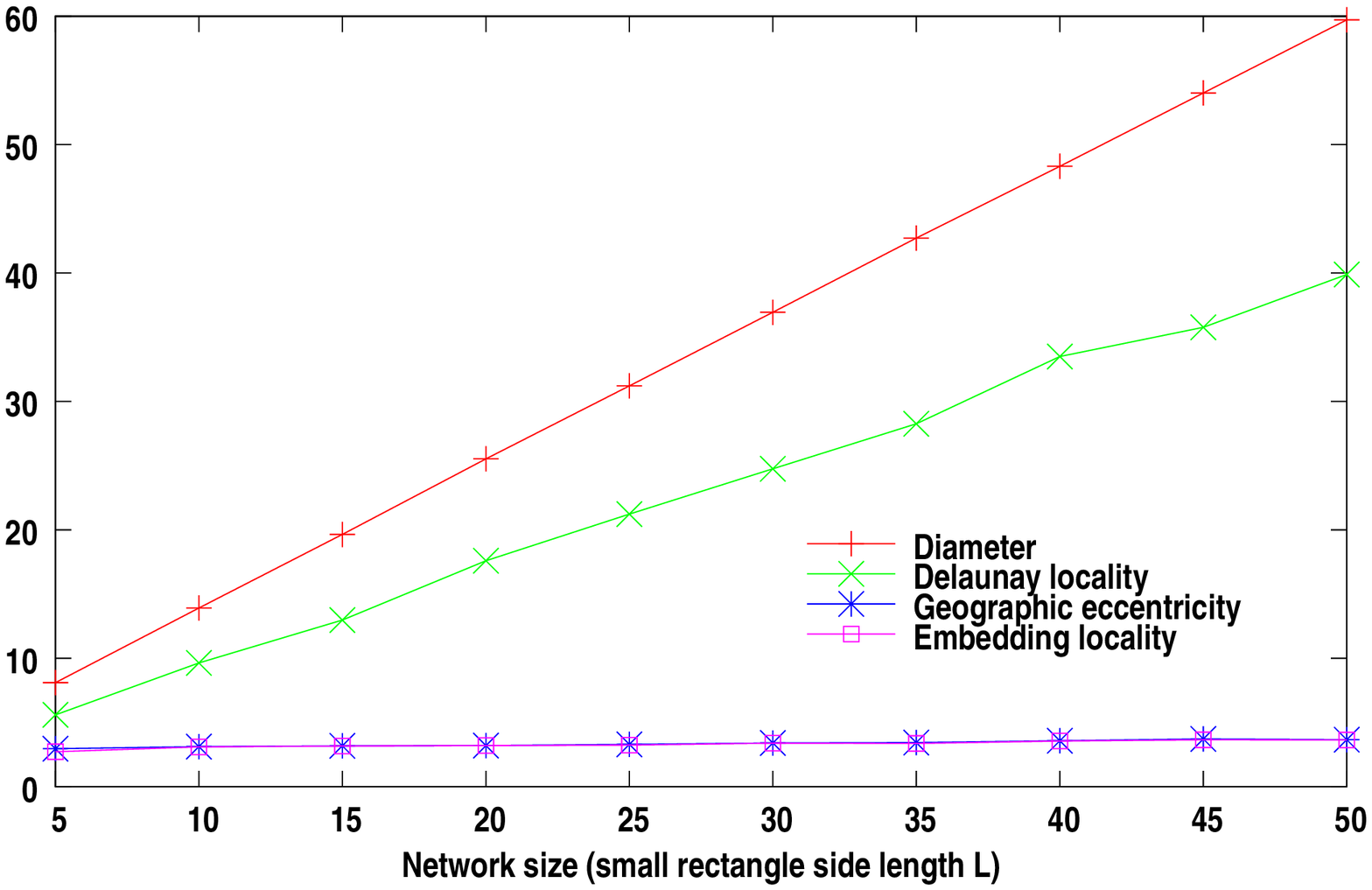}
~\\~\\
\begin{tabular}{|rr|rr|rr|rr|rr|}
\hline
\multicolumn{4}{|c|}{Simulation} & \multicolumn{6}{|c|}{Communication graph}\\
\multicolumn{2}{|c}{Net.} & \multicolumn{2}{c|}{Runs} & \multicolumn{2}{|c}{$D$} & \multicolumn{2}{c}{$N_1$} & \multicolumn{2}{c|}{$k_T$}\\
\hline
\multicolumn{1}{|c}{$L$} & \multicolumn{1}{c|}{$n$} & \multicolumn{1}{|c}{$\Sigma$} & \multicolumn{1}{c|}{$\Delta$} & \multicolumn{1}{|c}{$\overline{x}$} & \multicolumn{1}{c|}{$\sigma$} & \multicolumn{1}{|c}{$\overline{x}$} & \multicolumn{1}{c|}{$\sigma$} & \multicolumn{1}{|c}{$\overline{x}$} & \multicolumn{1}{c|}{$\sigma$}\\
5  &   100 & 102 &  2 &  8.10 & 0.54 & 11.65 & 0.70 &  5.59 & 0.99\\
10 &   400 & 102 &  2 & 13.91 & 0.60 & 13.85 & 0.32 &  9.63 & 1.77\\
15 &   900 & 104 &  4 & 19.64 & 0.57 & 14.58 & 0.23 & 12.97 & 2.31\\
20 &  1600 & 103 &  3 & 25.53 & 0.64 & 14.95 & 0.16 & 17.59 & 3.57\\
25 &  2500 & 105 &  5 & 31.21 & 0.71 & 15.23 & 0.14 & 21.23 & 4.13\\
30 &  3600 & 104 &  4 & 36.95 & 0.65 & 15.37 & 0.10 & 24.76 & 4.51\\
35 &  4900 & 109 &  9 & 42.72 & 0.78 & 15.48 & 0.09 & 28.26 & 6.36\\
40 &  6400 & 110 & 10 & 48.31 & 0.67 & 15.56 & 0.08 & 33.50 & 5.87\\
45 &  8100 & 110 & 10 & 54.02 & 0.72 & 15.64 & 0.06 & 35.77 & 6.95\\
50 & 10000 & 109 &  9 & 59.72 & 0.87 & 15.68 & 0.07 & 39.89 & 7.76\\
\hline
\end{tabular}\\
~\\~\\
\begin{tabular}{|r|rr|rr|rr|rr|rr|}
\hline
& \multicolumn{4}{|c|}{Embedding locality} & \multicolumn{4}{|c|}{Geographic eccentricity} & \multicolumn{2}{|c|}{Difference}\\
Net. & \multicolumn{2}{c}{$k_e$} & \multicolumn{2}{c|}{$N_{k_e}$} & \multicolumn{2}{c}{$k_g$} & \multicolumn{2}{c|}{$N_{k_g}$}& $dk$ & $dN$\\
\hline
\multicolumn{1}{|c|}{$L$} & \multicolumn{1}{|c}{$\overline{x}$} & \multicolumn{1}{c|}{$\sigma$} & \multicolumn{1}{|c}{$\overline{x}$} & \multicolumn{1}{c|}{$\sigma$} & \multicolumn{1}{|c}{$\overline{x}$} & \multicolumn{1}{c|}{$\sigma$} & \multicolumn{1}{|c}{$\overline{x}$} & \multicolumn{1}{c|}{$\sigma$} & \multicolumn{1}{|c}{$\sigma$} & \multicolumn{1}{c|}{$\sigma$}\\
5  & 2.72 & 0.49 &  55.20 &  11.20 & 2.95 & 0.30 &  60.49 &   7.43 & 0.42 & 9.76\\
10 & 3.09 & 0.32 & 130.93 &  19.83 & 3.12 & 0.32 & 132.69 &  19.83 & 0.17 & 10.08\\
15 & 3.17 & 0.40 & 176.52 &  42.71 & 3.18 & 0.41 & 177.58 &  43.63 & 0.10 & 10.54\\
20 & 3.20 & 0.45 & 202.04 &  59.63 & 3.20 & 0.45 & 202.04 &  59.63 & 0.00 & 0.00\\
25 & 3.24 & 0.43 & 221.40 &  62.29 & 3.29 & 0.45 & 228.90 &  66.79 & 0.22 & 32.73\\
30 & 3.39 & 0.49 & 254.76 &  77.79 & 3.40 & 0.49 & 256.35 &  78.08 & 0.10 & 15.78\\
35 & 3.36 & 0.56 & 259.65 & 100.31 & 3.43 & 0.57 & 271.42 & 102.35 & 0.26 & 42.90\\
40 & 3.55 & 0.55 & 298.53 &  99.95 & 3.58 & 0.55 & 303.69 &  99.24 & 0.17 & 29.32\\
45 & 3.64 & 0.52 & 319.82 &  95.46 & 3.71 & 0.50 & 332.37 &  91.25 & 0.26 & 45.73\\
50 & 3.63 & 0.48 & 322.13 &  88.43 & 3.66 & 0.47 & 327.59 &  86.80 & 0.17 & 31.08\\
\hline
\end{tabular}
\caption{Network metrics based on the SINR communication model with minimum range $1.2$ and maximum range $6$. A Gaussian localization error has been added with standard deviation $0.5$. Each network is composed of $n$ nodes scattered in a $L\times 4L$ rectangle.}
\label{exp_sinrScal}
\end{center}
\end{figure}